\documentclass[11pt,a4paper]{article}
\usepackage[margin=2.4cm]{geometry}
\usepackage{amsmath,amssymb,bm}
\usepackage{booktabs,tabularx,threeparttable}
\usepackage{graphicx}
\usepackage{float}
\usepackage{hyperref}
\usepackage{microtype}
\hypersetup{colorlinks=true,linkcolor=blue,citecolor=blue,urlcolor=blue}
\newcommand{\Rey}{\mathrm{Re}}

\usepackage{amsthm}
\usepackage{authblk}

\theoremstyle{plain}
\newtheorem{theorem}{Theorem}[section]
\newtheorem{lemma}[theorem]{Lemma}

\theoremstyle{definition}

\theoremstyle{remark}
\newtheorem{remark}{Remark}[section]

\title{Stochastic Representation-Inspired Physics-Informed Neural Networks for Incompressible Fluid Flows}

\author[1]{Zihao Guo}
\author[2,3]{Zhongmin Qian\thanks{Corresponding author.}}

\affil[1]{Institute for Advanced Research, Great Bay University, Dongguan, Guangdong, China\\ \href{mailto:guozh@gbu.edu.cn}{guozh@gbu.edu.cn}}

\affil[2]{Mathematical Institute, University of Oxford, Oxford OX2 6GG, United Kingdom}

\affil[3]{Oxford Suzhou Centre for Advanced Research, Suzhou, Jiangsu, China\\ \href{mailto:qianz@maths.ox.ac.uk}{qianz@maths.ox.ac.uk}}

\date{}

\begin{document}
\maketitle

\begin{abstract}
The effectiveness of neural networks in solving partial differential equations depends not only on their approximation capability but also on the suitability of their representations for the underlying solutions. Motivated by stochastic integral representations of incompressible fluid flows and the associated Biot--Savart kernel, we propose a Biot--Savart-informed physics-informed neural network (BS-PINN) for the incompressible Navier--Stokes equations. By extracting a rational profile from a regularized Biot--Savart kernel, we construct a smooth activation function with fixed and learnable kernel scales. Both variants retain the conventional velocity--pressure formulation and physics-informed training framework, allowing the influence of the activation function to be examined independently. We evaluate the method on five benchmarks, including Kovasznay flow, cylinder wake, three-dimensional Beltrami flow, Taylor--Green vortex, and double shear layer. Across 21 benchmark--architecture comparisons, both variants consistently reduce the combined velocity error relative to the conventional $\tanh$-based PINN, achieving mean relative error reductions of 31.7\% and 33.2\%, respectively. Improvements are also observed in reported pressure and vorticity errors. Notably, the double-shear-layer experiments demonstrate improved resolution of localized shear structures and, for BS-PINN-F, qualitatively clearer vortex-roll-up patterns in the long-time simulation. These findings highlight the potential of incorporating analytical flow structures into neural-network representations for incompressible flow simulations.
\end{abstract}


\section{Introduction}
\label{sec:introduction}

The use of neural networks for solving partial differential equations (PDEs) is fundamentally a problem of function approximation. Classical universal approximation theorems establish that feedforward neural networks can approximate continuous functions on compact domains to arbitrary accuracy under suitable assumptions~\cite{cybenko1989,hornik1989}. However, the existence of an accurate approximation does not guarantee that it can be achieved efficiently with a network of prescribed size or recovered through numerical optimization. In practice, the performance of neural PDE solvers depends not only on the expressive capacity of the network, but also on the optimization procedure and the suitability of its representation for the target solution~\cite{wang2021grad,krishnapriyan2021}. This suggests that knowledge of the underlying solution structure may help guide the choice of approximation functions and improve the effectiveness of neural-network representations.

This perspective is particularly relevant to scientific machine learning, where the functions to be approximated satisfy known physical laws. Physics-informed neural networks (PINNs) incorporate these laws into the training process by penalizing the residuals of governing equations together with initial and boundary constraints~\cite{raissi2019}. The incompressible Navier--Stokes equations have become an important application of this framework. Raissi et al.~\cite{raissi2020hfm} introduced hidden fluid mechanics to infer velocity and pressure fields from flow visualizations. Jin et al.~\cite{jin2021} developed Navier--Stokes flow networks (NSFnets) using velocity--pressure and vorticity--velocity formulations for laminar and turbulent flows. Xu et al.~\cite{xu2025preprocessing} investigated data preprocessing strategies for improving PINN-based flow-field reconstruction, while Lai et al.~\cite{lai2026} studied spatio-temporal super-resolution of velocity measurements using a vorticity--velocity physics-informed formulation. Generalization error estimates for PINNs have also been developed~\cite{mishra2023}, including results specifically for the incompressible Navier--Stokes equations~\cite{deryck2024}.

Despite these developments, accurately approximating flows with multiple spatial scales, strong velocity gradients, or evolving vortical structures remains challenging. Considerable effort has therefore been devoted to improving PINN training through loss balancing, optimization strategies, and adaptive sampling~\cite{wang2021grad,wang2022ntk,krishnapriyan2021,wu2023sampling,song2024loss}. Other studies have investigated the neural representation itself, including adaptive activation functions~\cite{jagtap2020adaptive}, Fourier features~\cite{wang2021fourier}, periodic activations~\cite{sitzmann2020}, physically motivated activation functions~\cite{abbasi2024paf}, and trainable sinusoidal activations for the Navier--Stokes equations~\cite{khademi2025tsa}. 
These developments highlight the importance of both optimization strategies and neural representation design. Nevertheless, the analytical structure of incompressible flow solutions offers an additional source of information for constructing problem-adapted neural features. In particular, stochastic integral representations reveal a connection between flow dynamics and the spatial kernels underlying the vorticity-to-velocity map.

Stochastic formulations of fluid equations have a substantial mathematical foundation. Constantin and Iyer~\cite{constantinIyer2008,constantinIyer2011} developed stochastic Lagrangian representations for the incompressible Navier--Stokes equations, including flows in domains with boundary. Eyink et al.~\cite{eyink2020part1,eyink2020part2} further investigated stochastic vorticity dynamics and their applications to wall-bounded turbulent flows. Cruzeiro and Qian~\cite{CruzeiroQian2014} established a backward stochastic differential equation formulation associated with the two-dimensional vorticity equation, providing a nonlinear Feynman--Kac-type representation. From a complementary forward perspective, Qian~\cite{Qian2022Wall} developed stochastic integral representations for incompressible viscous flows in wall-bounded regions, connecting the vorticity formulation with McKean--Vlasov-type diffusion processes. Qian, S\"uli, and Zhang~\cite{QianSuliZhang2022} derived a closed random-vortex dynamics system for three-dimensional incompressible flows by incorporating vortex stretching, together with conditional-law duality and forward stochastic representations. Related analytical tools were further developed by Qian and Shen~\cite{QianShen2023} in their study of vorticity, expansion rate, and stochastic representations for viscous compressible flows. These works establish mathematical connections between fluid dynamics and stochastic processes, allowing flow quantities to be characterized through integral representations involving diffusive particle trajectories.

Such representations have also provided the basis for a series of stochastic numerical methods. Qian et al.~\cite{QianQiuZhaoWu2022} developed Monte Carlo simulations for wall-bounded flows using the random vortex method. Cherepanov, Liu, and Qian~\cite{CherepanovLiuQian2024} investigated the dynamics of boundary vorticity and its numerical treatment, while Cherepanov et al.~\cite{CherepanovErtelQianWu2024} developed random-vortex dynamics and numerical schemes for wall-bounded viscous flows, including convergence analysis. Related developments have extended stochastic formulations and numerical methods to more general fluid systems. Li, Qian, and Xu~\cite{LiQianXu2023} introduced a twin Brownian particle method for Oberbeck--Boussinesq flows, and Guo, Qian, and Shen~\cite{GuoQianShen2025} developed a random-vortex and expansion-rate model for these flows. For three-dimensional incompressible flows, Guo and Qian~\cite{GuoQian2024LES} combined random-vortex methods with large-eddy simulation, and subsequently extended this approach to the tracking of Brownian fluid particles in large-eddy simulations~\cite{GuoQian2025Tracking}. These developments demonstrate how stochastic representations can be translated into numerical schemes for fluid-flow simulation. A related connection between stochastic vortex dynamics and neural computation was explored by Zhang et al.~\cite{zhang2022drvn}, who developed the deep random vortex network (DRVN) using a Monte Carlo-based learning objective derived from random-vortex dynamics. In contrast, our approach retains the standard Navier--Stokes residual-based training objective and uses the stochastic representation to motivate the choice of hidden-layer activation.

The motivation follows from the stochastic integral representation of the velocity field. Under appropriate regularity assumptions, the velocity of an incompressible flow can be expressed through integrals involving the Biot--Savart kernel and the vorticity transported along stochastic trajectories. In three dimensions, additional terms account for vortex stretching. Although the stochastic trajectories, vorticity distributions, and associated weights depend on the unknown flow, the spatial form of the Biot--Savart kernel is known explicitly. Thus, the integral representation reveals that the velocity field can be viewed, in an appropriate sense, as a superposition of structured kernel contributions.

This observation suggests a different perspective on neural approximation. Rather than relying exclusively on a generic activation function and allowing the network to learn the required spatial features through optimization, we may introduce a characteristic kernel profile directly into the approximation architecture. The trainable weights and biases can then determine the locations, scales, and combinations of these nonlinear features. The resulting network does not reproduce the stochastic integral representation exactly, but its functional form is motivated by the known structure of the governing fluid equations.

Following this idea, we propose a Biot--Savart-informed physics-informed neural network (BS-PINN). Starting from a regularized component of the Biot--Savart kernel, we construct the scalar activation
\begin{equation}
\phi_{\beta}(z)=\frac{2\beta z}{z^2+\beta^2+\delta}, \qquad \beta>0,\quad\delta>0,
\label{eq:intro_bs_activation}
\end{equation}
where $\beta$ controls the characteristic scale and $\delta$ is a small fixed stabilization parameter. This activation retains the odd rational profile of the selected kernel component. We consider two variants: BS-PINN-F, which uses a fixed kernel scale, and BS-PINN-L, which learns the scale jointly with the network parameters. Both variants retain the conventional velocity--pressure outputs and the Navier--Stokes residual-based training framework, so that the effect of the kernel-informed activation can be examined under otherwise matched experimental settings.

The proposed activation should be distinguished from a direct numerical implementation of the stochastic representation. It is motivated by a scalar profile extracted from a regularized two-dimensional kernel and does not evaluate the full Biot--Savart integral. Its applicability to three-dimensional flows is therefore examined numerically. The central objective of this study is to investigate whether information about the analytical structure of incompressible flows can improve their neural approximation.

The main contributions of this work are summarized as follows.

\begin{enumerate}
\item \textbf{Stochastic-representation-informed neural approximation.} We connect the stochastic integral representations of incompressible flows with the design of neural approximation functions. In particular, the Biot--Savart velocity kernel provides a mathematical motivation for constructing nonlinear features adapted to the Navier--Stokes equations.

\item \textbf{Biot--Savart-informed activation and network design.} We construct a smooth rational activation based on a regularized Biot--Savart kernel component and develop fixed-scale and learnable-scale variants, BS-PINN-F and BS-PINN-L. The proposed modification preserves the conventional PINN training framework, allowing the influence of the activation to be examined independently of changes to the governing-equation residuals.

\item \textbf{Numerical validation across incompressible flows.} We investigate five benchmark problems involving steady and unsteady dynamics, two- and three-dimensional flows, and smooth or localized vortical structures. With three double-shear-layer parameters and three network architectures, the experiments comprise 21 benchmark--architecture comparisons. Both BS-PINN variants achieve lower combined velocity errors than the conventional $\tanh$-based PINN in all 21 comparisons. We further examine pressure, vorticity, and flow-field reconstruction accuracy. Beyond error metrics, a separate long-time double-shear-layer experiment under sparse velocity supervision illustrates that BS-PINN-F resolves the characteristic development of spiral vortex-roll-up patterns more clearly than the conventional PINN.
\end{enumerate}

Table~\ref{tab:benchmark_overview} summarizes the five benchmark problems considered in this study: Kovasznay flow, flow past a circular cylinder, three-dimensional Beltrami flow, the Taylor--Green vortex, and the double shear layer. These benchmarks provide analytical or numerical reference solutions for evaluating the proposed method over a range of flow configurations.

\begin{table}[htbp]
\centering
\caption{Overview of the five incompressible-flow benchmarks considered in this study. For each benchmark, the references below the image indicate a classical source and representative neural-network-based studies.The schematic illustrations were generated using ChatGPT (OpenAI).}
\label{tab:benchmark_overview}
\renewcommand{\arraystretch}{1.3}
\setlength{\tabcolsep}{3pt}
\begin{tabular}{ccccc}
\toprule
\shortstack{Kovasznay flow} &
\shortstack{Cylinder wake} &
\shortstack{Beltrami flow} &
\shortstack{Taylor--Green vortex} &
\shortstack{Double shear layer} \\
\midrule
\includegraphics[width=0.18\textwidth,height=2.5cm,keepaspectratio]{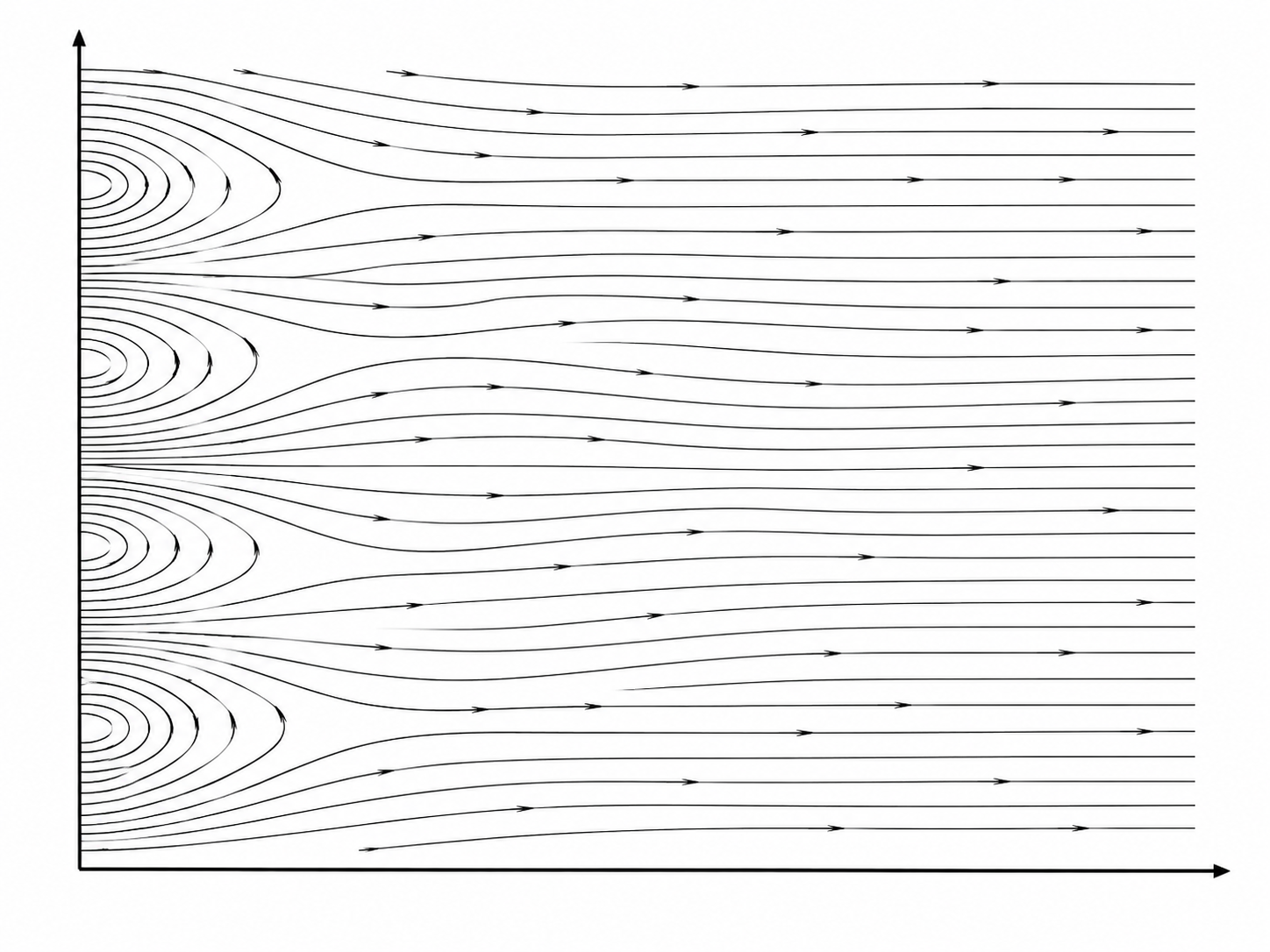} &
\includegraphics[width=0.18\textwidth,height=2.5cm,keepaspectratio]{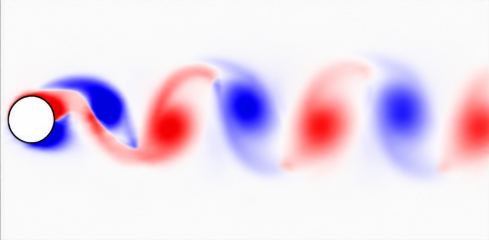} &
\includegraphics[width=0.18\textwidth,height=2.5cm,keepaspectratio]{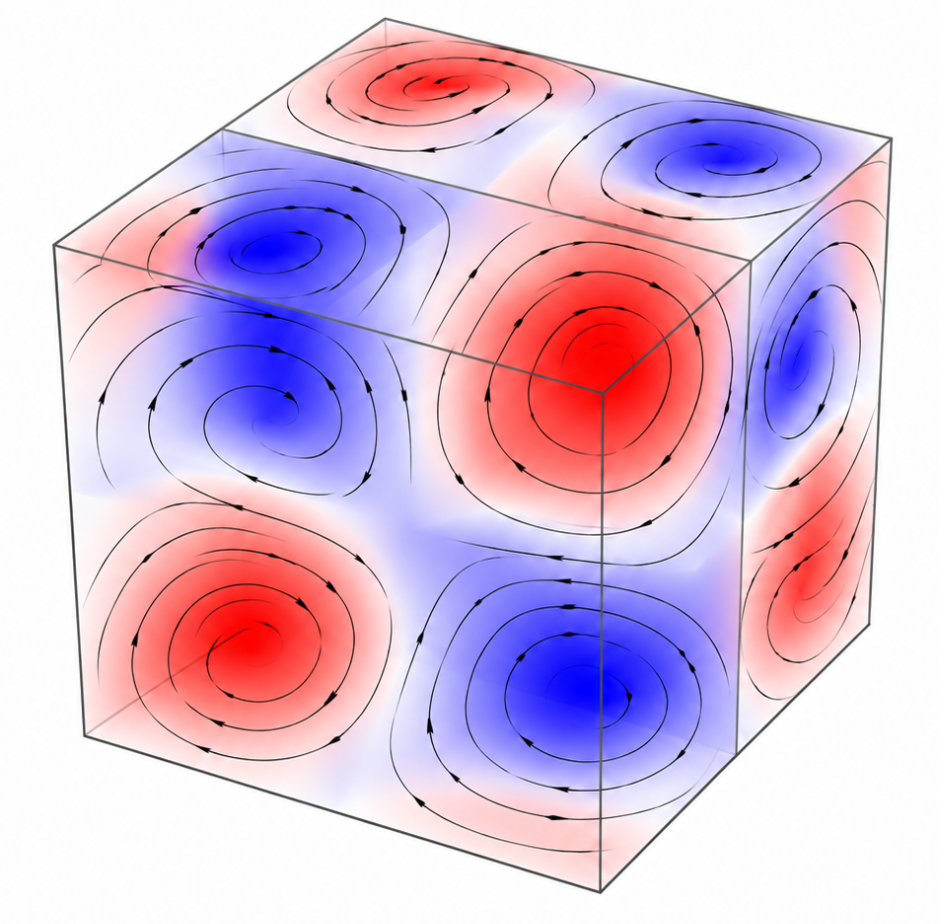} &
\includegraphics[width=0.18\textwidth,height=2.5cm,keepaspectratio]{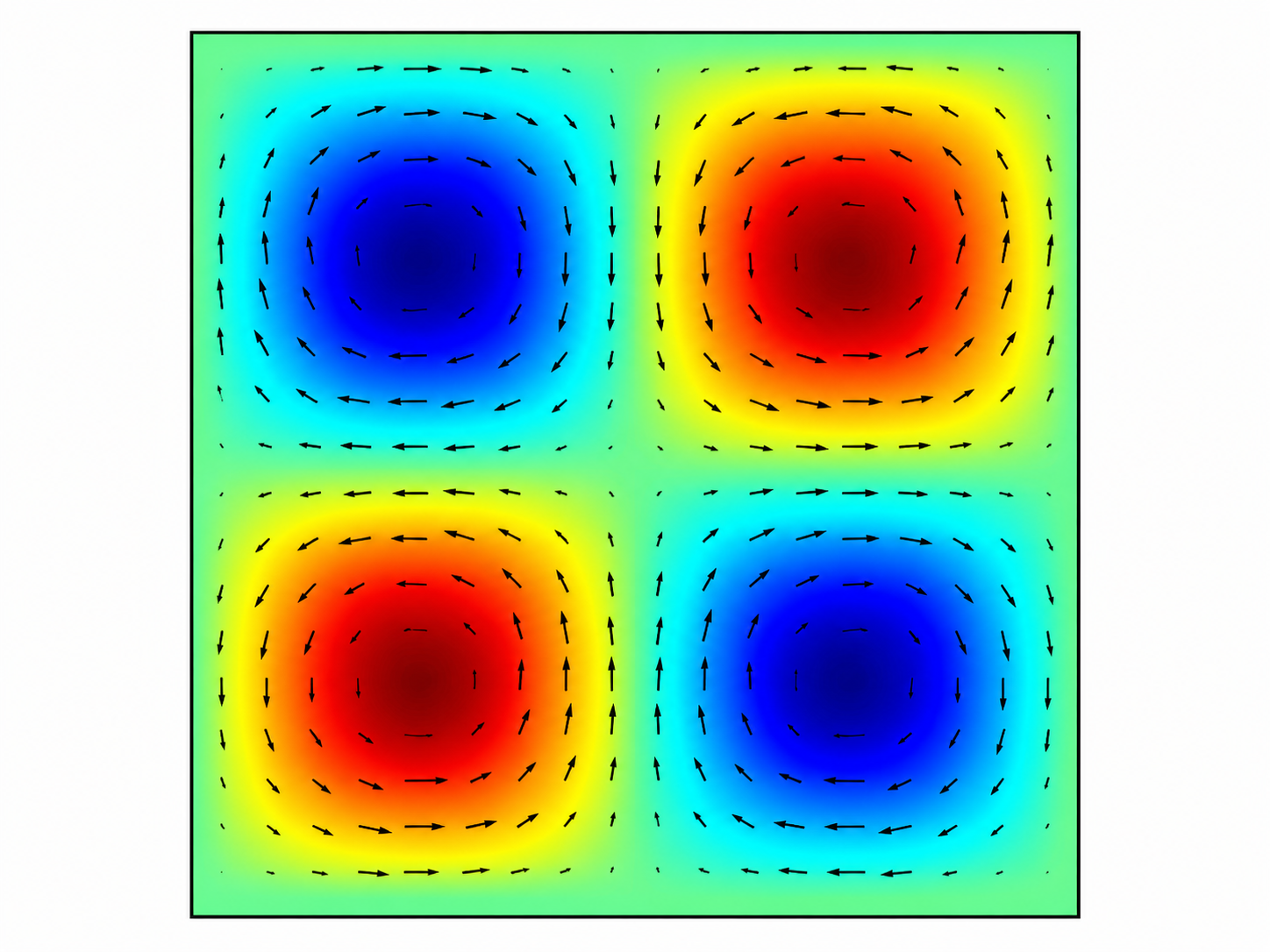} &
\includegraphics[width=0.18\textwidth,height=2.5cm,keepaspectratio]{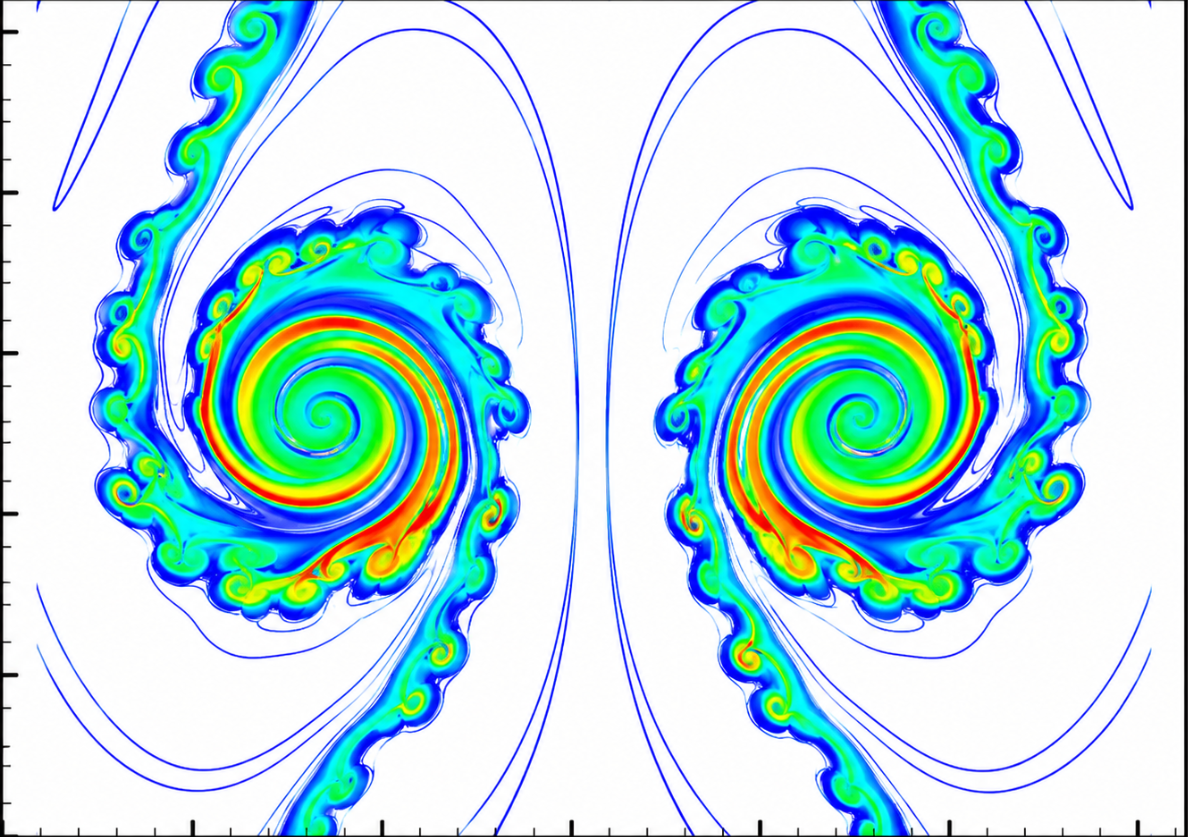} \\
\midrule

{\scriptsize \shortstack[c]{Proposed by \cite{kovasznay1948}\\ Studies: \cite{jin2021,wang2024iaw,lou2021bgk}}} &
{\scriptsize \shortstack[c]{Benchmark: \cite{schaefer1996}\\ Studies: \cite{raissi2019,jin2021,wang2024iaw,xu2023cylinder}}} &
{\scriptsize \shortstack[c]{Proposed by \cite{ethier1994}\\ Studies: \cite{jin2021,wang2024iaw,jnini2025rtnn}}} &
{\scriptsize \shortstack[c]{Classical: \cite{taylor1923}\\ Studies: \cite{lou2021bgk,wang2024iaw,puri2024tgv}}} &
{\scriptsize \shortstack[c]{Classical: \cite{bell1989}\\ Study: \cite{junk2026fno}}} \\

\bottomrule
\end{tabular}
\end{table}

\begin{figure}[H]
    \centering
    \includegraphics[width=\textwidth]{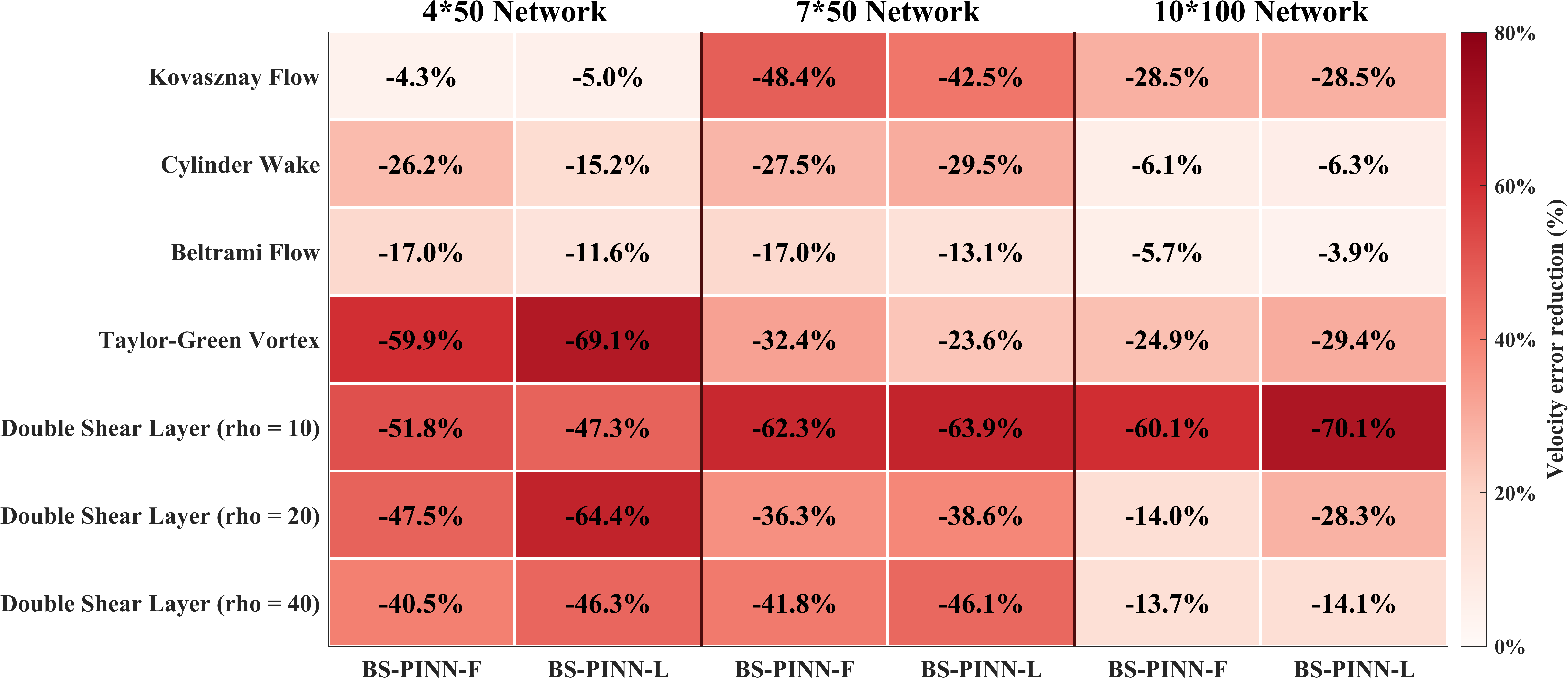}
    \caption{Overview of the relative changes in combined velocity error for BS-PINN-F and BS-PINN-L compared with the conventional PINN across the tested flow configurations and network architectures.}
    \label{fig:intro_velocity_overview}
\end{figure}

\begin{figure}[H]
    \centering
    \includegraphics[width=\textwidth]{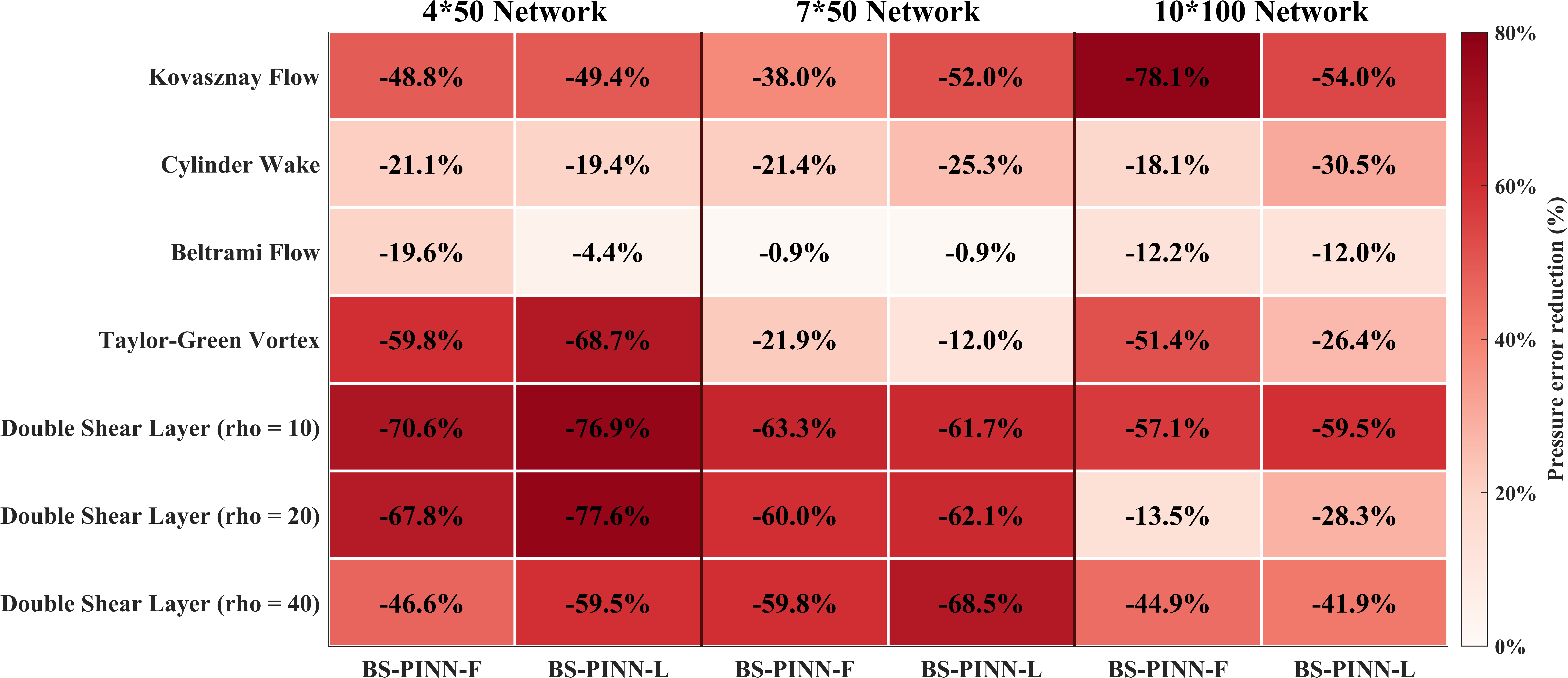}
    \caption{Overview of the relative changes in reported pressure error for BS-PINN-F and BS-PINN-L compared with the conventional PINN across the tested flow configurations and network architectures. The double-shear-layer configurations use reconstructed pressure, whereas the other benchmarks evaluate directly predicted pressure.}
    \label{fig:intro_pressure_overview}
\end{figure}

Figures~\ref{fig:intro_velocity_overview} and~\ref{fig:intro_pressure_overview} provide an overview of the numerical comparisons across the seven flow configurations and three network architectures. They summarize the relative performance of BS-PINN-F and BS-PINN-L against the conventional PINN in terms of combined velocity and reported pressure errors. Detailed quantitative results, including individual velocity components, vorticity, and temporal averages, are presented in Section~\ref{sec:results}.

The remainder of this paper is organized as follows. Section~\ref{sec:mathematical_foundations} reviews the stochastic integral representations of incompressible flows and develops the Biot--Savart-based motivation for the activation function. Section~\ref{sec:network_architecture} introduces the BS-PINN architecture and its physics-informed training formulation. Section~\ref{sec:benchmark} describes the benchmark problems and experimental settings. Section~\ref{sec:results} presents the numerical results and discusses the performance of the proposed networks. The final section concludes the paper, and additional flow-field visualizations are provided in \ref{representation}.

\section{Stochastic Representations of Incompressible Flows}
\label{sec:mathematical_foundations}

Stochastic formulations of incompressible fluid dynamics link the Navier--Stokes equations to forward diffusion processes and provide integral representations of the vorticity and velocity fields. The formulation is established using two principal analytical ingredients: conditional-law duality for diffusion bridges and a forward stochastic integral representation for parabolic systems~\cite{Qian2022Wall, QianSuliZhang2022}. Combined with the
Biot--Savart law, these results identify a singular spatial kernel that is determined by the governing equations rather than selected as an arbitrary approximation basis. This section recalls these established results and uses their kernel structure to motivate the activation function employed in our physics-informed neural representation.

Unless explicitly stated otherwise, we work on the whole space $\mathbb R^d$, $d\in\{2,3\}$, over a finite time interval $[0,T]$. The velocity and other coefficients are assumed sufficiently regular, and the fields have sufficient spatial decay and integrability to justify the probabilistic representations, integrations by parts, and interchanges of integrals below. Bounded and periodic domains require appropriate modifications of the spatial Green's function.

\subsection{Conditional-law duality for diffusion processes}
\label{subsec:conditional_duality}

Let $\boldsymbol b:\mathbb R^d\times[0,T]\to\mathbb R^d$ be a
sufficiently regular, bounded, divergence-free vector field, so that
$\nabla\cdot\boldsymbol b=0$. For $\nu>0$, consider the diffusion
generator
\begin{equation}
\mathcal L_{\boldsymbol b}=\nu\Delta+\boldsymbol b\cdot\nabla.
\label{eq:diffusion_generator}
\end{equation}
Integration by parts shows that its formal adjoint is
\begin{equation}
\mathcal L_{\boldsymbol b}^{*}
=\nu\Delta-\boldsymbol b\cdot\nabla
=\mathcal L_{-\boldsymbol b}.
\label{eq:adjoint_generator}
\end{equation}
The associated forward diffusion, initiated at $\boldsymbol\xi$ at
time $s$, is defined by
\begin{equation}
\mathrm d\boldsymbol X_r^{\boldsymbol\xi,s}
=\boldsymbol b(\boldsymbol X_r^{\boldsymbol\xi,s},r)\,\mathrm dr
+\sqrt{2\nu}\,\mathrm d\boldsymbol B_r, \quad \boldsymbol X_s^{\boldsymbol\xi,s}=\boldsymbol\xi,
\label{eq:general_diffusion}
\end{equation}
where $r\ge s$ and $\boldsymbol B$ is a standard $d$-dimensional Brownian motion. We write $p_{\boldsymbol b}(s,\boldsymbol\xi,t,\boldsymbol x)$ for its
transition density:
\begin{equation}
\mathbb P\bigl(\boldsymbol X_t^{\boldsymbol\xi,s}
\in\mathrm d\boldsymbol x\bigr)
=p_{\boldsymbol b}(s,\boldsymbol\xi,t,\boldsymbol x)
\,\mathrm d\boldsymbol x.
\end{equation}

For fixed endpoints $\boldsymbol\xi,\boldsymbol\eta\in\mathbb R^d$,
the pinned diffusion law (or diffusion-bridge law) is the
conditional path measure
\begin{equation}
\mathbb P_{\boldsymbol b}^{\boldsymbol\xi,0\to\boldsymbol\eta,T}
:=\mathbb P\bigl(
\,\cdot\,\bigm|\boldsymbol X_0=\boldsymbol\xi,
\boldsymbol X_T=\boldsymbol\eta\bigr).
\label{eq:pinned_measure}
\end{equation}
Define the time-reversed drift
$\boldsymbol b^T(\boldsymbol x,r)
=\boldsymbol b(\boldsymbol x,T-r)$ and the path-reversal map
$(\tau_T\gamma)(r)=\gamma(T-r)$. The conditional-law duality
of~\cite{QianSuliZhang2022} states that
\begin{equation}
\mathbb P_{\boldsymbol b}^{\boldsymbol\xi,0\to\boldsymbol\eta,T}
=
\mathbb P_{-\boldsymbol b^T}^{\boldsymbol\eta,0\to\boldsymbol\xi,T}
\circ\tau_T.
\label{eq:conditional_law_duality}
\end{equation}
In particular, the transition densities obey
\begin{equation}
p_{\boldsymbol b}(s,\boldsymbol\xi,t,\boldsymbol x)
=
p_{-\boldsymbol b^T}(T-t,\boldsymbol x,T-s,\boldsymbol\xi).
\label{eq:transition_density_duality}
\end{equation}

Applying \eqref{eq:transition_density_duality} to each factor which reverses their time order and gives the finite-dimensional distributions of the oppositely driven, time-reversed bridge. Thus the adjoint identity \eqref{eq:adjoint_generator} lifts to an identity of conditional path laws. Importantly, this permits the backward Feynman--Kac formulation of a forward parabolic equation to be re-expressed using forward diffusion trajectories.
\subsection{Stochastic integral representation theorem}
\label{subsec:vorticity_representation}

Consider the $m$-component parabolic system
\begin{equation}
\displaystyle
\partial_t\boldsymbol\Psi
=\nu\Delta\boldsymbol\Psi
-(\boldsymbol b\cdot\nabla)\boldsymbol\Psi
+\boldsymbol M\boldsymbol\Psi+\boldsymbol F,\quad
\boldsymbol\Psi(\boldsymbol x,0)=\boldsymbol\Psi_0(\boldsymbol x),
\label{eq:forward_parabolic_system}
\end{equation}
where $\boldsymbol M(\boldsymbol x,t)$ is an $m\times m$ matrix and
$\boldsymbol F(\boldsymbol x,t)$ is a source term. For notational
simplicity, let
$\boldsymbol X_r^{\boldsymbol\xi}
:=\boldsymbol X_r^{\boldsymbol\xi,0}$ be the forward diffusion
\eqref{eq:general_diffusion} starting at time 0. Along each trajectory define the
matrix-valued functional $Q^{\boldsymbol\xi,t}(r)$ by the
terminal-value ordinary differential equation
\begin{equation}
\frac{\mathrm d}{\mathrm dr}Q^{\boldsymbol\xi,t}(r)
=-Q^{\boldsymbol\xi,t}(r)
\boldsymbol M(\boldsymbol X_r^{\boldsymbol\xi},r),
\qquad Q^{\boldsymbol\xi,t}(t)=I_m.
\label{eq:matrix_functional_general}
\end{equation}
where $r<t$. 

Based on the conditional-law duality, the forward Feynman--Kac formula provides a probabilistic representation of the solution to the parabolic system. We recall the following result from~\cite{Qian2022Wall}.

\begin{lemma}[Stochastic integral representation]
\label{lem:forward_stochastic_representation}
Let $\boldsymbol\Psi$ be a sufficiently regular solution of
\eqref{eq:forward_parabolic_system} on $\mathbb R^d$,
with initial condition $\boldsymbol\Psi_0$.
Suppose that $\nabla\cdot\boldsymbol b=0$ and that
the preceding regularity and integrability assumptions hold.
Then, for any $\boldsymbol x\in\mathbb R^d$ and $t>0$,
the solution admits the following representation:
\begin{equation}
\begin{aligned}
\boldsymbol\Psi(\boldsymbol x,t)
={}&
\int_{\mathbb R^d}
p_{\boldsymbol b}(0,\boldsymbol\xi,t,\boldsymbol x)
\mathbb E\left[
Q^{\boldsymbol\xi,t}(0)\boldsymbol\Psi_0(\boldsymbol\xi)
\;\middle|\;
\boldsymbol X_t^{\boldsymbol\xi}=\boldsymbol x
\right]
\,\mathrm d\boldsymbol\xi
\\
&+
\int_0^t\int_{\mathbb R^d}
p_{\boldsymbol b}(0,\boldsymbol\xi,t,\boldsymbol x)
\mathbb E\left[
Q^{\boldsymbol\xi,t}(s)
\boldsymbol F(\boldsymbol X_s^{\boldsymbol\xi},s)
\;\middle|\;
\boldsymbol X_t^{\boldsymbol\xi}=\boldsymbol x
\right]
\,\mathrm d\boldsymbol\xi\,\mathrm ds.
\end{aligned}
\label{eq:stochastic_integral_representation}
\end{equation}
Here, $p_{\boldsymbol b}$ denotes the transition density
of the forward diffusion $\boldsymbol X^{\boldsymbol\xi}$,
and $Q^{\boldsymbol\xi,t}$ is the matrix-valued functional
defined above.
\end{lemma}

Lemma~\ref{lem:forward_stochastic_representation} is a forward Feynman--Kac representation underlying the random-vortex formulation~\cite{QianSuliZhang2022,Qian2022Wall}. The terminal-value matrix $Q$ accounts for the zeroth-order coupling
$\boldsymbol M\boldsymbol\Psi$, while the source appears through an
integral along the diffusion path. One way to obtain
\eqref{eq:stochastic_integral_representation} is to apply the
backward Feynman--Kac formula to the time-reversed problem, use
\eqref{eq:conditional_law_duality} to reverse the bridge, and
then apply Duhamel's principle. The identity
$\int_{\mathbb R^d}p_{\boldsymbol b}(0,\boldsymbol\xi,s,
\boldsymbol y)\,\mathrm d\boldsymbol\xi=1$ follows from
incompressibility and allows the source to be expressed in the same
forward-trajectory form.

We now apply this representation to incompressible Navier--Stokes flows:
\begin{equation}
\begin{cases}
\partial_t\boldsymbol u
 +(\boldsymbol u\cdot\nabla)\boldsymbol u
 =\nu\Delta\boldsymbol u-\nabla p+\boldsymbol g,\\
\nabla\cdot\boldsymbol u=0,
\end{cases}
\label{eq:ns_stochastic_foundation}
\end{equation}
where $\boldsymbol g$ is a prescribed external force (and may be set to zero). Let the Taylor diffusion be
\begin{equation}
\mathrm d\boldsymbol X_t^{\boldsymbol\xi}
=\boldsymbol u(\boldsymbol X_t^{\boldsymbol\xi},t)
\,\mathrm dt+\sqrt{2\nu}\,\mathrm d\boldsymbol B_t,
\qquad\boldsymbol X_0^{\boldsymbol\xi}=\boldsymbol\xi.
\label{eq:taylor_diffusion}
\end{equation}
Its transition density is denoted by $p_{\boldsymbol u}$.

\paragraph{Two-dimensional case.}
For $d=2$, define the scalar vorticity and the scalar source by
\begin{equation}
\omega=\partial_{x_1}u_2-\partial_{x_2}u_1,
\qquad f=\partial_{x_1}g_2-\partial_{x_2}g_1.
\end{equation}
Since vortex stretching is absent in two dimensions,
\begin{equation}
\partial_t\omega+\boldsymbol u\cdot\nabla\omega
=\nu\Delta\omega+f.
\label{eq:vorticity_2d}
\end{equation}
Taking $m=1$ and $M=0$ in
\eqref{eq:stochastic_integral_representation} gives
\begin{equation}
\begin{aligned}
\omega(\boldsymbol x,t)
={}&\int_{\mathbb R^2}p_{\boldsymbol u}
(0,\boldsymbol\xi,t,\boldsymbol x)\,\omega_0(\boldsymbol\xi)
\,\mathrm d\boldsymbol\xi+\int_0^t\!\int_{\mathbb R^2}
 p_{\boldsymbol u}(0,\boldsymbol\xi,t,\boldsymbol x)
 \mathbb E\!\left[
 f(\boldsymbol X_s^{\boldsymbol\xi},s)
 \mid\boldsymbol X_t^{\boldsymbol\xi}=\boldsymbol x\right]
 \mathrm d\boldsymbol\xi\,\mathrm ds.
\end{aligned}
\label{eq:vorticity_representation_2d}
\end{equation}
In the unforced case, $f=0$, so only the first integral remains.

\paragraph{Three-dimensional case.}
For $d=3$, set
$\boldsymbol\omega=\nabla\times\boldsymbol u$ and
$\boldsymbol f=\nabla\times\boldsymbol g$. Curling
\eqref{eq:ns_stochastic_foundation} yields
\begin{equation}
\partial_t\boldsymbol\omega
+(\boldsymbol u\cdot\nabla)\boldsymbol\omega
=\nu\Delta\boldsymbol\omega
+(\boldsymbol\omega\cdot\nabla)\boldsymbol u
+\boldsymbol f.
\label{eq:vorticity_equation_general}
\end{equation}

Let
\begin{equation}
A_{ij}=\partial_{x_j}u_i,
\qquad
S=\frac12(A+A^\top)
\end{equation}
denote the velocity-gradient tensor and
the rate-of-strain tensor, respectively.
Since the antisymmetric part of $A$
annihilates the vorticity vector,
the vortex-stretching term satisfies
\begin{equation}
(\boldsymbol\omega\cdot\nabla)\boldsymbol u
=A\boldsymbol\omega
=S\boldsymbol\omega.
\end{equation}

Following the strain-based random-vortex
formulation~\cite{QianSuliZhang2022},
we define the matrix-valued functional
$Q^{\boldsymbol\xi,t}(s)$ by
\begin{equation}
\frac{\mathrm d}{\mathrm ds}
Q^{\boldsymbol\xi,t}(s)
=
-Q^{\boldsymbol\xi,t}(s)
S(\bm{X}_s^{\bm{\xi}},s),
\qquad
Q^{\boldsymbol\xi,t}(t)=I_3.
\label{eq:gauge_functional}
\end{equation}

The resulting vorticity representation is
\begin{equation}
\begin{aligned}
\boldsymbol\omega(\boldsymbol x,t)
={}&\int_{\mathbb R^3}p_{\boldsymbol u}
(0,\boldsymbol\xi,t,\boldsymbol x)
\mathbb E\!\left[
Q^{\boldsymbol\xi,t}(0)\boldsymbol\omega_0(\boldsymbol\xi)
\mid\boldsymbol X_t^{\boldsymbol\xi}=\boldsymbol x
\right]\mathrm d\boldsymbol\xi\\
&+\int_0^t\!\int_{\mathbb R^3}p_{\boldsymbol u}
(0,\boldsymbol\xi,t,\boldsymbol x)
\mathbb E\!\left[
 Q^{\boldsymbol\xi,t}(s)
 \boldsymbol f(\boldsymbol X_s^{\boldsymbol\xi},s)
 \mid\boldsymbol X_t^{\boldsymbol\xi}=\boldsymbol x
\right]\mathrm d\boldsymbol\xi\,\mathrm ds.
\end{aligned}
\label{eq:vorticity_representation_3d}
\end{equation}
In contrast with the two-dimensional case, the matrix functional
$Q$ captures the deformation and stretching of vorticity along
random fluid trajectories. These formulas are exact representations
for the corresponding sufficiently regular solutions; they do not,
by themselves, constitute a neural-network approximation.

\subsection{Biot--Savart law and velocity representation }
\label{subsec:velocity_representation}

The Biot--Savart law provides an integral representation of
the incompressible velocity field in terms of its vorticity.
The corresponding kernels are determined by the fundamental
solutions of the Laplacian:
\begin{equation}
\Gamma_2(\boldsymbol z)
=\frac{1}{2\pi}\log|\boldsymbol z|,
\qquad
\Gamma_3(\boldsymbol z)
=-\frac{1}{4\pi|\boldsymbol z|},
\label{eq:laplacian_fundamental_solutions}
\end{equation}
Here, $\Delta\Gamma_d=\delta_0$ holds in the distributional sense, where $\delta_0$ denotes the Dirac distribution at the origin and is unrelated to the stabilization parameter $\delta$ introduced later.

We first recall the classical Biot--Savart law in two and
three dimensions.

\begin{lemma}[Biot--Savart law in the whole space]
\label{lem:biot_savart}
Let $d=2$ or $3$, and suppose that
$\boldsymbol u:\mathbb R^d\to\mathbb R^d$ is a sufficiently
smooth, divergence-free velocity field satisfying appropriate
decay and integrability conditions at infinity.

\noindent
(i) In two dimensions, let
$\omega=\partial_{x_1}u_2-\partial_{x_2}u_1$.
Then the velocity field satisfies
\begin{equation}
\boldsymbol u(\boldsymbol x)
=\int_{\mathbb R^2}
\boldsymbol K_2(\boldsymbol x-\boldsymbol y)
\omega(\boldsymbol y)\,\mathrm d\boldsymbol y,
\label{eq:biot_savart_2d_foundation}
\end{equation}
where
\begin{equation}
\boldsymbol K_2(\boldsymbol z)
=\frac{1}{2\pi}
\frac{\boldsymbol z^\perp}{|\boldsymbol z|^2},
\qquad
\boldsymbol z^\perp=(-z_2,z_1),
\quad \boldsymbol z\neq0.
\label{eq:biot_savart_kernel_2d}
\end{equation}

\noindent
(ii) In three dimensions, let
$\boldsymbol\omega=\nabla\times\boldsymbol u$.
Then the velocity field satisfies
\begin{equation}
\boldsymbol u(\boldsymbol x)
=
\int_{\mathbb R^3}
\mathcal K_3(\boldsymbol x-\boldsymbol y)
\boldsymbol\omega(\boldsymbol y)
\,\mathrm d\boldsymbol y,
\label{eq:biot_savart_3d_foundation}
\end{equation}
where the matrix-valued Biot--Savart kernel is defined by
\begin{equation}
\mathcal K_3(\boldsymbol z)
=
\frac{1}{4\pi|\boldsymbol z|^3}
\begin{pmatrix}
0 & z_3 & -z_2\\
-z_3 & 0 & z_1\\
z_2 & -z_1 & 0
\end{pmatrix},
\qquad \boldsymbol z\neq0.
\label{eq:biot_savart_kernel_3d}
\end{equation}
Equivalently, for any $\boldsymbol a\in\mathbb R^3$,
\begin{equation}
\mathcal K_3(\boldsymbol z)\boldsymbol a
=
\frac{\boldsymbol a\times\boldsymbol z}
{4\pi|\boldsymbol z|^3}.
\end{equation}

\begin{proof}
In two dimensions, introducing the stream function
$\psi$ such that
$\boldsymbol u=\nabla^\perp\psi$ and
$\Delta\psi=\omega$ gives
$\psi=\Gamma_2*\omega$.
Consequently,
$\boldsymbol u=\nabla^\perp\Gamma_2*\omega$,
which yields \eqref{eq:biot_savart_2d_foundation}.

In three dimensions, the identities
$\nabla\cdot\boldsymbol u=0$ and
$\boldsymbol\omega=\nabla\times\boldsymbol u$
imply
\begin{equation}
-\Delta\boldsymbol u
=\nabla\times\boldsymbol\omega.
\end{equation}
Applying the fundamental solution $\Gamma_3$ and
integrating by parts yields
\eqref{eq:biot_savart_3d_foundation}.
\end{proof}
\end{lemma}

Combining the Biot--Savart law with the stochastic
representations of the vorticity field derived above,
we obtain explicit stochastic integral representations
of the velocity field.

\begin{lemma}[Two-dimensional stochastic velocity representation]
\label{lem:stochastic_velocity_2d}
Let $\boldsymbol u$ be a sufficiently regular solution
of the two-dimensional incompressible Navier--Stokes
equations on $\mathbb R^2$, with initial vorticity
$\omega_0$ and scalar vorticity source term $f$.
Suppose that the assumptions of
Lemma~\ref{lem:biot_savart} and the forward stochastic
vorticity representation hold.

Then, for any $\boldsymbol x\in\mathbb R^2$ and $t>0$,
the velocity field admits the representation
\begin{equation}
\begin{aligned}
\boldsymbol u(\boldsymbol x,t)
={}&
\int_{\mathbb R^2}
\mathbb E\left[
\boldsymbol K_2
(\boldsymbol x-\boldsymbol X_t^{\boldsymbol\xi})
\right]
\omega_0(\boldsymbol\xi)
\,\mathrm d\boldsymbol\xi
\\
&+
\int_0^t\int_{\mathbb R^2}
\mathbb E\left[
\boldsymbol K_2
(\boldsymbol x-\boldsymbol X_t^{\boldsymbol\xi})
f(\boldsymbol X_s^{\boldsymbol\xi},s)
\right]
\,\mathrm d\boldsymbol\xi\,\mathrm ds,
\end{aligned}
\label{eq:stochastic_velocity_2d}
\end{equation}
where $\boldsymbol X^{\boldsymbol\xi}$ denotes the
forward Taylor diffusion driven by the velocity field.
\end{lemma}

\begin{proof}
Substituting the two-dimensional vorticity representation
\eqref{eq:vorticity_representation_2d}
into the Biot--Savart law
\eqref{eq:biot_savart_2d_foundation},
and applying Fubini's theorem together with the
transition-density identity and incompressibility,
we obtain \eqref{eq:stochastic_velocity_2d}.
\end{proof}

\begin{lemma}[Three-dimensional stochastic velocity representation]
\label{lem:stochastic_velocity_3d}
Let $\boldsymbol u$ be a sufficiently regular solution
of the three-dimensional incompressible Navier--Stokes
equations on $\mathbb R^3$, with initial vorticity
$\boldsymbol\omega_0$ and vector-valued vorticity
source term $\boldsymbol f$.
Suppose that the assumptions of
Lemma~\ref{lem:biot_savart} and the forward stochastic
vorticity representation hold.

Then, for any $\boldsymbol x\in\mathbb R^3$ and $t>0$,
the velocity field admits the representation
\begin{equation}
\begin{aligned}
\boldsymbol u(\boldsymbol x,t)
={}&
\int_{\mathbb R^3}
\mathbb E\Big[
\mathcal K_3
(\boldsymbol x-\boldsymbol X_t^{\boldsymbol\xi})
Q^{\boldsymbol\xi,t}(0)
\boldsymbol\omega_0(\boldsymbol\xi)
\Big]
\,\mathrm d\boldsymbol\xi
\\
&+
\int_0^t\int_{\mathbb R^3}
\mathbb E\Big[
\mathcal K_3
(\boldsymbol x-\boldsymbol X_t^{\boldsymbol\xi})
Q^{\boldsymbol\xi,t}(s)
\boldsymbol f(\boldsymbol X_s^{\boldsymbol\xi},s)
\Big]
\,\mathrm d\boldsymbol\xi\,\mathrm ds,
\end{aligned}
\label{eq:stochastic_velocity_3d}
\end{equation}
where $Q^{\boldsymbol\xi,t}(s)$ is the matrix-valued
functional defined in the three-dimensional vorticity
representation, accounting for vortex stretching.
\end{lemma}

\begin{proof}
The result follows by substituting the three-dimensional
vorticity representation
\eqref{eq:vorticity_representation_3d}
into the Biot--Savart law
\eqref{eq:biot_savart_3d_foundation},
followed by an interchange of integration and
conditional expectation under the stated assumptions.
\end{proof}

\begin{remark}
\label{rem:biot_savart_scope}
The preceding stochastic integral representations and
Biot--Savart laws are formulated in the whole-space setting.
Related stochastic formulations for wall-bounded incompressible flows and coupled fluid--thermal systems have also been studied, see \cite{constantinIyer2011,Qian2022Wall,LiQianXu2023}.
Since our primary objective is to motivate the construction
of kernel-informed neural networks, we restrict our
discussion to the whole-space setting, where the
integral kernels admit explicit analytical expressions.
Interested readers are referred to the aforementioned
works for further details on the corresponding
boundary representations.
\end{remark}

\subsection{Derivation of the Biot--Savart-informed activation}
\label{subsec:kernel_motivation}

The stochastic velocity representations established above express the incompressible velocity field through superpositions of
Biot--Savart kernels. These formulas suggest that the spatial profile of the kernel may provide a useful source of nonlinear features for approximating incompressible flows.

This observation motivates our activation
design. Rather than implementing the full
Biot--Savart operator, we extract a scalar
profile from a regularized component of the
two-dimensional kernel and use it as a
hidden-layer nonlinearity. The resulting
activation retains the odd rational structure
of the selected kernel component while
remaining suitable for standard fully
connected neural networks.

The proposed construction is therefore
motivated by the analytical representation
of the flow, but does not constitute an
exact discretization of the stochastic
integral formula. Whether this kernel-shaped
nonlinearity offers numerical advantages
over conventional activations is investigated
through the benchmark experiments below.

Because the singularity of $\boldsymbol K_2$ at the origin is
undesirable for a neural activation, introduce the regularized
planar kernel
\begin{equation}
\boldsymbol K_{2,\beta}(\boldsymbol z)
=\frac{1}{2\pi}
\frac{\boldsymbol z^\perp}{|\boldsymbol z|^2+\beta^2},
\qquad\beta>0.
\label{eq:regularized_biot_savart_foundation}
\end{equation}
This regularization is an approximation of the singular kernel,
not an identity for the original Biot--Savart operator.

For a scalar activation, consider the one-dimensional slice
$\boldsymbol z=(z,0)$. Its nonzero component is
\begin{equation}
[\boldsymbol K_{2,\beta}(z,0)]_2
=\frac{1}{2\pi}\frac{z}{z^2+\beta^2}.
\label{eq:kernel_component_slice}
\end{equation}
Multiplying this component by $4\pi\beta$ removes the physical
prefactor and normalizes its extrema to $\pm1$ when no further
regularization is present. Hence the normalized profile is
\begin{equation}
\phi_{\beta}^{(0)}(z)
=4\pi\beta[\boldsymbol K_{2,\beta}(z,0)]_2
=\frac{2\beta z}{z^2+\beta^2}.
\label{eq:normalized_kernel_profile}
\end{equation}
For numerical stability we include a small $\delta>0$ in the
denominator and use the activation
\begin{equation}
\phi_{\beta}(z)
=\frac{2\beta z}{z^2+\beta^2+\delta},
\qquad\beta>0.
\label{eq:kernel_activation_foundation}
\end{equation}
The function is smooth, odd, bounded, and has algebraic tails.
In particular,
\begin{equation}
\phi_{\beta}'(z)
=\frac{2\beta(\beta^2+\delta-z^2)}
{(z^2+\beta^2+\delta)^2},
\qquad
\sup_{z\in\mathbb R}|\phi_{\beta}(z)|
=\frac{\beta}{\sqrt{\beta^2+\delta}}\le1.
\label{eq:kernel_activation_properties}
\end{equation}
Thus $\beta$ controls the characteristic width of the profile,
it can be fixed or treated as a trainable positive parameter.
The three-dimensional Biot--Savart kernel has a different radial
scaling, proportional to
$z/(z^2+\beta^2)^{3/2}$ on a corresponding regularized
one-dimensional slice. Therefore
\eqref{eq:kernel_activation_foundation} is specifically derived
from the two-dimensional kernel component, not claimed to be an
exact one-dimensional reduction of the three-dimensional law.

In the proposed physics-informed neural network, the usual
componentwise activation (e.g., $\tanh$) is replaced by
$\phi_{\beta}$ in the selected hidden layers. This replaces a
generic nonlinearity by an odd kernel-shaped rational profile
suggested by the vorticity-to-velocity map. Crucially, the
resulting finite-dimensional network does not evaluate the
stochastic expectations in
\eqref{eq:stochastic_velocity_2d}--\eqref{eq:stochastic_velocity_3d},
and the activation itself neither guarantees incompressibility
nor imposes boundary conditions. Those properties are enforced
through the governing-equation, divergence, and boundary losses
in the subsequent PINN formulation. The stochastic representation
provides the mathematical motivation for the architecture,
its approximation accuracy and computational effectiveness must
be assessed independently.

\section{Physics-Informed Neural Network Architecture}
\label{sec:network_architecture}

In this section, we first introduce the conventional
physics-informed neural network (PINN) formulation
for the incompressible Navier--Stokes equations.
We then present the proposed Biot--Savart-informed
physics-informed neural network (BS-PINN), in which
the standard nonlinear activation is replaced by
a kernel-inspired activation motivated by
Section~\ref{subsec:kernel_motivation}.
Two variants are considered, depending on whether
the kernel scale parameter $\beta$ is fixed or trainable.

\subsection{Physics-informed neural networks}
\label{subsec:standard_pinn}

We consider the incompressible
Navier--Stokes equations
\begin{equation}
\partial_t\bm{u}
+(\bm{u}\cdot\nabla)\bm{u}
-\nu\Delta\bm{u}+\nabla p=\bm{0},\quad
\nabla\cdot\bm{u}=0,
\label{eq:ns}
\end{equation}
where $\bm{u}(\bm{x},t)\in\mathbb R^d$ and
$p(\bm{x},t)\in\mathbb R$ denote the velocity
and pressure fields, respectively,
$d\in\{2,3\}$ is the spatial dimension,
and $\nu$ is the nondimensional kinematic viscosity, with $\nu=1/\Rey$ under the adopted reference scales. For the steady Kovasznay flow, the time derivative in \eqref{eq:ns} is omitted.

We approximate the velocity and pressure
simultaneously using a fully connected
neural network:
\begin{equation}
N_{\theta}(\bm{\xi})
=
\left(
\bm{u}_{\theta}(\bm{\xi}),
p_{\theta}(\bm{\xi})
\right)
\in\mathbb R^{d+1},
\label{eq:vp_network}
\end{equation}
where $\theta$ denotes the trainable
network parameters. The network input is
$\bm{\xi}=\bm{x}$ for steady flows and
$\bm{\xi}=(\bm{x},t)$ for unsteady flows.

For a network with $L$ hidden layers,
the forward propagation is defined as
\begin{align}
\bm{h}^{(0)}
&=\bm{\xi},\\
\bm{h}^{(\ell)}
&=\sigma\left(
W^{(\ell)}\bm{h}^{(\ell-1)}
+\bm{b}^{(\ell)}
\right),
\quad \ell=1,\ldots,L,\\
N_{\theta}(\bm{\xi})
&=
W^{(L+1)}\bm{h}^{(L)}
+\bm{b}^{(L+1)},
\label{eq:pinn_architecture}
\end{align}
where $W^{(\ell)}$ and $\bm{b}^{(\ell)}$
represent the weight matrices and bias
vectors, respectively, and $\sigma$ denotes
the componentwise nonlinear activation function.
For the conventional PINN, we adopt the
hyperbolic tangent activation
$\sigma(z)=\tanh(z)$.

The Navier--Stokes equations are enforced
through the continuity and momentum
residuals, defined by
\begin{equation}
r_{\mathrm{div}}(\bm{\xi};\theta)
=
\nabla\cdot\bm{u}_{\theta},
\label{eq:continuity_residual}
\end{equation}
and
\begin{equation}
\bm{r}_{\mathrm{mom}}(\bm{\xi};\theta)
=
\partial_t\bm{u}_{\theta}
+
(\bm{u}_{\theta}\cdot\nabla)\bm{u}_{\theta}
+
\nabla p_{\theta}
-
\nu\Delta\bm{u}_{\theta}.
\label{eq:momentum_residual}
\end{equation}
For steady flows, the temporal derivative
is omitted. All derivatives are evaluated
using automatic differentiation.

Let $\mathcal X_r$ denote the set of interior
collocation points. We define the mean-squared
residual losses as
\begin{align}
\mathcal L_{\mathrm{mom}}
&=
\frac{1}{|\mathcal X_r|}
\sum_{\bm{\xi}\in\mathcal X_r}
\left\|
\bm{r}_{\mathrm{mom}}(\bm{\xi};\theta)
\right\|_2^2,\\
\mathcal L_{\mathrm{div}}
&=
\frac{1}{|\mathcal X_r|}
\sum_{\bm{\xi}\in\mathcal X_r}
\left|
r_{\mathrm{div}}(\bm{\xi};\theta)
\right|^2.
\end{align}
The PDE residual loss is their sum:
\begin{equation}
\mathcal L_{\mathrm{PDE}}
=
\mathcal L_{\mathrm{mom}}
+
\mathcal L_{\mathrm{div}}.
\label{eq:pde_loss}
\end{equation}

The initial and boundary conditions are
enforced, where applicable, through
\begin{align}
\mathcal L_{\mathrm{IC}}
&=
\frac{1}{|\mathcal X_0|}
\sum_{\bm{x}\in\mathcal X_0}
\left\|
\bm{u}_{\theta}(\bm{x},0)
-
\bm{u}_0(\bm{x})
\right\|_2^2,
\label{eq:initial_loss}\\
\mathcal L_{\mathrm{BC}}
&=
\frac{1}{|\mathcal X_b|}
\sum_{\bm{\xi}\in\mathcal X_b}
\left\|
\bm{u}_{\theta}(\bm{\xi})
-
\bm{u}_b(\bm{\xi})
\right\|_2^2,
\label{eq:boundary_loss}
\end{align}
where $\mathcal X_0$ and $\mathcal X_b$
denote the initial and boundary training
points, while $\bm{u}_0$ and $\bm{u}_b$
are the prescribed initial and boundary
velocity fields, respectively. The boundary
loss above corresponds to prescribed
velocity conditions.

When additional reference velocity observations
are used for training, we include the data loss
\begin{equation}
\mathcal L_{\mathrm{data}}
=
\frac{1}{|\mathcal X_{\mathrm{data}}|}
\sum_{\bm{\xi}\in\mathcal X_{\mathrm{data}}}
\left\|
\bm{u}_{\theta}(\bm{\xi})
-
\bm{u}^{\mathrm{data}}(\bm{\xi})
\right\|_2^2,
\label{eq:data_loss}
\end{equation}
where $\mathcal X_{\mathrm{data}}$ denotes the set of training data points and $\bm{u}^{\mathrm{data}}$ represents the corresponding reference velocity values.

The general training objective is
\begin{equation}
\begin{aligned}
\mathcal L_{\mathrm{total}}
={}&
\lambda_{\mathrm{PDE}}
\mathcal L_{\mathrm{PDE}}
+
\lambda_{\mathrm{IC}}
\mathcal L_{\mathrm{IC}}
+
\lambda_{\mathrm{BC}}
\mathcal L_{\mathrm{BC}}
\\
&+
\lambda_{\mathrm{per}}
\mathcal L_{\mathrm{per}}
+
\lambda_{\mathrm{data}}
\mathcal L_{\mathrm{data}},
\end{aligned}
\label{eq:total_loss}
\end{equation}

The nonnegative coefficients
$\lambda_{\mathrm{PDE}}$,
$\lambda_{\mathrm{IC}}$,
$\lambda_{\mathrm{BC}}$,
$\lambda_{\mathrm{per}}$, and
$\lambda_{\mathrm{data}}$
control the contributions of the
corresponding loss terms.

The periodicity loss
$\mathcal L_{\mathrm{per}}$
penalizes velocity mismatches at
corresponding points on opposite
periodic boundaries. Its explicit
form for the two-dimensional periodic
benchmarks is given in
Eq.~\eqref{eq:tgv_periodic_loss}.

Only the loss terms applicable to
each benchmark are included.
In particular, $\mathcal L_{\mathrm{data}}$
is used only when additional velocity
observations are available for training.

The same network outputs, governing residuals,
and training objective are used for the
conventional PINN and the proposed BS-PINN
variants. Their principal difference is
the choice of hidden-layer activation.

\subsection{Biot--Savart-informed physics-informed neural networks}
\label{subsec:bs_pinn}

Motivated by the stochastic representation
of incompressible flow and the associated
Biot--Savart kernel, we introduce the
Biot--Savart-informed physics-informed
neural network (BS-PINN).
It retains the network outputs and training
objective of Section~\ref{subsec:standard_pinn}
but replaces the hidden-layer activation.

Specifically, instead of employing the
conventional hyperbolic tangent activation,
we introduce the Biot--Savart-informed
activation
\begin{equation}
\phi_{\beta}(z)
=
\frac{2\beta z}
{z^2+\beta^2+\delta},
\qquad
\beta>0,\quad\delta>0,
\label{eq:bs_pinn_activation}
\end{equation}
motivated in
Section~\ref{subsec:kernel_motivation}.
Here, $\beta$ determines the characteristic
scale of the activation, while $\delta$
is a small, fixed numerical stabilization
parameter. Throughout the numerical experiments, the stabilization parameter is fixed at $\delta=10^{-12}$.

Accordingly, the hidden-layer transformation
becomes
\begin{equation}
\bm{h}^{(\ell)}
=
\phi_{\beta}\left(
W^{(\ell)}\bm{h}^{(\ell-1)}
+
\bm{b}^{(\ell)}
\right),
\qquad
\ell=1,\ldots,L,
\label{eq:bs_pinn_hidden_layer}
\end{equation}
where $\phi_{\beta}$ is applied componentwise.

The odd rational profile is motivated by
a regularized component of the Biot--Savart
kernel. The network combines such nonlinear
features through its trainable affine
transformations and output weights,
it does not directly evaluate a
Biot--Savart integral.

To investigate the influence of the kernel
scale, we consider two variants of BS-PINN.

\paragraph{BS-PINN-F.}
In the fixed-parameter variant, the kernel
scale $\beta=\beta_0>0$ is prescribed
before training and remains unchanged
throughout optimization.
The trainable parameters consist only
of the network weights and biases,
\begin{equation}
\theta_{\mathrm{F}}
=
\{W^{(\ell)},\bm{b}^{(\ell)}\}_{\ell=1}^{L+1}.
\label{eq:bs_pinn_fixed_parameters}
\end{equation}
This variant isolates the effect of replacing
the conventional activation with a
Biot--Savart-informed nonlinear profile.

\paragraph{BS-PINN-L.}
In the learnable-parameter variant,
the scale parameter $\beta$ is optimized
jointly with the network weights and biases.
To ensure its positivity, we introduce
an unconstrained trainable parameter
$\eta\in\mathbb R$ and define
\begin{equation}
\beta
=
\operatorname{softplus}(\eta)
=
\log(1+e^\eta)>0.
\label{eq:bs_pinn_trainable_beta}
\end{equation}
The trainable parameter set becomes
\begin{equation}
\theta_{\mathrm{L}}
=
\{W^{(\ell)},\bm{b}^{(\ell)}\}_{\ell=1}^{L+1}
\cup\{\eta\}.
\label{eq:bs_pinn_adaptive_parameters}
\end{equation}
The scale is therefore optimized together
with the network parameters.
If the desired initial scale is $\beta_0$,
the corresponding initialization is
$\eta_0=\log(e^{\beta_0}-1)$.

Both BS-PINN variants employ the residuals
and training objective in
Eqs.~\eqref{eq:pde_loss}--\eqref{eq:total_loss}.
Thus, the comparisons isolate the activation
profile and, within BS-PINN, the effect
of learning its scale.

\begin{figure}[H]
    \centering
    \includegraphics[width=\textwidth]{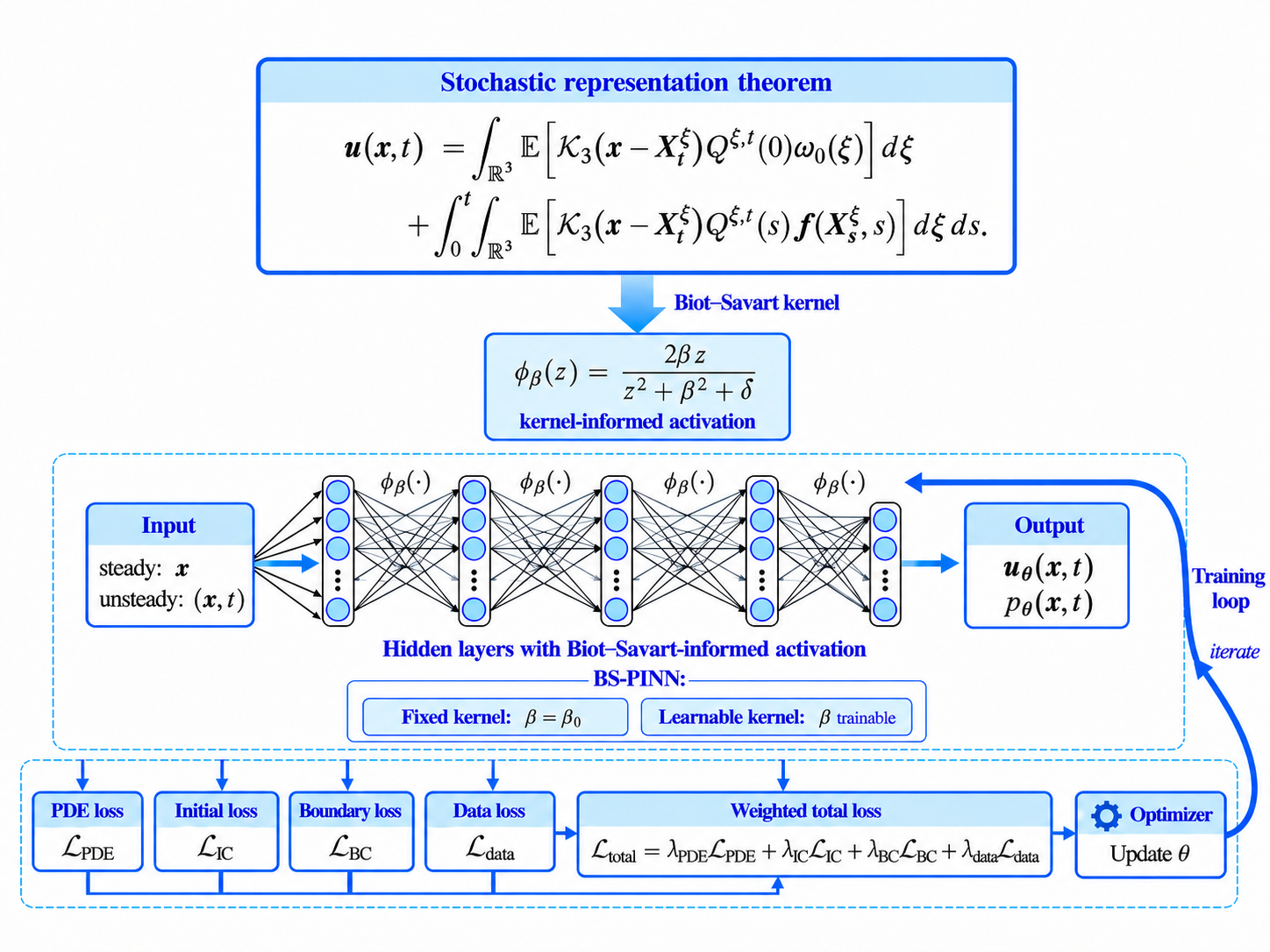}
    
\caption{Schematic of the BS-PINN architecture and its physics-informed training framework. The stochastic velocity representation provides the analytical motivation, while the scalar activation is derived from a regularized two-dimensional Biot--Savart kernel component. The displayed loss terms illustrate the general training procedure, periodicity constraints are additionally included when required by the benchmark.}

\end{figure}

\section{Benchmark tasks and experimental settings}
\label{sec:benchmark}

We evaluate the proposed Biot--Savart-informed
physics-informed neural network (BS-PINN) on
five incompressible-flow benchmarks:
the steady Kovasznay flow, the unsteady cylinder
wake, a three-dimensional Beltrami flow,
the Taylor--Green vortex, and a double shear layer.
These problems cover steady and unsteady dynamics,
analytical and numerical reference solutions,
two- and three-dimensional flows,
periodic boundary conditions, vortical structures,
and localized shear gradients.

\subsection{Common formulation and comparison protocol}
\label{par:benchmark_protocol}

All benchmarks solve the incompressible
Navier--Stokes equations in
Eq.~\eqref{eq:ns}, with zero external forcing.
For the steady Kovasznay flow, the
temporal derivative is omitted.

The network takes the spatial coordinates
as inputs for steady flows and both spatial
coordinates and time for unsteady flows.
It predicts the velocity components and
pressure simultaneously, yielding
$(u,v,p)$ in two dimensions and
$(u,v,w,p)$ in three dimensions.

We compare the conventional PINN with
the fixed-kernel and learnable-kernel
variants, denoted by BS-PINN-F and
BS-PINN-L, respectively.
The notation $L\times W$ denotes a
fully connected network with $L$
hidden layers and $W$ neurons per layer.
Three architectures are considered:
$4\times50$, $7\times50$, and $10\times100$.

The training objective follows Eq.~\eqref{eq:total_loss}. For each benchmark, all five loss weights are specified explicitly, with zero indicating that the corresponding loss term is omitted from the training objective. The PDE loss weight is fixed at $\lambda_{\mathrm{PDE}}=1$ throughout the five main benchmark comparisons. The separate long-time double-shear-layer experiment additionally incorporates velocity-data supervision and is described independently.

Within each benchmark and architecture,
the three models use identical sampling
points, loss weights, and optimization
schedules. The hidden-layer activation
is the architectural modification,
with BS-PINN-L additionally optimizing
its kernel scale.

BS-PINN-F uses $\beta=2$, while
BS-PINN-L is initialized at
$\beta=2$ before its scale is updated
during training. The activation-scale
parameter $\beta$ is distinct from
the loss-weight coefficients.

\paragraph{Error metrics.}

The accuracy of the predicted flow fields is evaluated
using the relative discrete $L^2$ error. Let
$\{\bm{x}_i\}_{i=1}^{N}$ denote the spatial evaluation
points at time $t$, and let $q_\theta$ and $q^*$
represent the predicted and reference fields, respectively.
For a scalar quantity $q$, the relative error is defined as
\begin{equation}
E_q(t)=
\frac{
\left(
\sum_{i=1}^{N}
|q_\theta(\bm{x}_i,t)-q^*(\bm{x}_i,t)|^2
\right)^{1/2}
}{
\left(
\sum_{i=1}^{N}
|q^*(\bm{x}_i,t)|^2
\right)^{1/2}
}.
\label{eq:benchmark_scalar_error}
\end{equation}
This definition applies to the individual velocity
components $u$, $v$, and $w$ (in three dimensions),
as well as to pressure and scalar vorticity.

Since pressure is determined only up to an additive constant, a gauge correction is applied before evaluating its error. The predicted pressure is shifted to match the spatial mean of the reference pressure. For time-dependent problems, this alignment is performed independently at each evaluation snapshot. Vorticity is computed from spatial derivatives of the predicted velocity using automatic differentiation. For the double-shear-layer benchmark, the pressure error is evaluated using pressure reconstructed from the predicted velocity through the periodic pressure Poisson equation, rather than the network's direct pressure output. The corresponding error is denoted by $E_p^{\mathrm{rec}}$.

For vector-valued quantities, the relative error
is computed by combining all components:
\begin{equation}
E_{\bm q}(t)=
\frac{
\left(
\sum_{i=1}^{N}
\|\bm q_\theta(\bm{x}_i,t)
-\bm q^*(\bm{x}_i,t)\|_2^2
\right)^{1/2}
}{
\left(
\sum_{i=1}^{N}
\|\bm q^*(\bm{x}_i,t)\|_2^2
\right)^{1/2}
}.
\label{eq:benchmark_vector_error}
\end{equation}
This definition is used for the combined velocity
error $E_{\bm u}$, with $\bm u=(u,v)$ in two
dimensions and $\bm u=(u,v,w)$ in three dimensions.
It also applies to three-dimensional vorticity
when its vector error is reported. In particular,
the combined velocity error is not the arithmetic
mean of the component-wise velocity errors.

For unsteady problems, we report errors at selected
time instants and their arithmetic mean over the
evaluation snapshots:
\begin{equation}
\overline{E}_q
=
\frac{1}{N_t}
\sum_{j=1}^{N_t}E_q(t_j),
\label{eq:benchmark_error}
\end{equation}
where $N_t$ denotes the number of evaluation
snapshots and $q$ may represent either a scalar
or vector-valued field. All errors are reported
as percentages. The average error is calculated
from the snapshot-wise relative errors rather
than from a single relative $L^2$ norm over the
entire space--time domain.

\subsection{Benchmark tasks}
\label{subsec:benchmark_tasks}

\paragraph{Kovasznay flow.}
\label{par:case_kovasznay}

The first benchmark is a two-dimensional, steady
analytical solution at $\Rey=40$ ($\nu=1/40$),
following \cite{jin2021}. Its exact fields are
\begin{equation}
\begin{aligned}
&u(x,y)
=1-e^{\lambda x}\cos(2\pi y),\\
&v(x,y)
=\frac{\lambda}{2\pi}e^{\lambda x}\sin(2\pi y),\\
&p(x,y)
=\frac12(1-e^{2\lambda x}),\\
&\lambda
=\frac{\Rey}{2}
-\sqrt{\frac{\Rey^2}{4}+4\pi^2}.
\end{aligned}
\label{eq:kovasznay_exact}
\end{equation}

The spatial domain is
$\Omega=[-0.5,1]\times[-0.5,1.5]$.
Exact velocity values prescribe the boundary
conditions, the steady problem has no initial
condition. The network maps $(x,y)$ to $(u,v,p)$. Training uses 2,601 interior collocation points and 400 boundary points. The loss weights are $\lambda_{\mathrm{PDE}}=1$, $\lambda_{\mathrm{IC}}=0$, $\lambda_{\mathrm{BC}}=1$, $\lambda_{\mathrm{per}}=0$, and $\lambda_{\mathrm{data}}=0$.

Optimization first uses 110,000 Adam updates,
with learning rates $10^{-3}$, $10^{-4}$,
$10^{-5}$, and $10^{-6}$ over successive
blocks of 5,000, 5,000, 50,000, and 50,000
updates. After Adam, the $\tanh$ PINN and both
BS-PINN variants are further trained with
L-BFGS for 5,000 epochs. The L-BFGS stage
is part of the Kovasznay protocol.

\paragraph{Cylinder wake.}
\label{par:case_cylinder}

The second benchmark considers the two-dimensional
unsteady flow past a circular cylinder at
$\Rey=100$ ($\nu=0.01$), following the cylinder-wake
benchmark in \cite{jin2021}.
The reference velocity and pressure fields are obtained
from the publicly available dataset%
\footnote{\url{https://github.com/maziarraissi/PINNs/blob/master/main/Data/cylinder_nektar_wake.mat}},
provided in the PINNs repository of Raissi et al.
\cite{raissi2019}.
The dataset contains numerical flow-field snapshots
characterizing the unsteady wake and vortex shedding
behind the cylinder.
We consider the downstream computational domain $\Omega=[1,8]\times[-2,2]$ over the time interval $t\in[0,7]$. This domain is a subregion of the original cylinder-wake dataset and does not explicitly include the cylinder surface. The network takes $(x,y,t)$ as inputs and predicts $(u,v,p)$.

Reference velocity data are prescribed only at the
initial time and on the four boundaries of the
computational domain, while no interior velocity
or pressure labels are used during training.

The training set consists of 140,000 space--time
collocation points for the PDE residuals,
21,300 boundary samples, and 5,000 initial samples.
The loss weights are $\lambda_{\mathrm{PDE}}=1$, $\lambda_{\mathrm{IC}}=1$, $\lambda_{\mathrm{BC}}=1$, $\lambda_{\mathrm{per}}=0$, and $\lambda_{\mathrm{data}}=0$.
For evaluation, we use 70 temporal snapshots at
$t=0,0.1,\ldots,6.9$.
The predicted velocity components and pressure are
compared against the reference fields using the
relative $L^2$ errors defined above, with pressure
gauge correction applied at each snapshot.
We report errors at selected time instants
and their arithmetic averages over all 70 snapshots.
This benchmark assesses the ability of the models
to reconstruct the evolving wake dynamics and
vortex-shedding structures using the governing
equations and velocity data restricted to the
initial and boundary conditions.

All models are trained using 110,000 Adam updates.
The learning rate is decreased successively from
$10^{-3}$ to $10^{-4}$, $10^{-5}$, and $10^{-6}$,
with 5,000, 5,000, 50,000, and 50,000 updates
at the respective learning rates.
No L-BFGS refinement is performed for this benchmark.

\paragraph{Three-dimensional Beltrami flow.}
\label{par:case_beltrami}

The third benchmark is a three-dimensional
unsteady analytical solution at
$\Rey=1$ ($\nu=1$).
Write
$\bm{u}(x,y,z,t)=e^{-t}\bm{F}(x,y,z)$,
where
\begin{equation}
\bm{F}
=
-\begin{pmatrix}
e^x\sin(y+z)+e^z\cos(x+y)\\
e^y\sin(z+x)+e^x\cos(y+z)\\
e^z\sin(x+y)+e^y\cos(z+x)
\end{pmatrix},
\qquad
p=-\frac12 e^{-2t}\|\bm{F}\|^2.
\label{eq:beltrami_exact}
\end{equation}

The problem is posed on
$[-1,1]^3\times[0,1]$.
Analytical velocity supplies the initial
and prescribed boundary data.
The network maps $(x,y,z,t)$ to
$(u,v,w,p)$.
Training uses 70,000 PDE points,
59,400 boundary points, and 29,791
initial points. The loss weights are $\lambda_{\mathrm{PDE}}=1$, $\lambda_{\mathrm{IC}}=100$, $\lambda_{\mathrm{BC}}=100$, $\lambda_{\mathrm{per}}=0$, and $\lambda_{\mathrm{data}}=0$. The PDE loss includes all three momentum residuals and the continuity residual. Predictions are evaluated against the analytical velocity and mean-aligned pressure at 11 equally spaced time instants, $t=0,0.1,\ldots,1.0$. The average errors are computed as the arithmetic mean of the snapshot-wise relative $L^2$ errors, following Eq.~\eqref{eq:benchmark_error}.

The BS activation is used as a nonlinear
representation in this three-dimensional
task, its two-dimensional kernel motivation
does not imply an exact three-dimensional
Biot--Savart inversion.

Optimization uses 110,000 Adam updates,
with learning rates $10^{-3}$, $10^{-4}$,
$10^{-5}$, and $10^{-6}$ over successive
blocks of 5,000, 5,000, 50,000, and 50,000
updates. No L-BFGS stage is used in this task.

\paragraph{Taylor--Green vortex.}
\label{par:case_taylor_green}

The fourth benchmark uses the two-dimensional
Taylor--Green vortex on the periodic domain
$\Omega=[0,2\pi]^2$.
Its exact solution is
\begin{equation}
\begin{aligned}
u(x,y,t)
&=\sin x\cos y\,e^{-2\nu t},\\
v(x,y,t)
&=-\cos x\sin y\,e^{-2\nu t},\\
p(x,y,t)
&=\frac14(\cos 2x+\cos 2y)e^{-4\nu t},\\
\omega(x,y,t)
&=\partial_x v-\partial_y u
=2\sin x\sin y\,e^{-2\nu t}.
\end{aligned}
\label{eq:tgv_exact}
\end{equation}

The principal comparison uses
$\nu=0.05$ ($\Rey=20$ when $U_{\mathrm{ref}}=L_{\mathrm{ref}}=1$)
and $t\in[0,1]$.
The network maps $(x,y,t)$ to $(u,v,p)$.
We sample 20,000 interior space--time points,
2,000 initial points, and 2,000 matched
boundary pairs in each periodic direction.

The loss weights are $\lambda_{\mathrm{PDE}}=1$, $\lambda_{\mathrm{IC}}=10$, $\lambda_{\mathrm{BC}}=0$, $\lambda_{\mathrm{per}}=1$, and $\lambda_{\mathrm{data}}=0$. The periodicity loss $\mathcal L_{\mathrm{per}}$ enforces velocity matching at opposite periodic boundaries and is defined by

\begin{equation}
\begin{aligned}
\mathcal L_{\mathrm{per}}
={}&
\frac{1}{N_x}
\sum_{j=1}^{N_x}
\left\|
\bm{u}_{\theta}(0,y_j,t_j)
-
\bm{u}_{\theta}(2\pi,y_j,t_j)
\right\|_2^2
\\
&+
\frac{1}{N_y}
\sum_{j=1}^{N_y}
\left\|
\bm{u}_{\theta}(x_j,0,t_j)
-
\bm{u}_{\theta}(x_j,2\pi,t_j)
\right\|_2^2,
\end{aligned}
\label{eq:tgv_periodic_loss}
\end{equation}
where $N_x=N_y=2000$ denote the numbers
of paired boundary points in the two
spatial directions. Optimization uses 110,000 Adam updates
with mini-batches of 1,024 interior points,
256 initial points, and 256 pairs per
periodic direction.
The learning rates are $10^{-3}$,
$10^{-4}$, $10^{-5}$, and $10^{-6}$
over successive blocks of 5,000, 5,000,
50,000, and 50,000 updates.
No L-BFGS stage is used in this task.

The exact solution is used for the initial
velocities and evaluation, but not as a
later-time interior training target.
We evaluate on a $256\times256$ grid,
excluding the repeated periodic endpoint,
at $t=0,0.25,0.5,0.75,1$.
Reported quantities include velocity, mean-aligned pressure, and vorticity errors, with vorticity evaluated from the spatial derivatives of the predicted velocity.

\paragraph{Double shear layer.}
\label{par:case_double_shear}

The fifth benchmark is a viscous
two-dimensional double shear layer
on the periodic domain
$\Omega=[0,2\pi]^2$.
Its initial velocity is
\begin{equation}
u(x,y,0)=
\begin{cases}
\tanh\!\bigl(\rho(y-\pi/2)\bigr),
& y\leq\pi,\\
\tanh\!\bigl(\rho(3\pi/2-y)\bigr),
& y>\pi,
\end{cases}
\qquad
v(x,y,0)
=
\varepsilon_{\mathrm{pert}}\sin x.
\label{eq:dsl_initial}
\end{equation}

The parameter $\rho$ controls the
initial shear-layer steepness,
whereas $\varepsilon_{\mathrm{pert}}$
specifies the amplitude of the
transverse perturbation.
In the short-time experiments,
we set $\nu=0.01$,
$\varepsilon_{\mathrm{pert}}=0.05$,
$\rho\in\{10,20,40\}$, and
$t\in[0,0.25]$.

The network takes $(x,y,t)$ as inputs
and predicts $(u,v,p)$.
The three compared models share
20,000 interior points, 2,000 initial
points, and 2,000 matched boundary
pairs in each periodic direction,
as well as the same initial linear-layer
weights. The loss weights are $\lambda_{\mathrm{PDE}}=1$, $\lambda_{\mathrm{IC}}=10$, $\lambda_{\mathrm{BC}}=0$, $\lambda_{\mathrm{per}}=1$, and $\lambda_{\mathrm{data}}=0$, with the periodicity loss $\mathcal L_{\mathrm{per}}$ defined in Eq.~\eqref{eq:tgv_periodic_loss}.

The fixed kernel uses $\beta=2$,
and the learnable kernel is initialized
at $\beta=2$.

Optimization uses 110,000 Adam updates
with mini-batches of 1,024 interior points,
256 initial points, and 256 pairs per
periodic direction.
The learning rates are $10^{-3}$,
$10^{-4}$, $10^{-5}$, and $10^{-6}$
over successive blocks of 5,000, 5,000,
50,000, and 50,000 updates.
No L-BFGS stage is used in this
short-time study.

\noindent\textbf{DNS reference and convergence assessment.}
No analytical solution is available after
$t=0$, so we generate the reference fields
independently with a Fourier pseudospectral
solver in the vorticity--streamfunction
formulation,
\begin{equation}
\partial_t\omega+\bm{u}\cdot\nabla\omega
=\nu\Delta\omega,\qquad
\Delta\psi=\omega,\qquad
\bm{u}=(-\partial_y\psi,\partial_x\psi).
\label{eq:dsl_dns_vorticity}
\end{equation}
The solver evaluates spatial derivatives
spectrally, applies the two-thirds truncation
to the nonlinear term, and advances in time
with a three-stage low-storage Runge--Kutta
scheme and a CFL-controlled time step.
It obtains a zero-mean reference pressure
from the periodic pressure Poisson equation
\begin{equation}
\Delta p
=
-\sum_{i,j=1}^{2}
(\partial_i u_j)(\partial_j u_i),
\qquad
\langle p\rangle_{\Omega}=0.
\label{eq:dsl_dns_pressure}
\end{equation}

The saved reference contains
$u,v,p,\omega$ and the four first-order
velocity derivatives at
$t=0,0.05,0.10,0.15,0.20,0.25$.
The reference simulations use $512^2$
grid points for $\rho=10$ and $1024^2$
grid points for $\rho=20,40$,
with a maximum time step of
$4\times10^{-4}$.
These DNS snapshots are used exclusively
for post-training evaluation.

We assess the spatial and temporal convergence
of the DNS reference separately.
At each saved time, we compare
$u$, $v$, $\omega$, and mean-aligned $p$
using the relative $L^2$ error defined
in Eq.~\eqref{eq:benchmark_scalar_error}.
For spatial refinement, the fine-grid
solution is restricted to the coarse grid
before computing the relative difference.
We compare $256^2$ and $512^2$ grid points
for $\rho=10$, and $512^2$ and $1024^2$
grid points for $\rho=20,40$,
with the maximum time step fixed at
$8\times10^{-4}$.
For temporal refinement, we fix the
fine spatial grid and reduce the
maximum time step from $8\times10^{-4}$
to $4\times10^{-4}$.

A relative difference below $0.5\%$
is adopted as the convergence criterion
for each evaluated field and snapshot.
All spatial- and temporal-refinement
comparisons satisfy this criterion,
indicating that the DNS reference
solutions are sufficiently resolved
at the selected spatial and temporal
resolutions. These refinement differences
assess the numerical convergence of
the reference solutions and are distinct
from the prediction errors of the
neural-network models.

We compare the network predictions with the DNS reference using velocity and vorticity errors, reconstructed pressure errors, and spatial flow-field visualizations.
When evaluating pressure reconstructed
from the predicted velocity, we solve
the periodic pressure Poisson equation
and align the spatial mean of the
reconstructed pressure with that of
the DNS reference at each snapshot.
The corresponding relative error
is denoted by $E_p^{\mathrm{rec}}$.

\section{Results}
\label{sec:results}

\subsection{Overall comparison}
\label{subsec:results_overall}

We first compare the performance of the proposed
BS-PINN variants with the conventional PINN across
the five benchmark problems described in
Section~\ref{sec:benchmark}. Including three
shear-layer steepness parameters, the comparison
comprises seven flow configurations, each evaluated
using $4\times50$, $7\times50$, and $10\times100$
networks. The conventional PINN with a $\tanh$
activation serves as the baseline, while BS-PINN-F
and BS-PINN-L employ fixed and learnable kernel
parameters, respectively. Within each comparison,
the network architecture, training samples,
physical constraints, and optimization budget
are kept identical.

To quantify the improvement over the baseline,
we define the relative error reduction as
\begin{equation}
\mathcal R_q^{(m)}
=
\frac{
E_q^{\mathrm{PINN}}-E_q^{(m)}
}{
E_q^{\mathrm{PINN}}
}
\times 100\%,
\qquad
m\in\{\mathrm{F},\mathrm{L}\},
\label{eq:relative_error_reduction}
\end{equation}
where $E_q^{(m)}$ and $E_q^{\mathrm{PINN}}$ denote the reported errors of model $m$ and the conventional PINN, respectively. For time-dependent benchmarks, these errors correspond to the snapshot averages defined in Eq.~\eqref{eq:benchmark_error}. In the pressure comparisons, the reported metric is the mean-aligned direct pressure error $E_p$ for the first four benchmarks and the reconstructed pressure error $E_p^{\mathrm{rec}}$ for the double-shear-layer configurations. A positive value of $\mathcal R_q^{(m)}$ indicates improved accuracy relative to the conventional PINN.

\begin{figure}[H]
    \centering
    \includegraphics[width=\textwidth]{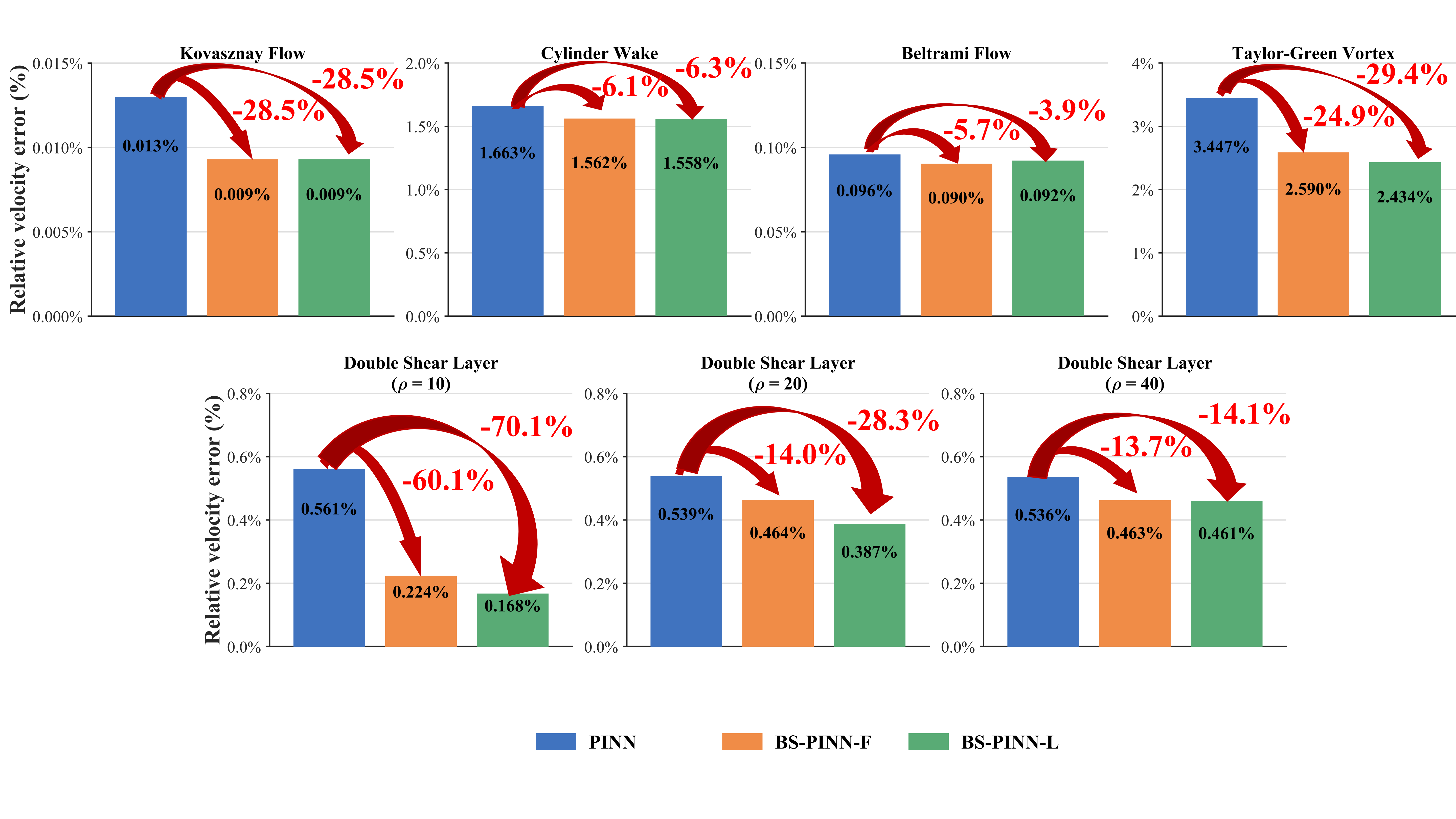}
    \caption{Comparison of combined velocity errors for the $10\times100$ architecture. Percentage annotations show the signed relative error changes with respect to the conventional PINN; negative values indicate error reductions.}
    \label{fig:velocity_1000}
\end{figure}

\begin{figure}[H]
    \centering
    \includegraphics[width=\textwidth]{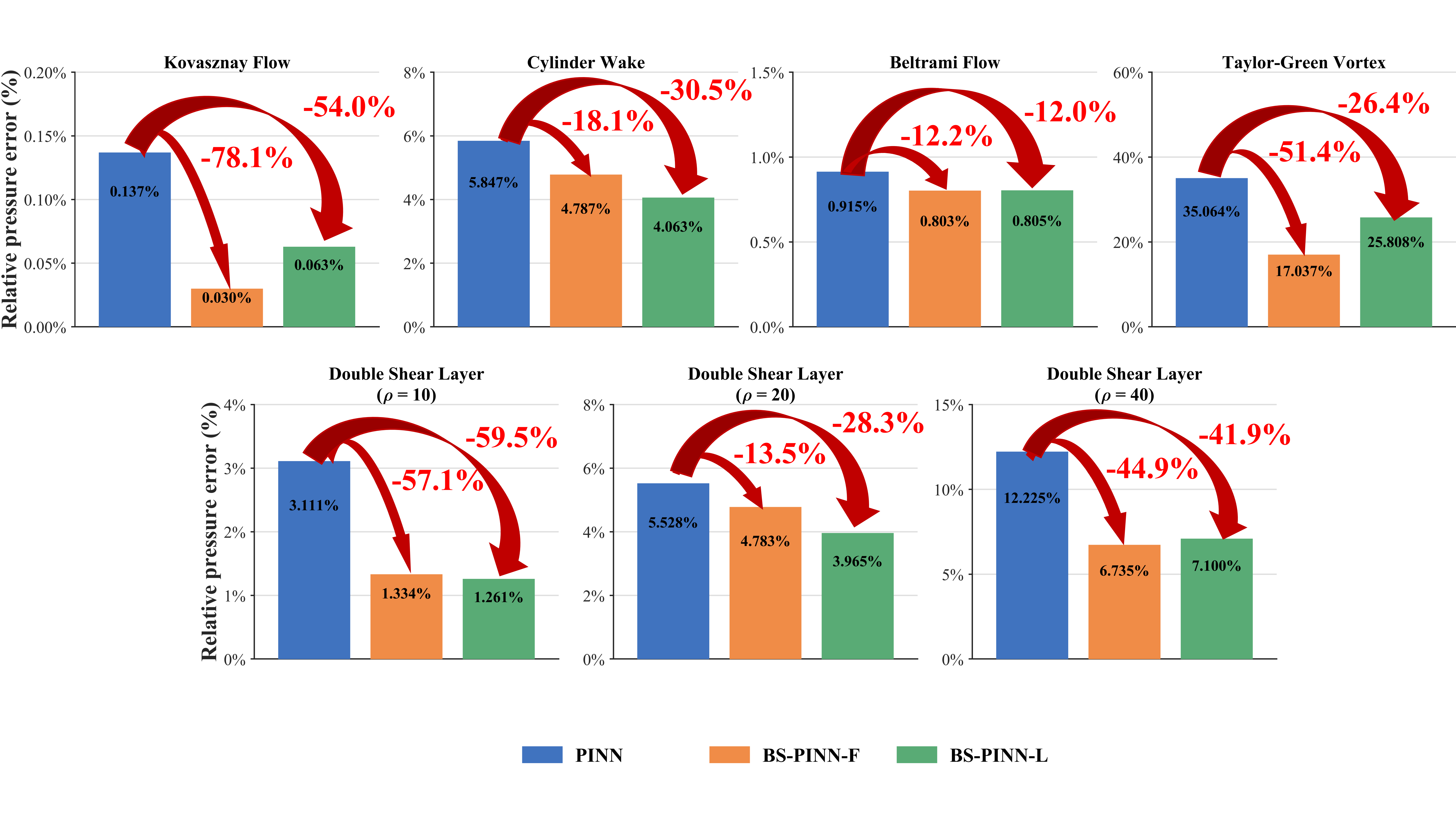}
    \caption{Comparison of reported pressure errors for the $10\times100$ architecture. Percentage annotations show the signed relative error changes with respect to the conventional PINN; negative values indicate error reductions.}
    \label{fig:pressure_1000}
\end{figure}

Figures~\ref{fig:velocity_1000} and~\ref{fig:pressure_1000} present the comparisons of combined velocity and reported pressure errors, respectively, for the $10\times100$ network architecture. The corresponding results for the $4\times50$ and $7\times50$ architectures are provided in \ref{representation}. The percentage annotations indicate signed relative error changes, corresponding to $-\mathcal R_q^{(m)}$ as defined in Eq.~\eqref{eq:relative_error_reduction}, negative values therefore indicate error reductions relative to the conventional PINN. For the double-shear-layer configurations, the reported pressure errors are based on pressure reconstructed from the predicted velocity rather than the network's direct pressure output.

For the combined velocity error, both BS-PINN
variants achieve positive $\mathcal R_{\bm u}^{(m)}$
in all 21 benchmark--architecture combinations.
The same holds for $\mathcal R_p^{(m)}$.
To summarize the overall improvements, we compute
the unweighted mean of the relative error reductions
across these 21 combinations:
\begin{equation}
\left\langle \mathcal R_q^{(m)} \right\rangle
=
\frac{1}{21}
\sum_{k=1}^{21}
\mathcal R_{q,k}^{(m)}.
\label{eq:mean_error_reduction}
\end{equation}
For BS-PINN-F and BS-PINN-L, the resulting mean
reductions in combined velocity error are
31.7\% and 33.2\%, respectively, while the
corresponding mean reductions in reported pressure
error are 41.7\% and 42.4\%.

These averages assign equal weight to each
benchmark--architecture combination rather
than pooling errors from different physical
problems. The improvements are particularly
pronounced in several periodic-flow configurations,
although the relative performance of fixed
and learnable kernels varies across different
problems and network architectures.

For conciseness, the flow-field visualizations in the following subsections focus on BS-PINN-F with the $10\times100$ architecture. Additional visualizations for the $4\times50$ and $7\times50$ architectures are provided in \ref{representation}. Nevertheless, all three architectures and activation functions are included in the quantitative comparisons throughout this section.

\subsection{Kovasznay flow}
\label{subsec:results_kovasznay}

The steady Kovasznay flow offers a first test
against an analytical solution at $\Rey=40$.
Table~\ref{tab:kovasznay} shows that all three
models approximate the velocity accurately,
with combined velocity errors below $0.07\%$
for every architecture. Nevertheless, both
BS-PINN variants consistently improve upon
the $\tanh$ baseline.

\begin{table}[htbp]
\centering
\small
\begin{threeparttable}
\caption{Kovasznay flow. Relative $L^2$ errors (\%)
for different network architectures and
activation functions.}
\label{tab:kovasznay}

\begin{tabular}{llrrrr}
\toprule
Architecture & Activation
& $E_u$ & $E_v$ & $E_{\bm{u}}$ & $E_p$ \\
\midrule

$4\times50$ & PINN
& 0.0570\% & 0.3990\% & 0.0656\%
& 1.1230\% \\
& BS-PINN-F
& \textbf{0.0560\%} & 0.3480\%
& 0.0628\% & 0.5750\% \\
& BS-PINN-L
& \textbf{0.0560\%} & \textbf{0.3360\%}
& \textbf{0.0623\%} & \textbf{0.5680\%} \\

\midrule

$7\times50$ & PINN
& 0.0480\% & 0.3530\%
& 0.0560\% & 0.9110\% \\
& BS-PINN-F
& \textbf{0.0250\%} & \textbf{0.1770\%}
& \textbf{0.0289\%} & 0.5650\% \\
& BS-PINN-L
& 0.0280\% & 0.1950\%
& 0.0322\% & \textbf{0.4370\%} \\

\midrule

$10\times100$ & PINN
& 0.0110\% & 0.0850\%
& 0.0130\% & 0.1370\% \\
& BS-PINN-F
& \textbf{0.0080\%} & \textbf{0.0570\%}
& \textbf{0.0093\%} & \textbf{0.0300\%} \\
& BS-PINN-L
& \textbf{0.0080\%} & 0.0580\%
& \textbf{0.0093\%} & 0.0630\% \\

\bottomrule
\end{tabular}

\begin{tablenotes}[flushleft]
\footnotesize
\item Errors are relative $L^2$ errors in percent.
$E_{\bm{u}}$ denotes the combined two-component
velocity error. Pressure is mean-aligned before
computing $E_p$. Bold values indicate the lowest
error among the three activation functions for
the same architecture and error metric.
Values tied at the reported precision
are highlighted together.
\end{tablenotes}

\end{threeparttable}
\end{table}

For the $4\times50$ network, the combined velocity
error decreases from $0.0656\%$ for PINN to
$0.0628\%$ and $0.0623\%$ for BS-PINN-F and
BS-PINN-L. The advantage is more substantial for
$7\times50$: the fixed kernel reduces the error
from $0.0560\%$ to $0.0289\%$ (48.4\%), while
the learnable kernel gives $0.0322\%$ (42.5\%).
At $10\times100$, both variants attain $0.0093\%$,
compared with $0.0130\%$ for PINN.

Improvements in pressure are also evident.
The lowest pressure errors for the three
architectures are $0.5680\%$, $0.4370\%$,
and $0.0300\%$, compared with baseline errors
of $1.1230\%$, $0.9110\%$, and $0.1370\%$,
respectively.

\begin{figure}[H]
    \centering
    \includegraphics[width=0.8\textwidth]{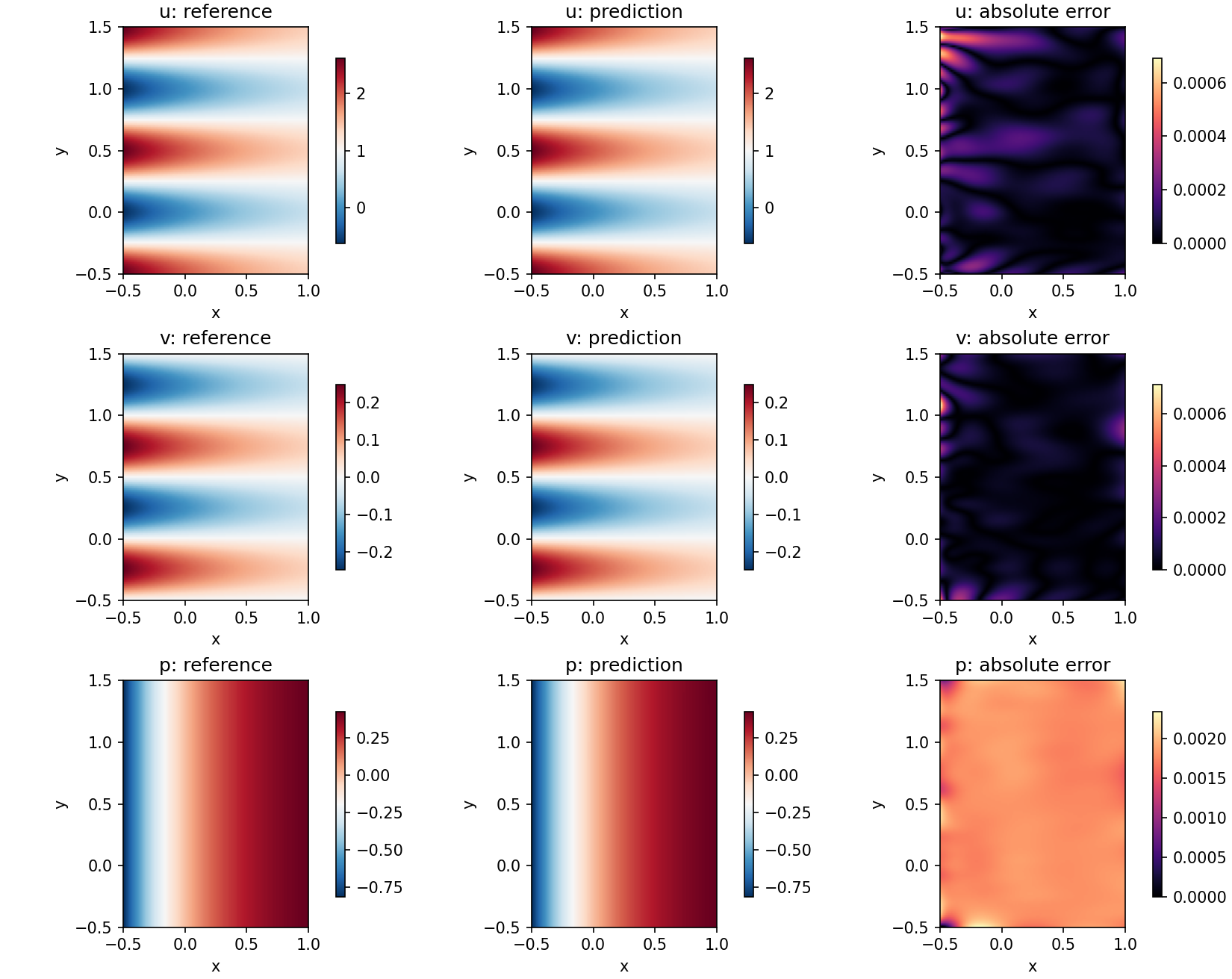}
    \caption{Kovasznay flow: analytical reference,
    BS-PINN-F prediction, and pointwise absolute
    errors using the $10\times100$ architecture.
    The three rows correspond to $u$, $v$,
    and $p$, respectively.}
    \label{fig:kovasznay1000}
\end{figure}

Figure~\ref{fig:kovasznay1000} compares the
analytical reference, the BS-PINN-F prediction,
and the pointwise absolute errors for the
$10\times100$ architecture. The predicted
velocity and pressure fields closely follow
their analytical counterparts. Because the
baseline itself is already accurate in this
smooth steady setting, the numerical improvement
is most appropriately viewed as a further
reduction of approximation error rather than
recovery of a flow structure absent from the
baseline.

\subsection{Cylinder wake}
\label{subsec:results_cylinder}

The cylinder-wake benchmark evaluates the ability
of the proposed models to capture unsteady flow
dynamics and evolving vortical structures.
Table~\ref{tab:case2} presents the relative $L^2$
errors at $t=3.5$ and $t=6.9$, together with
the average errors over 70 evaluation snapshots.
Results are reported for all three network
architectures and activation functions.

\begin{table}[htbp]
\centering
\small
\begin{threeparttable}
\caption{Cylinder wake. Relative $L^2$ errors (\%)
at representative time instants and averaged
over all evaluation snapshots.}
\label{tab:case2}

\begin{tabular}{lllrrrr}
\toprule
Architecture & Activation & Time
& $E_u$ & $E_v$ & $E_{\bm{u}}$ & $E_p$ \\
\midrule

$4\times50$ & PINN & $t=3.5$
& 1.7760\% & 6.8820\% & 2.5939\% & 7.2440\% \\
& & $t=6.9$
& 2.8270\% & 10.5800\% & 4.0658\% & 14.4200\% \\
& & Average
& 2.1117\% & 8.1727\% & 3.1126\% & 10.6155\% \\
\addlinespace

& BS-PINN-F & $t=3.5$
& \textbf{1.3660\%} & \textbf{4.6740\%}
& \textbf{1.8658\%} & \textbf{4.4670\%} \\
& & $t=6.9$
& 1.7030\% & \textbf{5.7610\%}
& \textbf{2.3213\%} & 11.6300\% \\
& & Average
& \textbf{1.5944\%} & \textbf{5.9260\%}
& \textbf{2.2970\%} & \textbf{8.3792\%} \\
\addlinespace

& BS-PINN-L & $t=3.5$
& 1.3870\% & 4.9480\%
& 1.9359\% & 5.3310\% \\
& & $t=6.9$
& \textbf{1.6470\%} & 5.9410\%
& 2.3215\% & \textbf{10.6000\%} \\
& & Average
& 1.6242\% & 6.1464\%
& 2.6380\% & 8.5598\% \\

\midrule

$7\times50$ & PINN & $t=3.5$
& 1.4360\% & 5.1410\%
& 2.0080\% & 5.3530\% \\
& & $t=6.9$
& 2.0670\% & 7.8240\%
& 2.9917\% & 10.9700\% \\
& & Average
& 1.7445\% & 6.6355\%
& 2.5450\% & 8.5367\% \\
\addlinespace

& BS-PINN-F & $t=3.5$
& 1.0240\% & 3.5700\%
& 1.4122\% & 3.6590\% \\
& & $t=6.9$
& 0.9489\% & 3.0270\%
& 1.2566\% & 6.6010\% \\
& & Average
& 1.2716\% & 4.7845\%
& 1.8460\% & 6.7110\% \\
\addlinespace

& BS-PINN-L & $t=3.5$
& \textbf{0.9960\%} & \textbf{3.5170\%}
& \textbf{1.3827\%} & \textbf{3.5140\%} \\
& & $t=6.9$
& \textbf{0.8652\%} & \textbf{2.7180\%}
& \textbf{1.1375\%} & \textbf{4.6510\%} \\
& & Average
& \textbf{1.2310\%} & \textbf{4.6650\%}
& \textbf{1.7940\%} & \textbf{6.3740\%} \\

\midrule

$10\times100$ & PINN & $t=3.5$
& 0.9065\% & 3.5380\%
& 1.3294\% & 3.1840\% \\
& & $t=6.9$
& 0.5592\% & 2.0250\%
& 0.7899\% & 3.9330\% \\
& & Average
& 1.0991\% & 4.4500\%
& 1.6630\% & 5.8468\% \\
\addlinespace

& BS-PINN-F & $t=3.5$
& 0.9038\% & \textbf{3.4900\%}
& 1.3174\% & 3.1180\% \\
& & $t=6.9$
& 0.5233\% & 1.9480\%
& 0.7504\% & 2.1990\% \\
& & Average
& \textbf{1.0743\%} & \textbf{4.3426\%}
& 1.5620\% & 4.7875\% \\
\addlinespace

& BS-PINN-L & $t=3.5$
& \textbf{0.8950\%} & 3.5030\%
& \textbf{1.3147\%} & \textbf{3.0490\%} \\
& & $t=6.9$
& \textbf{0.5130\%} & \textbf{1.9160\%}
& \textbf{0.7370\%} & \textbf{2.1070\%} \\
& & Average
& 1.0910\% & 4.3480\%
& \textbf{1.5577\%} & \textbf{4.0630\%} \\

\bottomrule
\end{tabular}

\begin{tablenotes}[flushleft]
\footnotesize
\item Errors are reported as relative $L^2$
errors in percent. $E_{\bm{u}}$ denotes the
combined velocity error. Pressure is mean-aligned
with the reference at each snapshot before
evaluating $E_p$. Average denotes the arithmetic
mean of the snapshot-wise relative errors over
70 evaluation times. Bold values indicate the
lowest error among the three activation functions
for each architecture, time, and error metric.
\end{tablenotes}
\end{threeparttable}
\end{table}

As shown in Table~\ref{tab:case2}, both BS-PINN
variants reduce the average velocity and pressure
errors relative to the conventional PINN across
all three architectures. For the $4\times50$
network, BS-PINN-F decreases the average combined
velocity error from $3.1126\%$ to $2.2970\%$,
corresponding to a relative reduction
$\mathcal R_{\bm u}^{(\mathrm F)}=26.2\%$.
BS-PINN-L achieves a reduction of $15.2\%$,
with an average error of $2.6380\%$.

For the $7\times50$ architecture, BS-PINN-L
achieves the lowest average combined velocity
error of $1.7940\%$, compared with $2.5450\%$
for PINN and $1.8460\%$ for BS-PINN-F.
The corresponding pressure errors are
$6.3740\%$, $8.5367\%$, and $6.7110\%$,
respectively. Increasing the architecture
to $10\times100$ further reduces the overall
prediction errors. In this setting, the average
combined velocity errors are $1.6630\%$,
$1.5620\%$, and $1.5577\%$ for PINN,
BS-PINN-F, and BS-PINN-L, respectively.
Although the relative improvements in velocity
are smaller for this larger network, the
pressure error remains noticeably lower:
BS-PINN-L achieves $4.0630\%$, compared with
$5.8468\%$ for the conventional PINN.

The improvements are also evident at individual
evaluation times. At $t=6.9$, for example,
BS-PINN-L with the $7\times50$ architecture
reduces the combined velocity error from
$2.9917\%$ to $1.1375\%$, while the pressure
error decreases from $10.9700\%$ to $4.6510\%$.
These results indicate that the benefits of the
proposed activation are not restricted to the
earlier stages of the wake evolution.

\begin{figure}[H]
    \centering
    \includegraphics[width=0.7\textwidth]
    {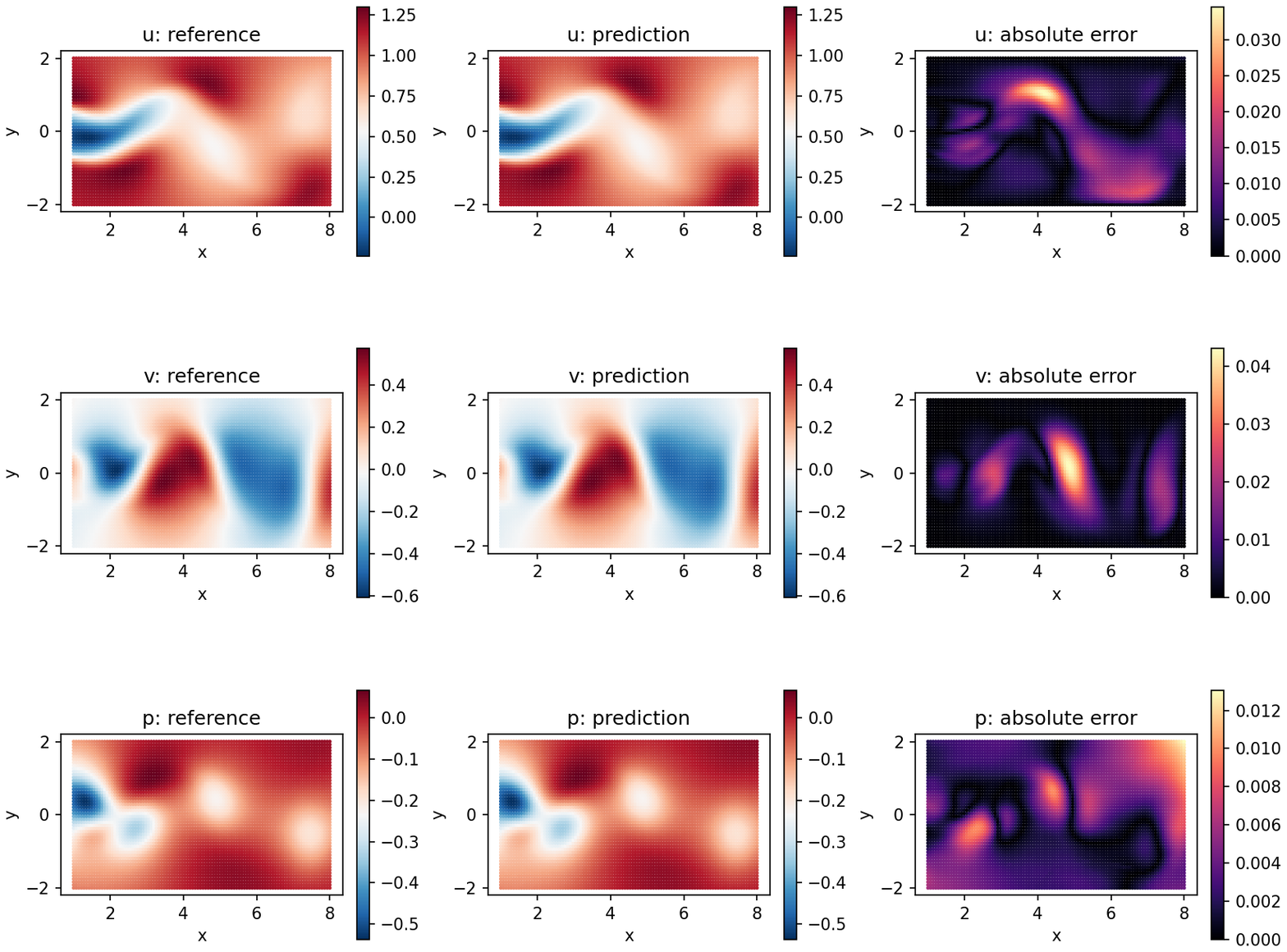}
    \caption{Reference fields, BS-PINN-F predictions,
    and absolute errors for the cylinder wake at
    $t=3.5$, using the $10\times100$ architecture.}
    \label{fig:cylinder_s35}
\end{figure}

\begin{figure}[H]
    \centering
    \includegraphics[width=0.7\textwidth]
    {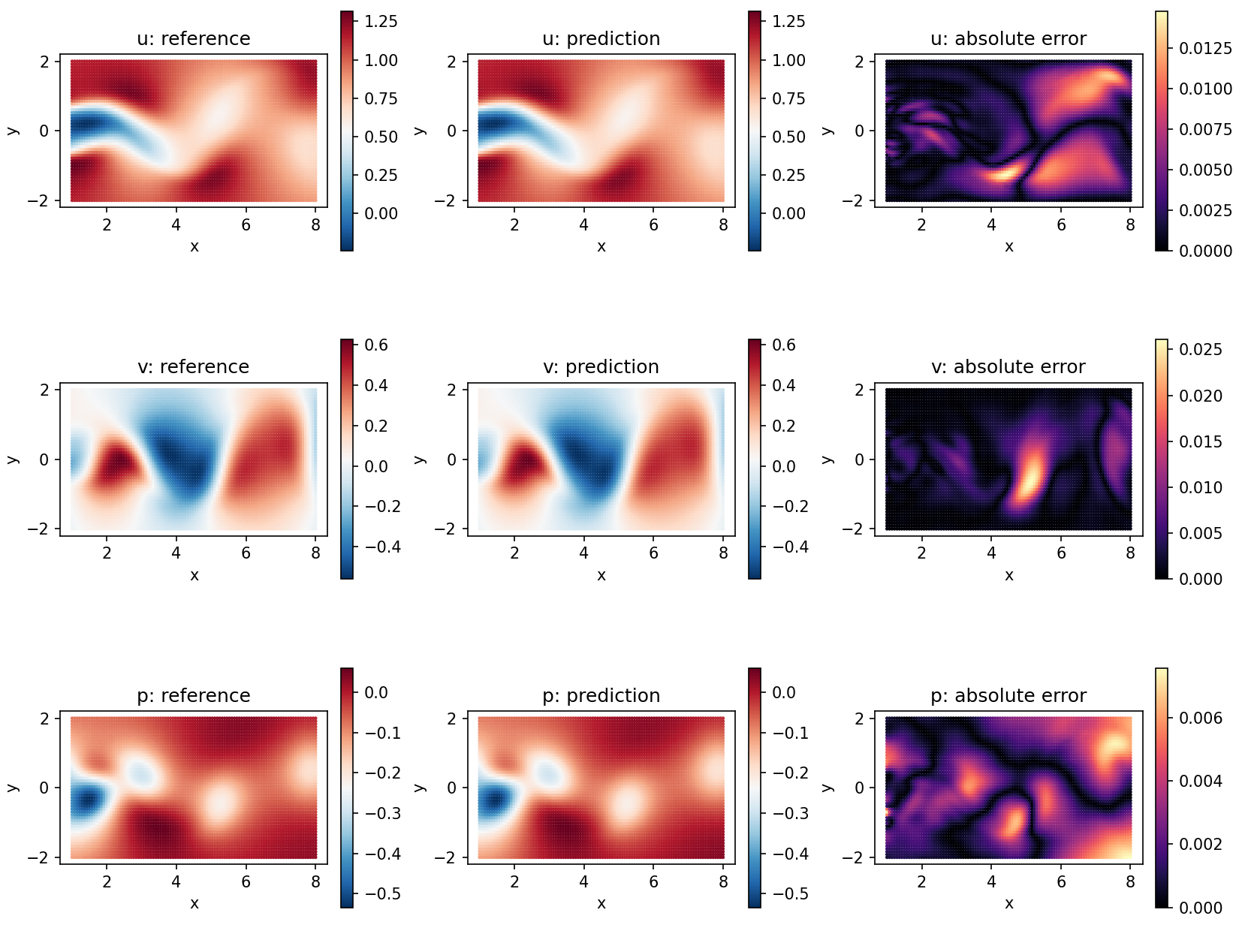}
    \caption{Reference fields, BS-PINN-F predictions,
    and absolute errors for the cylinder wake at
    $t=6.9$, using the $10\times100$ architecture.}
    \label{fig:cylinder_s69}
\end{figure}

\begin{figure}[H]
    \centering
    \includegraphics[width=0.7\textwidth]
    {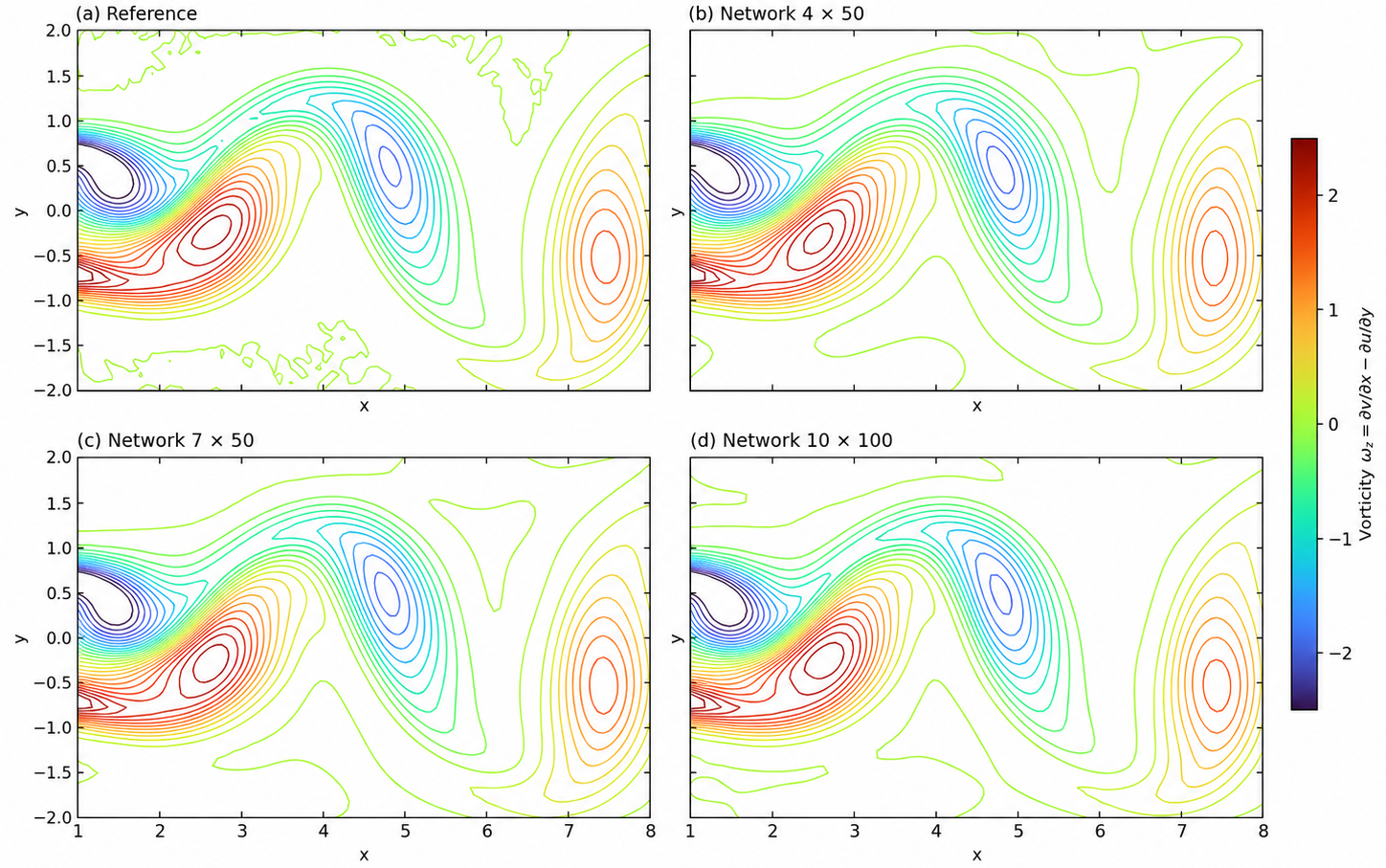}
    \caption{Comparison of vorticity contours for
    the cylinder wake at $t=3.5$, using the
    $10\times100$ architecture.}
    \label{fig:cylinder_vorticity_s35}
\end{figure}

\begin{figure}[H]
    \centering
    \includegraphics[width=0.7\textwidth]
    {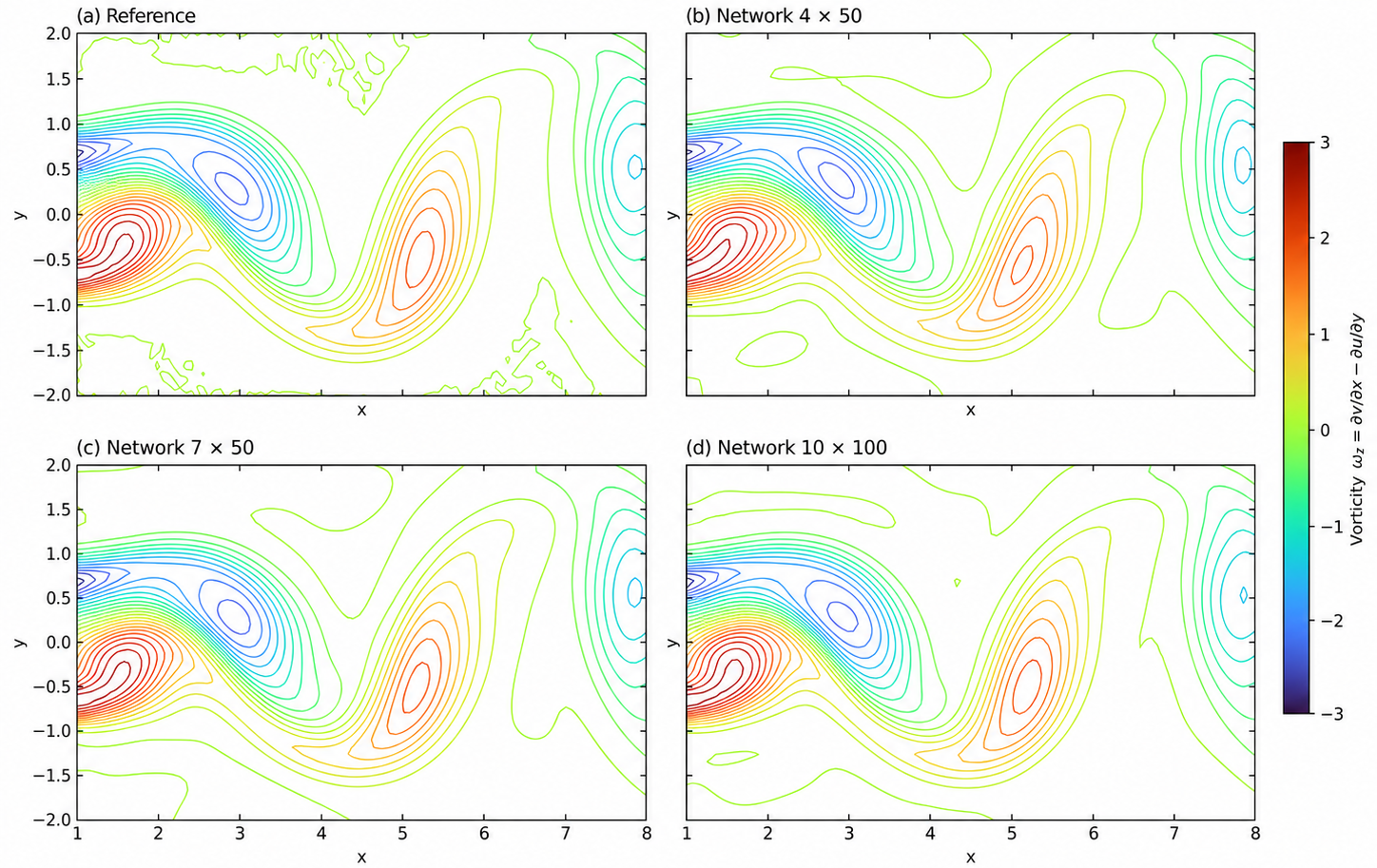}
    \caption{Comparison of vorticity contours for
    the cylinder wake at $t=6.9$, using the
    $10\times100$ architecture.}
    \label{fig:cylinder_vorticity_s69}
\end{figure}

To further examine the spatial accuracy of the
predictions, Figs.~\ref{fig:cylinder_s35}
and~\ref{fig:cylinder_s69} compare the reference
velocity and pressure fields with the predictions
of BS-PINN-F using the $10\times100$ architecture.
The corresponding absolute-error distributions
provide a spatial view of the discrepancies
quantified in Table~\ref{tab:case2}.
In addition, Figs.~\ref{fig:cylinder_vorticity_s35}
and~\ref{fig:cylinder_vorticity_s69} present
vorticity-contour comparisons at the same time
instants, illustrating the representation of
the evolving vortical structures in the wake.
Together, these results demonstrate that the
Biot--Savart-inspired activation improves
the accuracy of the predicted unsteady flow
fields without changing the underlying
Navier--Stokes constraints.

\subsection{Three-dimensional Beltrami flow}
\label{subsec:results_beltrami}

The three-dimensional Beltrami benchmark evaluates
the applicability of the proposed activation to
unsteady flows involving three velocity components
and pressure. Table~\ref{tab:case3} reports the
relative $L^2$ errors at $t=0.5$ and $t=1.0$,
together with the average errors over all
evaluation snapshots, for the three network
architectures.

\begin{table}[htbp]
\centering
\small
\begin{threeparttable}
\caption{Beltrami flow. Relative $L^2$ errors (\%)
at representative time instants and averaged
over all evaluation snapshots.}
\label{tab:case3}

\begin{tabular}{lllrrrrr}
\toprule
Architecture & Activation & Time
& $E_u$ & $E_v$ & $E_w$
& $E_{\bm{u}}$ & $E_p$ \\
\midrule

$4\times50$ & PINN & $t=0.5$
& 0.2223\% & 0.1910\% & 0.1690\%
& 0.1953\% & 1.3330\% \\
& & $t=1.0$
& 0.6207\% & 0.3965\% & 0.3877\%
& 0.4806\% & 5.4810\% \\
& & Average
& 0.2820\% & 0.1944\% & 0.1955\%
& 0.2286\% & 2.0879\% \\
\addlinespace

& BS-PINN-F & $t=0.5$
& \textbf{0.2118\%} & \textbf{0.1727\%}
& \textbf{0.1033\%} & \textbf{0.1687\%}
& \textbf{1.1020\%} \\
& & $t=1.0$
& 0.4125\% & 0.3654\%
& \textbf{0.3310\%} & 0.3711\%
& \textbf{3.8850\%} \\
& & Average
& 0.2276\% & \textbf{0.1825\%}
& \textbf{0.1473\%} & \textbf{0.1898\%}
& \textbf{1.6784\%} \\
\addlinespace

& BS-PINN-L & $t=0.5$
& 0.2221\% & 0.1929\% & 0.1603\%
& 0.1934\% & 1.1355\% \\
& & $t=1.0$
& \textbf{0.3534\%} & \textbf{0.2911\%}
& 0.3823\% & \textbf{0.3444\%}
& 4.8380\% \\
& & Average
& \textbf{0.2133\%} & 0.1924\%
& 0.1867\% & 0.2021\% & 1.9966\% \\

\midrule

$7\times50$ & PINN & $t=0.5$
& 0.2274\% & 0.1617\% & 0.1736\%
& 0.1897\% & 1.5750\% \\
& & $t=1.0$
& 0.7310\% & 0.6721\% & 0.5664\%
& 0.6600\% & 7.1970\% \\
& & Average
& 0.3165\% & 0.2855\% & 0.2668\%
& 0.2997\% & 2.6233\% \\
\addlinespace

& BS-PINN-F & $t=0.5$
& \textbf{0.1686\%} & 0.1125\%
& \textbf{0.1456\%} & \textbf{0.1441\%}
& \textbf{1.3690\%} \\
& & $t=1.0$
& \textbf{0.7090\%} & 0.4307\%
& 0.5234\% & 0.5663\% & 7.1000\% \\
& & Average
& \textbf{0.2919\%} & 0.1769\%
& 0.2611\% & \textbf{0.2488\%}
& \textbf{2.5988\%} \\
\addlinespace

& BS-PINN-L & $t=0.5$
& 0.1950\% & \textbf{0.1054\%}
& 0.1661\% & 0.1599\% & 1.4910\% \\
& & $t=1.0$
& 0.7300\% & \textbf{0.3718\%}
& \textbf{0.5225\%} & \textbf{0.5610\%}
& \textbf{7.0090\%} \\
& & Average
& 0.3137\% & \textbf{0.1540\%}
& \textbf{0.2430\%} & 0.2605\%
& 2.6000\% \\

\midrule

$10\times100$ & PINN & $t=0.5$
& 0.1150\% & 0.0828\% & 0.0864\%
& 0.0958\% & 0.7137\% \\
& & $t=1.0$
& 0.1765\% & 0.1571\% & 0.1683\%
& 0.1675\% & 1.9150\% \\
& & Average
& 0.1095\% & 0.0834\% & 0.0907\%
& 0.0959\% & 0.9147\% \\
\addlinespace

& BS-PINN-F & $t=0.5$
& \textbf{0.1000\%} & \textbf{0.0675\%}
& \textbf{0.0720\%} & \textbf{0.0811\%}
& \textbf{0.5170\%} \\
& & $t=1.0$
& \textbf{0.1694\%} & \textbf{0.1276\%}
& \textbf{0.1122\%} & \textbf{0.1385\%}
& \textbf{1.4380\%} \\
& & Average
& \textbf{0.0949\%} & \textbf{0.0703\%}
& \textbf{0.0755\%} & \textbf{0.0904\%}
& \textbf{0.8031\%} \\
\addlinespace

& BS-PINN-L & $t=0.5$
& 0.1054\% & 0.0732\% & 0.0918\%
& 0.0911\% & 0.5211\% \\
& & $t=1.0$
& 0.1845\% & 0.1357\% & 0.1385\%
& 0.1545\% & 1.7390\% \\
& & Average
& 0.1024\% & 0.0760\% & 0.0919\%
& 0.0922\% & 0.8047\% \\

\bottomrule
\end{tabular}

\begin{tablenotes}[flushleft]
\footnotesize
\item Errors are reported as relative $L^2$
errors in percent. $E_{\bm{u}}$ denotes the
combined three-component velocity error.
Pressure is mean-aligned with the reference
at each snapshot before evaluating $E_p$.
Average denotes the arithmetic mean of the snapshot-wise relative errors over 11 evaluation times, $t=0,0.1,\ldots,1.0$. Bold values indicate
the lowest error among the three models
for each architecture, evaluation time,
and error metric.
\end{tablenotes}
\end{threeparttable}
\end{table}

As shown in Table~\ref{tab:case3}, BS-PINN-F
achieves the lowest average combined velocity
error across all three architectures.
For the $4\times50$ and $7\times50$ networks,
the average combined velocity errors decrease
from $0.2286\%$ to $0.1898\%$ and from
$0.2997\%$ to $0.2488\%$, respectively,
corresponding to relative reductions
$\mathcal R_{\bm u}^{(\mathrm F)}$
of approximately $17.0\%$ in both cases.
For the $10\times100$ architecture, the
corresponding error decreases from $0.0959\%$
to $0.0904\%$, giving a smaller reduction
of approximately $5.7\%$.

The fixed-kernel variant also achieves
the lowest average pressure error
for all three architectures.
In particular, the pressure error decreases
from $2.0879\%$ to $1.6784\%$ for the
$4\times50$ network and from $0.9147\%$
to $0.8031\%$ for the $10\times100$ network.
The learnable-kernel variant remains
competitive at individual evaluation times.
For example, at $t=1.0$ with the
$4\times50$ architecture, BS-PINN-L
achieves the lowest combined velocity
error of $0.3444\%$, whereas BS-PINN-F
gives the lowest pressure error of
$3.8850\%$.

\begin{figure}[H]
    \centering
    \includegraphics[width=0.8\textwidth]
    {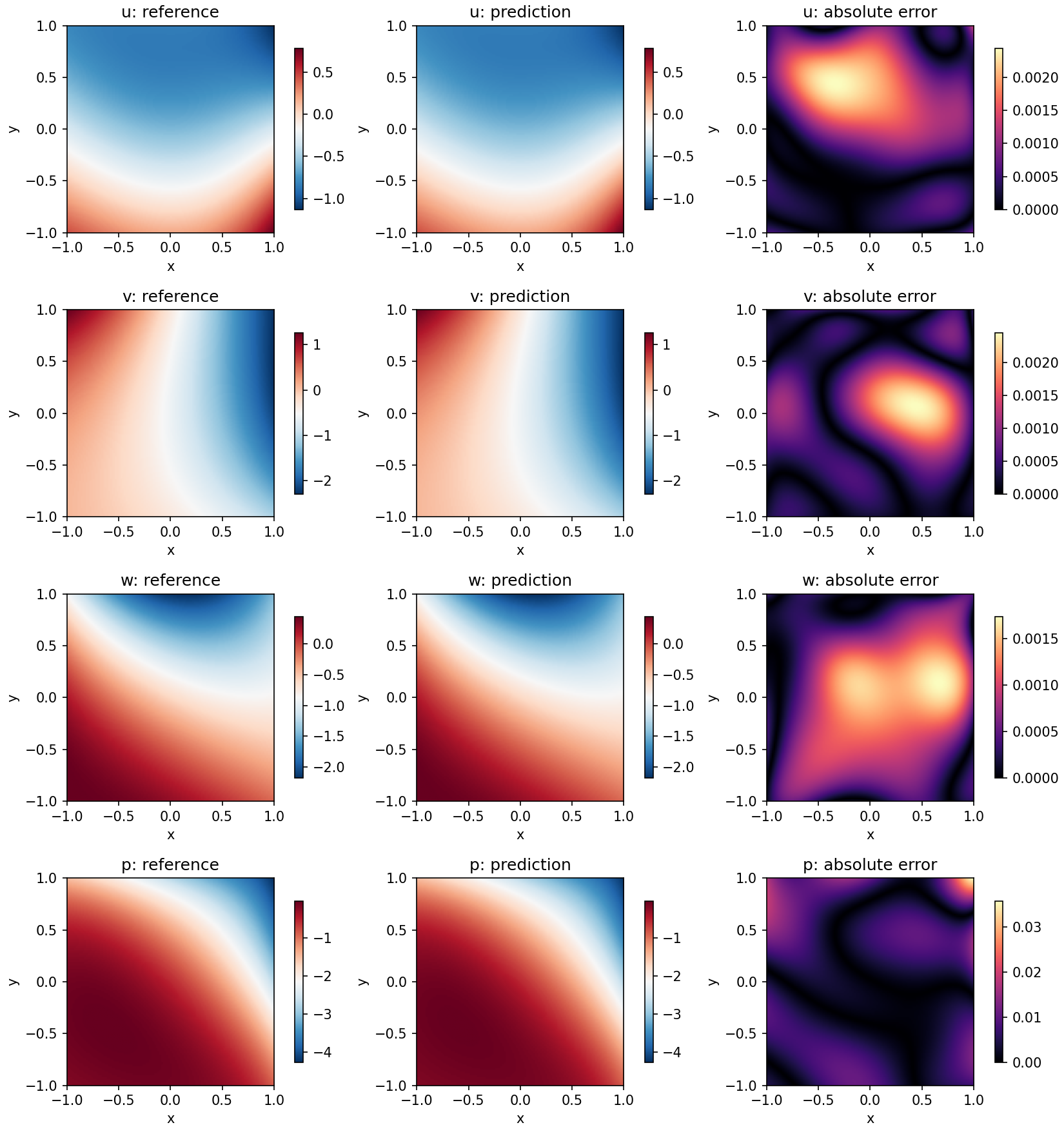}
    \caption{Three-dimensional Beltrami flow:
    reference fields, BS-PINN-F predictions,
    and absolute errors on the $z=0$ plane
    at $t=0.5$, using the $10\times100$
    architecture.}
    \label{fig:beltrami_t05}
\end{figure}

\begin{figure}[H]
    \centering
    \includegraphics[width=0.8\textwidth]
    {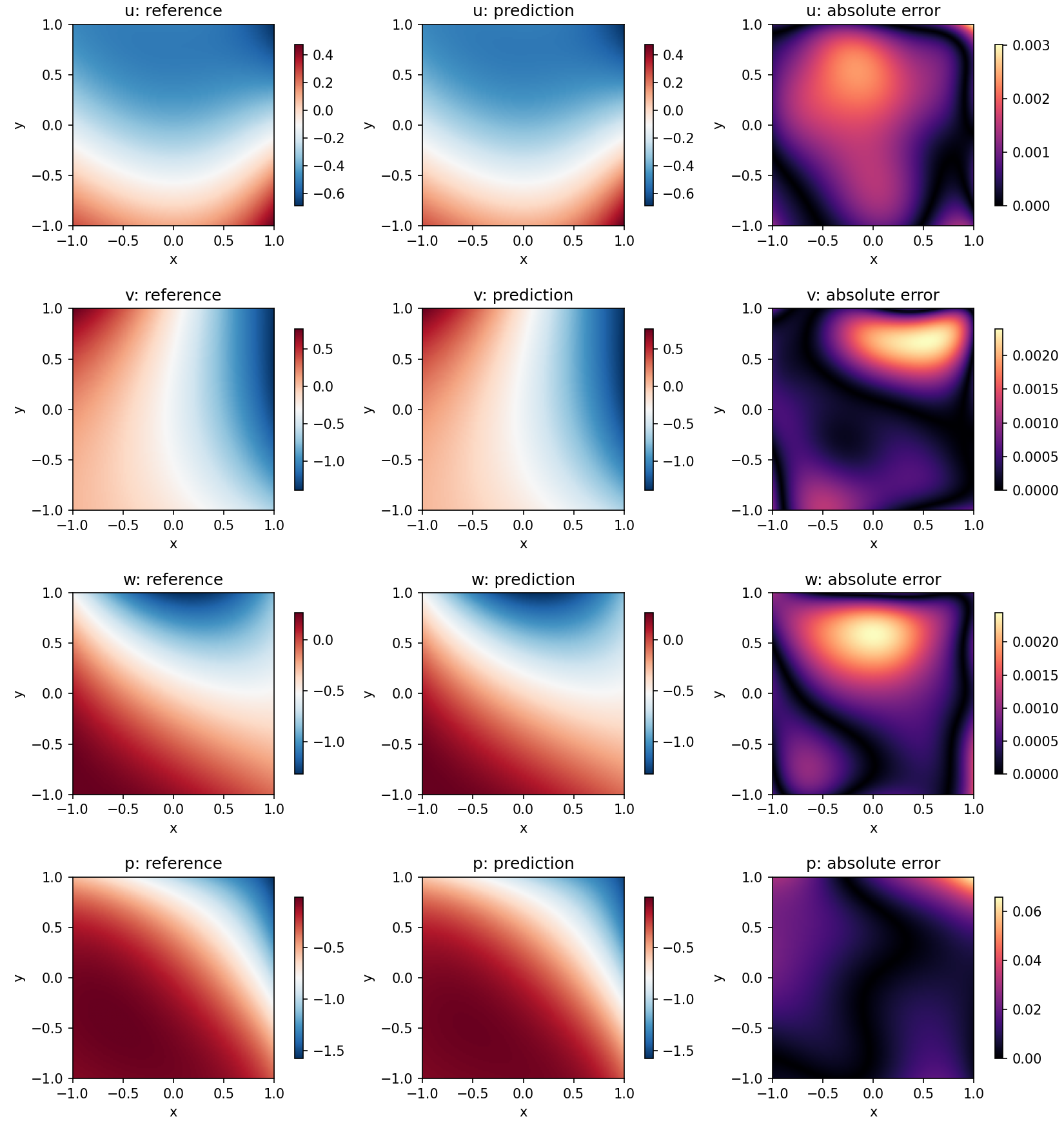}
    \caption{Three-dimensional Beltrami flow:
    reference fields, BS-PINN-F predictions,
    and absolute errors on the $z=0$ plane
    at $t=1.0$, using the $10\times100$
    architecture.}
    \label{fig:beltrami_t10}
\end{figure}

Figures~\ref{fig:beltrami_t05}
and~\ref{fig:beltrami_t10} compare the
reference fields, BS-PINN-F predictions,
and absolute errors at $t=0.5$ and
$t=1.0$, respectively, using the
$10\times100$ architecture.
For visualization, the three-dimensional
flow fields are displayed on the
two-dimensional plane $z=0$.
These plots therefore show cross-sectional
views of the three-dimensional solution,
whereas the relative $L^2$ errors reported
in Table~\ref{tab:case3} are computed
using the three-dimensional evaluation points.

Overall, the results indicate that the
proposed scalar activation is effective
for the tested three-dimensional flow.
Its use in this benchmark does not,
however, constitute a direct implementation
of the three-dimensional Biot--Savart
integral operator.

\subsection{Taylor--Green vortex}
\label{subsec:results_taylor_green}

The Taylor--Green vortex provides an analytical
benchmark for evaluating the accuracy of the
predicted flow fields under periodic boundary
conditions. In addition to velocity and pressure,
we examine vorticity, which is computed from
spatial derivatives of the predicted velocity.
This allows us to assess the accuracy of both
the flow-field predictions and their spatial
derivatives.

Table~\ref{tab:tgv_re20} reports the relative
$L^2$ errors at $t=0.5$ and $t=1.0$, together
with the average errors over five evaluation
snapshots, for all three network architectures.

\begin{table}[htbp]
\centering
\small
\begin{threeparttable}
\caption{Taylor--Green vortex. Relative $L^2$
errors (\%) at representative time instants
and averaged over five evaluation snapshots.}
\label{tab:tgv_re20}

\begin{tabular}{lllrrrrr}
\toprule
Architecture & Activation & Time
& $E_u$ & $E_v$ & $E_{\bm{u}}$
& $E_{\omega}$ & $E_p$ \\
\midrule

$4\times50$ & PINN & $t=0.5$
& 8.7861\% & 9.1427\% & 8.9662\%
& 2.4224\% & 94.7751\% \\
& & $t=1.0$
& 18.6127\% & 19.6731\% & 19.1502\%
& 6.5229\% & 102.6720\% \\
& & Average
& 9.1372\% & 9.5882\% & 9.3655\%
& 3.2377\% & 94.4737\% \\
\addlinespace

& BS-PINN-F & $t=0.5$
& 3.9677\% & 3.1196\% & 3.5690\%
& 1.4025\% & 37.9008\% \\
& & $t=1.0$
& 8.4022\% & 6.9131\% & 7.6938\%
& 3.1290\% & 43.2035\% \\
& & Average
& 4.1311\% & 3.3336\% & 3.7540\%
& 1.7081\% & 38.0144\% \\
\addlinespace

& BS-PINN-L & $t=0.5$
& \textbf{3.1627\%} & \textbf{2.1912\%}
& \textbf{2.7206\%}
& \textbf{1.2910\%} & \textbf{29.2495\%} \\
& & $t=1.0$
& \textbf{6.7392\%} & \textbf{5.1143\%}
& \textbf{5.9822\%}
& \textbf{2.7069\%} & \textbf{35.6129\%} \\
& & Average
& \textbf{3.3078\%} & \textbf{2.4172\%}
& \textbf{2.8982\%}
& \textbf{1.5385\%} & \textbf{29.5754\%} \\

\midrule

$7\times50$ & PINN & $t=0.5$
& 2.9810\% & 4.1725\% & 3.6260\%
& 1.4473\% & 33.0317\% \\
& & $t=1.0$
& 6.1037\% & 8.0372\% & 7.1362\%
& 2.8667\% & 30.6591\% \\
& & Average
& 3.0455\% & 4.1575\% & 3.9141\%
& 1.7230\% & 32.5274\% \\
\addlinespace

& BS-PINN-F & $t=0.5$
& \textbf{2.5495\%} & \textbf{2.7004\%}
& \textbf{2.6260\%}
& \textbf{0.8995\%} & \textbf{25.0400\%} \\
& & $t=1.0$
& \textbf{5.0526\%} & \textbf{5.4288\%}
& \textbf{5.2441\%}
& \textbf{2.2217\%} & \textbf{23.4119\%} \\
& & Average
& \textbf{2.5586\%} & \textbf{2.7313\%}
& \textbf{2.6464\%}
& \textbf{1.1505\%} & \textbf{25.4110\%} \\
\addlinespace

& BS-PINN-L & $t=0.5$
& 2.9903\% & 2.8757\% & 2.9335\%
& 0.9222\% & 28.2981\% \\
& & $t=1.0$
& 5.9028\% & 5.8803\% & 5.8915\%
& 2.3411\% & 26.8405\% \\
& & Average
& 2.9915\% & 2.9251\% & 2.9920\%
& 1.2005\% & 28.6390\% \\

\midrule

$10\times100$ & PINN & $t=0.5$
& 3.1951\% & 3.4610\% & 3.3307\%
& 0.8024\% & 34.9088\% \\
& & $t=1.0$
& 6.7697\% & 7.3820\% & 7.0824\%
& 2.3576\% & 38.0239\% \\
& & Average
& 3.3007\% & 3.5871\% & 3.4469\%
& 1.0007\% & 35.0640\% \\
\addlinespace

& BS-PINN-F & $t=0.5$
& \textbf{2.8825\%} & 1.8780\%
& 2.4327\% & 0.7867\%
& \textbf{16.3167\%} \\
& & $t=1.0$
& 6.7261\% & 3.7311\%
& 5.4388\% & \textbf{2.1019\%}
& \textbf{21.9657\%} \\
& & Average
& \textbf{3.1356\%} & 1.8862\%
& 2.5899\% & \textbf{0.9972\%}
& \textbf{17.0365\%} \\
\addlinespace

& BS-PINN-L & $t=0.5$
& 2.9457\% & \textbf{1.1973\%}
& \textbf{2.2484\%}
& \textbf{0.6759\%} & 25.7207\% \\
& & $t=1.0$
& \textbf{6.7247\%} & \textbf{2.6869\%}
& \textbf{5.1206\%}
& 2.1971\% & 32.1760\% \\
& & Average
& 3.1926\% & \textbf{1.2837\%}
& \textbf{2.4343\%}
& 0.9976\% & 25.8083\% \\

\bottomrule
\end{tabular}

\begin{tablenotes}[flushleft]
\footnotesize
\item Errors are reported as relative $L^2$
errors in percent. $E_{\bm{u}}$ denotes the
combined velocity error, and $E_{\omega}$
denotes the vorticity error.
Pressure is mean-aligned with the reference
at each snapshot before evaluating $E_p$.
Average denotes the arithmetic mean of the
snapshot-wise relative errors at
$t=0,0.25,0.5,0.75,1$.
Bold values indicate the lowest error among
the three models for each architecture,
evaluation time, and error metric.
\end{tablenotes}
\end{threeparttable}
\end{table}

As shown in Table~\ref{tab:tgv_re20},
the proposed BS-PINN variants substantially
improve the prediction accuracy for the
$4\times50$ architecture.
The average combined velocity error decreases
from $9.3655\%$ for PINN to $3.7540\%$
for BS-PINN-F and $2.8982\%$ for BS-PINN-L,
corresponding to relative error reductions
$\mathcal R_{\bm u}^{(\mathrm F)}=59.9\%$
and $\mathcal R_{\bm u}^{(\mathrm L)}=69.1\%$,
respectively.
The corresponding average pressure errors
decrease from $94.4737\%$ to $38.0144\%$
and $29.5754\%$.
Notably, both variants also improve the
vorticity prediction, reducing the average
vorticity error from $3.2377\%$ to
$1.7081\%$ and $1.5385\%$, respectively.
These improvements indicate that the proposed activation benefits not only the predicted velocity fields but also the reconstruction of vorticity, which depends on spatial derivatives of the velocity.

For the $7\times50$ architecture,
BS-PINN-F achieves the lowest average
combined velocity error of $2.6464\%$,
compared with $3.9141\%$ for PINN
and $2.9920\%$ for BS-PINN-L.
The fixed-kernel variant also provides
the lowest average pressure and
vorticity errors, reaching $25.4110\%$
and $1.1505\%$, respectively.

For the $10\times100$ architecture,
BS-PINN-L gives the lowest average
combined velocity error of $2.4343\%$,
compared with $3.4469\%$ for PINN.
In contrast, BS-PINN-F achieves the
lowest average pressure error of
$17.0365\%$, substantially below
the baseline value of $35.0640\%$.
Both BS-PINN variants also improve
the average vorticity error, although
the reductions are less pronounced
than those obtained with the
smaller network architectures.

\begin{figure}[H]
    \centering
    \includegraphics[width=0.8\textwidth]
    {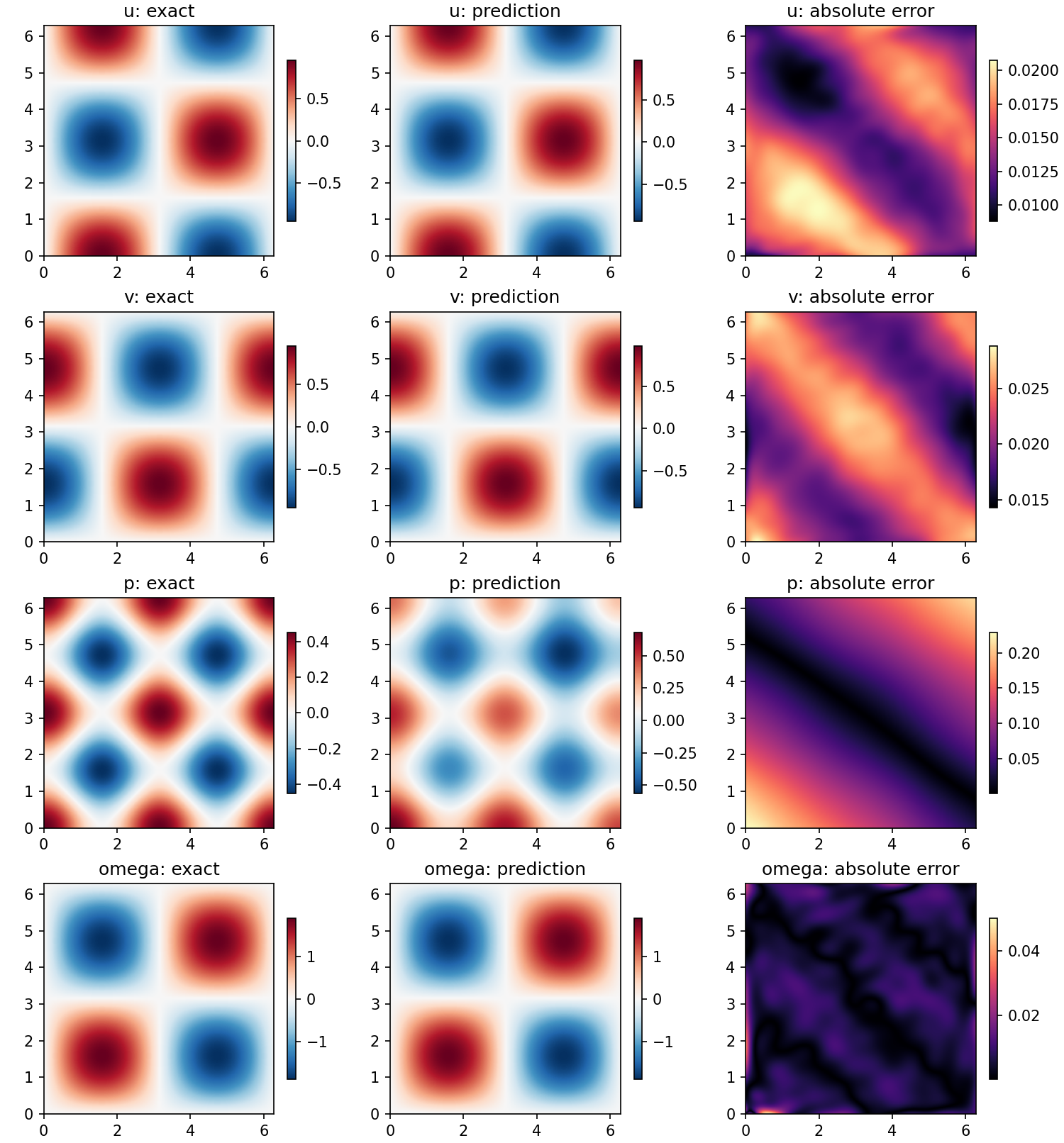}
    \caption{Taylor--Green vortex:
    reference fields, BS-PINN-F predictions,
    and absolute errors at $t=0.5$,
    using the $10\times100$ architecture.}
    \label{fig:tgv_t05}
\end{figure}

\begin{figure}[H]
    \centering
    \includegraphics[width=0.8\textwidth]
    {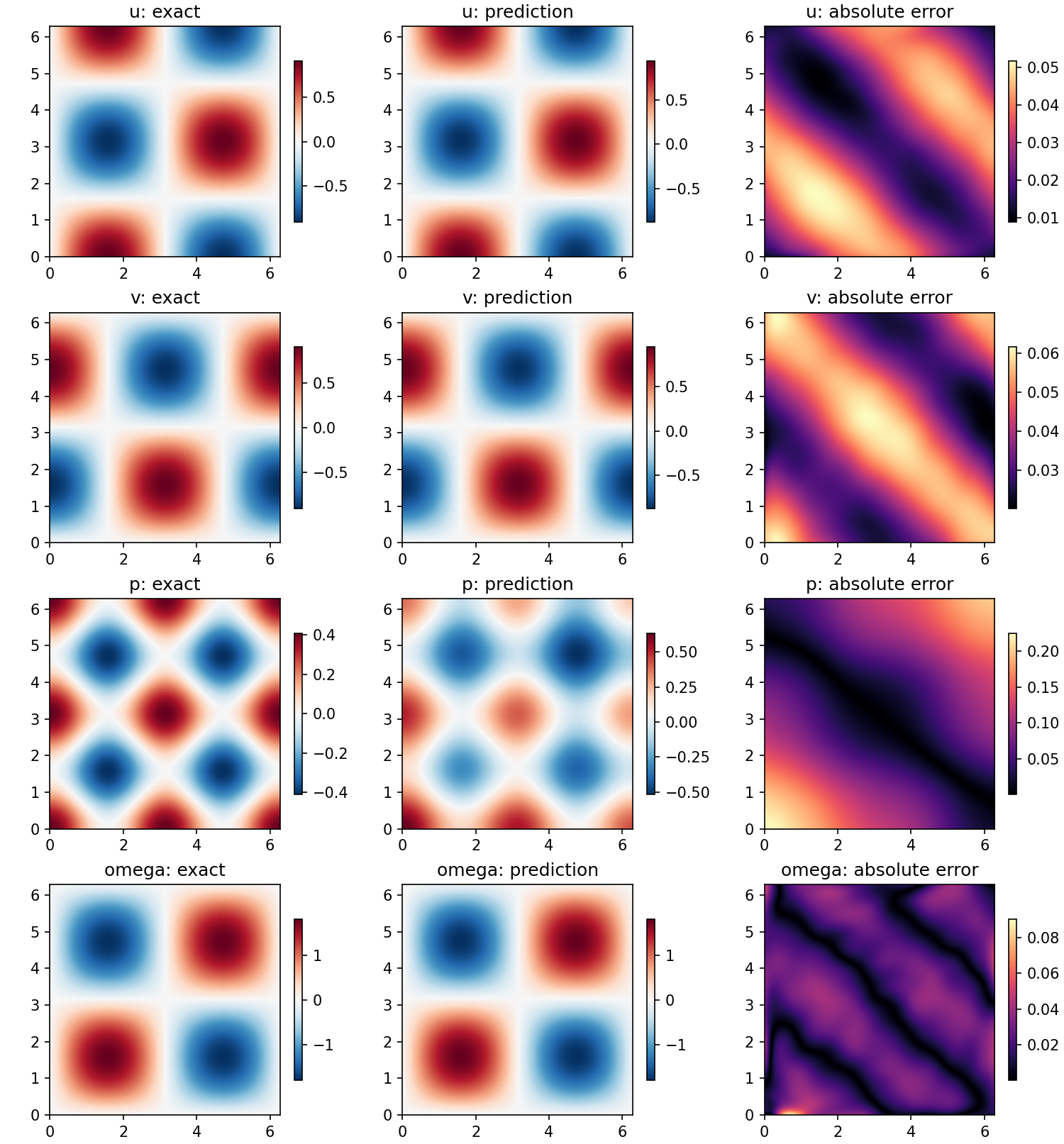}
    \caption{Taylor--Green vortex:
    reference fields, BS-PINN-F predictions,
    and absolute errors at $t=1.0$,
    using the $10\times100$ architecture.}
    \label{fig:tgv_t10}
\end{figure}

Figures~\ref{fig:tgv_t05}
and~\ref{fig:tgv_t10} compare the
reference velocity and pressure fields
with the BS-PINN-F predictions and
corresponding absolute errors at
$t=0.5$ and $t=1.0$, respectively,
using the $10\times100$ architecture.
These spatial comparisons complement
the quantitative results in
Table~\ref{tab:tgv_re20},
illustrating the accuracy of the
predicted flow structures over time.

Overall, the results demonstrate
the effectiveness of the
Biot--Savart-inspired activation
for this periodic-flow benchmark.
The relative advantages of the
fixed and learnable kernels vary
with the network architecture and
the evaluated quantity, indicating
that learning the kernel parameter
does not consistently outperform
the fixed-kernel formulation.

\subsection{Double shear layer}
\label{subsec:results_double_shear}

The double-shear-layer benchmark is particularly informative because its initially perturbed, oppositely signed shear layers undergo deformation and, over longer times, develop characteristic vortex-roll-up structures. It therefore assesses not only the accuracy of velocity and vorticity predictions but also the ability of a neural representation to capture recognizable flow patterns. The parameter $\rho$ controls the steepness of the initial shear layers, with larger values producing sharper velocity transitions. We first examine DNS-referenced short-time evolution over $t\in[0,0.25]$ for $\rho\in\{10,20,40\}$, followed by a separate qualitative long-time experiment with sparse velocity-data supervision.

\paragraph{Short-time evolution.}

Tables~\ref{tab:dsl_rho10_t025}--
\ref{tab:dsl_rho40_t025} report the relative
$L^2$ errors for the three shear-layer
configurations. Both BS-PINN variants reduce
the average combined velocity error relative
to the conventional PINN for every tested
network architecture and value of $\rho$.

For $\rho=10$, BS-PINN-F reduces the average
combined velocity error of the $4\times50$
network from $1.5053\%$ to $0.7250\%$,
corresponding to
$\mathcal R_{\bm u}^{(\mathrm F)}=51.8\%$.
For the $7\times50$ architecture,
BS-PINN-L achieves an average error of
$0.3169\%$, compared with $0.8786\%$
for PINN, giving a reduction of $63.9\%$.
The largest improvement in this configuration
is observed with the $10\times100$ architecture,
where BS-PINN-L reduces the average error
from $0.5608\%$ to $0.1676\%$,
corresponding to a reduction of $70.1\%$.

The improvements persist as the initial shear
layers become steeper. For $\rho=20$,
BS-PINN-L reduces the average combined velocity
error of the $4\times50$ network from
$2.1447\%$ to $0.7635\%$, a reduction of
$64.4\%$. For $\rho=40$, the corresponding
errors decrease from $3.1878\%$ to
$1.7118\%$, yielding a reduction of $46.3\%$.
Increasing the network size substantially
reduces the absolute errors in the more
challenging $\rho=40$ configuration.
For example, with the $10\times100$
architecture, the average combined velocity
errors are $0.5364\%$ for PINN and
$0.4610\%$ for BS-PINN-L. Although the
relative improvement is smaller for this
larger network, the proposed activation
continues to improve prediction accuracy.

The velocity improvements are accompanied by
reductions in vorticity error. For $\rho=40$
with the $4\times50$ architecture, the
average vorticity error decreases from
$21.6803\%$ for PINN to $12.8727\%$
for BS-PINN-F and $11.0086\%$ for
BS-PINN-L. Since vorticity is computed
from spatial derivatives of the predicted
velocity, this comparison provides an
additional assessment of the models'
ability to resolve localized shear-layer
structures.

For this benchmark, the reported pressure
error $E_p^{\mathrm{rec}}$ is evaluated using
pressure reconstructed from the predicted
velocity through the periodic pressure
Poisson equation, rather than the network's
direct pressure output. The reconstructed
pressure errors are also substantially
reduced by the proposed models.
For example, at $\rho=20$ with the
$4\times50$ architecture, the average
error decreases from $46.0901\%$ for
PINN to $14.83\%$ for BS-PINN-F and
$10.3360\%$ for BS-PINN-L, corresponding
to relative reductions of $67.8\%$
and $77.6\%$, respectively.
These results demonstrate improved
accuracy of the reconstructed pressure
fields, without implying the same
improvement in the network's direct
pressure prediction.

\begin{figure}[H]
    \centering
    \includegraphics[width=0.8\textwidth]
    {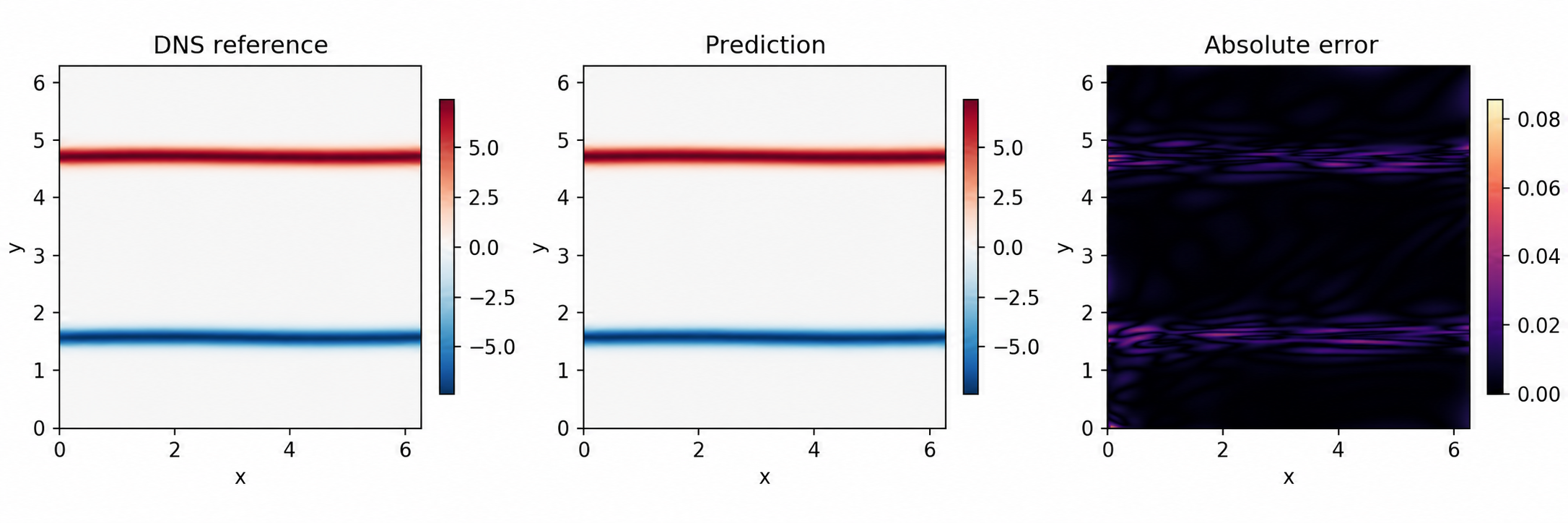}
    \caption{Double shear layer with $\rho=10$:
    DNS reference fields, BS-PINN-F predictions,
    and pointwise absolute errors at $t=0.25$,
    using the $10\times100$ architecture.}
    \label{fig:dsl_rho10}
\end{figure}

\begin{figure}[H]
    \centering
    \includegraphics[width=0.8\textwidth]
    {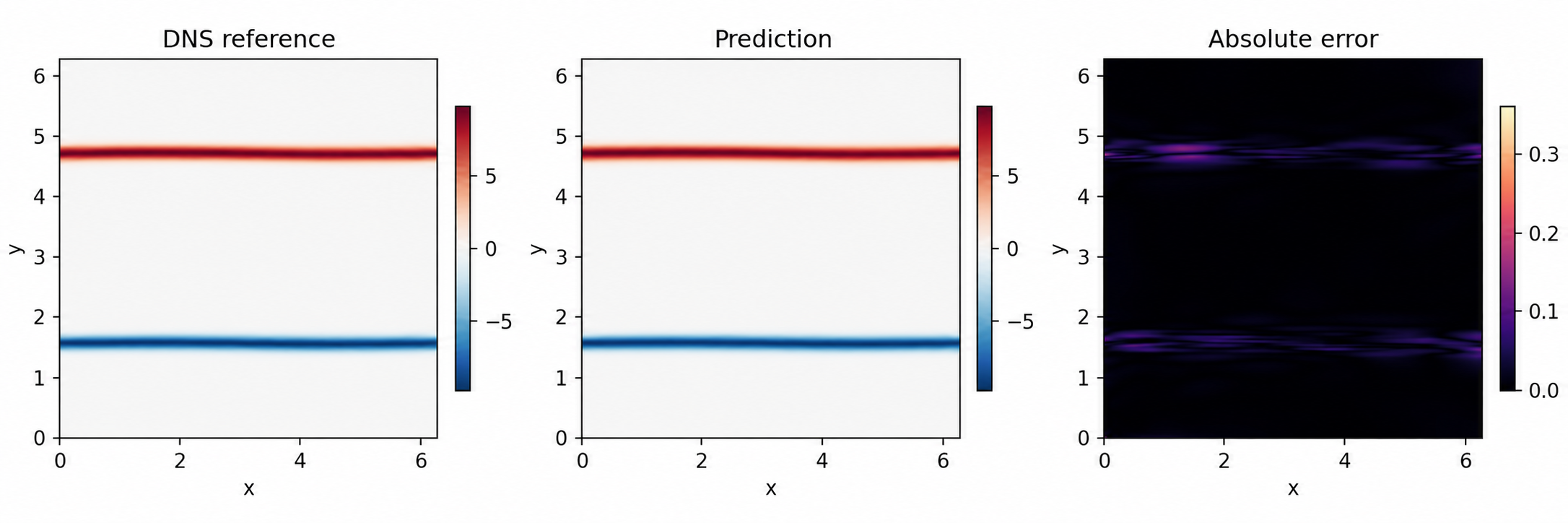}
    \caption{Double shear layer with $\rho=20$:
    DNS reference fields, BS-PINN-F predictions,
    and pointwise absolute errors at $t=0.25$,
    using the $10\times100$ architecture.}
    \label{fig:dsl_rho20}
\end{figure}

\begin{figure}[H]
    \centering
    \includegraphics[width=0.8\textwidth]
    {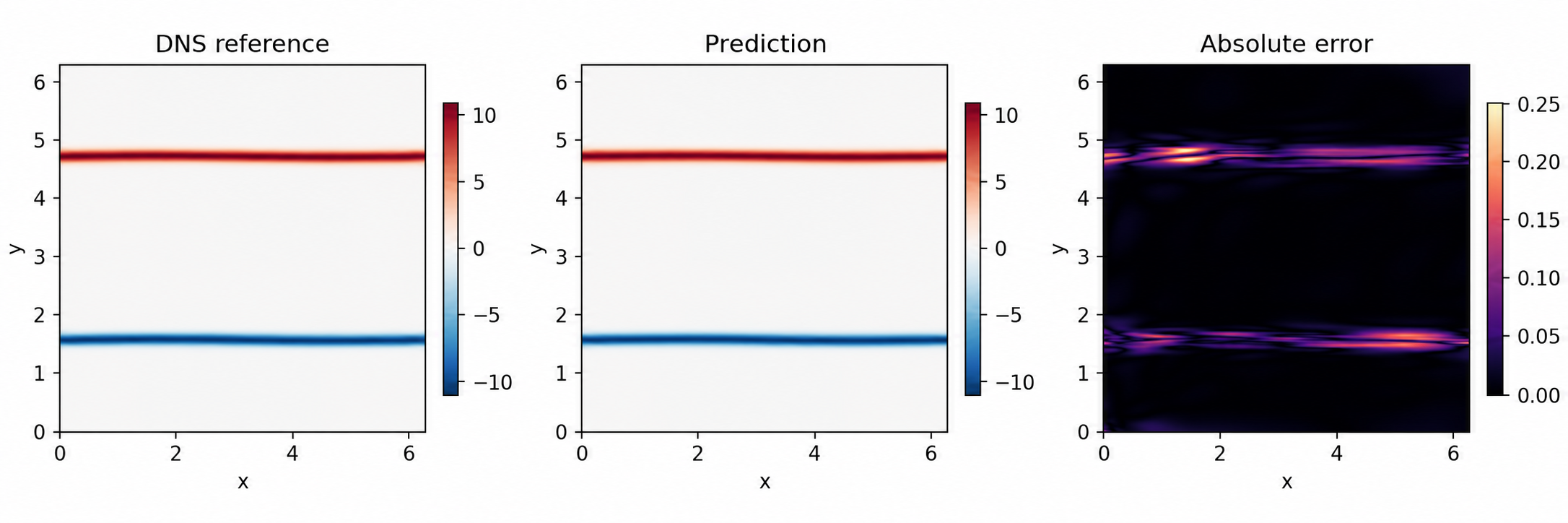}
    \caption{Double shear layer with $\rho=40$:
    DNS reference fields, BS-PINN-F predictions,
    and pointwise absolute errors at $t=0.25$,
    using the $10\times100$ architecture.}
    \label{fig:dsl_rho40}
\end{figure}

Figures~\ref{fig:dsl_rho10}--\ref{fig:dsl_rho40} compare the DNS reference fields, BS-PINN-F predictions, and pointwise absolute errors at $t=0.25$ for the $10\times100$ network. Across the three shear-layer steepness parameters, the DNS fields exhibit two narrow bands of opposite sign corresponding to the localized shear layers. BS-PINN-F reproduces their positions and overall spatial patterns, while the visible discrepancies are concentrated near the regions of strongest variation. In combination with the reduced vorticity errors in Tables~\ref{tab:dsl_rho10_t025}--\ref{tab:dsl_rho40_t025}, these visualizations show that the improvements extend beyond global velocity norms to the representation of localized shear-layer structures. At this early time, the principal feature is the resolution of the thin shear bands, the later vortex-roll-up process is examined separately below.

\begin{table}[htbp]
\centering
\small
\begin{threeparttable}
\caption{Double shear layer with $\rho=10$
and $\nu=0.01$. Relative $L^2$ errors (\%)
at representative time instants and averaged
over six evaluation snapshots.}
\label{tab:dsl_rho10_t025}

\begin{tabular}{lllrrrrr}
\toprule
Architecture & Activation & Time
& $E_u$ & $E_v$ & $E_{\bm{u}}$
& $E_{\omega}$ & $E_p^{\mathrm{rec}}$ \\
\midrule

$4\times50$ & PINN & $t=0.15$
& 0.7413\% & 35.9873\% & 1.5164\%
& 3.1213\% & 27.9183\% \\
& & $t=0.25$
& 1.1231\% & 57.5794\% & 2.4082\%
& 3.8822\% & 37.0412\% \\
& & Average
& 0.8270\% & 33.4694\% & 1.5053\%
& 4.2388\% & 27.3242\% \\
\addlinespace

& BS-PINN-F & $t=0.15$
& \textbf{0.3695\%} & \textbf{17.7410\%}
& \textbf{0.7495\%}
& \textbf{1.3183\%} & 7.8707\% \\
& & $t=0.25$
& \textbf{0.5776\%} & \textbf{28.6936\%}
& \textbf{1.2085\%}
& \textbf{1.6374\%} & 10.7577\% \\
& & Average
& \textbf{0.4056\%} & \textbf{15.8207\%}
& \textbf{0.7250\%}
& \textbf{2.0427\%} & 8.0304\% \\
\addlinespace

& BS-PINN-L & $t=0.15$
& 0.4130\% & 20.2636\% & 0.8517\%
& 1.5608\% & \textbf{6.3499\%} \\
& & $t=0.25$
& 0.6475\% & 33.4168\% & 1.3956\%
& 1.9622\% & \textbf{9.1673\%} \\
& & Average
& 0.4234\% & 17.7762\% & 0.7936\%
& 2.0439\% & \textbf{6.3165\%} \\

\midrule

$7\times50$ & PINN & $t=0.15$
& 0.3490\% & 23.8336\% & 0.9431\%
& 1.3079\% & 9.5700\% \\
& & $t=0.25$
& 0.6000\% & 37.6132\% & 1.5155\%
& 1.6240\% & 11.5480\% \\
& & Average
& 0.3618\% & 21.6630\% & 0.8786\%
& 1.6522\% & 9.6900\% \\
\addlinespace

& BS-PINN-F & $t=0.15$
& 0.2276\% & 6.8936\% & 0.3406\%
& 0.8034\% & \textbf{3.7600\%} \\
& & $t=0.25$
& \textbf{0.3630\%} & 11.5374\%
& 0.5603\% & 0.8378\%
& \textbf{5.5360\%} \\
& & Average
& 0.2330\% & 6.2525\% & 0.3309\%
& 1.1397\% & \textbf{3.5610\%} \\
\addlinespace

& BS-PINN-L & $t=0.15$
& \textbf{0.2267\%} & \textbf{6.3416\%}
& \textbf{0.3251\%}
& \textbf{0.7580\%} & 4.0980\% \\
& & $t=0.25$
& 0.3656\% & \textbf{10.7793\%}
& \textbf{0.5409\%}
& \textbf{0.7517\%} & 5.6750\% \\
& & Average
& \textbf{0.2307\%} & \textbf{5.8004\%}
& \textbf{0.3169\%}
& \textbf{1.0878\%} & 3.7140\% \\

\midrule

$10\times100$ & PINN & $t=0.15$
& 0.1923\% & 16.5389\% & 0.6377\%
& 0.5041\% & 2.8090\% \\
& & $t=0.25$
& 0.3603\% & 27.4682\% & 1.0783\%
& 0.8385\% & 4.7470\% \\
& & Average
& 0.1904\% & 14.2356\% & 0.5608\%
& 0.6418\% & 3.1110\% \\
\addlinespace

& BS-PINN-F & $t=0.15$
& 0.1580\% & \textbf{3.0233\%}
& 0.1931\%
& \textbf{0.2901\%} & 1.2560\% \\
& & $t=0.25$
& 0.3565\% & \textbf{5.1099\%}
& 0.4033\%
& 0.3083\% & 2.5810\% \\
& & Average
& 0.1505\% & \textbf{2.7113\%}
& 0.2240\%
& \textbf{0.3436\%} & 1.3340\% \\
\addlinespace

& BS-PINN-L & $t=0.15$
& \textbf{0.1351\%} & 3.3572\%
& \textbf{0.1830\%}
& 0.2955\% & \textbf{1.0140\%} \\
& & $t=0.25$
& \textbf{0.2273\%} & 5.8949\%
& \textbf{0.3149\%}
& \textbf{0.2993\%} & \textbf{2.0630\%} \\
& & Average
& \textbf{0.1256\%} & 2.9954\%
& \textbf{0.1676\%}
& 0.3963\% & \textbf{1.2610\%} \\

\bottomrule
\end{tabular}

\begin{tablenotes}[flushleft]
\footnotesize
\item Errors are reported as relative $L^2$
errors in percent. $E_{\bm{u}}$ denotes
the combined velocity error, and
$E_{\omega}$ denotes the vorticity error.
$E_p^{\mathrm{rec}}$ denotes the error in
pressure reconstructed from the predicted
velocity using the periodic pressure
Poisson equation, with pressure means
aligned at each snapshot.
Average denotes the arithmetic mean
of snapshot-wise relative errors at
$t=0,0.05,0.10,0.15,0.20,0.25$.
Bold values indicate the lowest error
among the three models for each
architecture, time, and error metric.
\end{tablenotes}

\end{threeparttable}
\end{table}

\begin{table}[htbp]
\centering
\small
\begin{threeparttable}
\caption{Double shear layer with $\rho=20$
and $\nu=0.01$. Relative $L^2$ errors (\%)
at representative time instants and averaged
over six evaluation snapshots.}
\label{tab:dsl_rho20_t025}

\begin{tabular}{lllrrrrr}
\toprule
Architecture & Activation & Time
& $E_u$ & $E_v$ & $E_{\bm{u}}$
& $E_{\omega}$ & $E_p^{\mathrm{rec}}$ \\
\midrule

$4\times50$ & PINN & $t=0.15$
& 1.3604\% & 34.2102\% & 1.8415\%
& 8.1887\% & 41.1818\% \\
& & $t=0.25$
& 1.5123\% & 45.0053\% & 2.2359\%
& 7.5718\% & 69.8232\% \\
& & Average
& 1.7222\% & 33.2907\% & 2.1447\%
& 12.4098\% & 46.0901\% \\
\addlinespace

& BS-PINN-F & $t=0.15$
& 0.6722\% & 21.8917\% & 1.0407\%
& 3.2317\% & 13.3353\% \\
& & $t=0.25$
& 0.8997\% & 34.5727\% & 1.5526\%
& 2.9459\% & 21.1322\% \\
& & Average
& 0.8142\% & 19.7473\% & 1.1260\%
& 5.6471\% & 14.8300\% \\
\addlinespace

& BS-PINN-L & $t=0.15$
& \textbf{0.4552\%} & \textbf{12.8914\%}
& \textbf{0.6527\%}
& \textbf{2.3699\%} & \textbf{9.9039\%} \\
& & $t=0.25$
& \textbf{0.5641\%} & \textbf{20.9407\%}
& \textbf{0.9516\%}
& \textbf{2.5679\%} & \textbf{12.5557\%} \\
& & Average
& \textbf{0.5774\%} & \textbf{12.7746\%}
& \textbf{0.7635\%}
& \textbf{4.3475\%} & \textbf{10.3360\%} \\

\midrule

$7\times50$ & PINN & $t=0.15$
& 0.5002\% & 19.4856\% & 0.8663\%
& 2.4885\% & 17.6510\% \\
& & $t=0.25$
& 0.7119\% & 27.1804\% & 1.2233\%
& 2.4906\% & 23.8890\% \\
& & Average
& 0.5750\% & 18.9797\% & 0.9056\%
& 3.9661\% & 19.2090\% \\
\addlinespace

& BS-PINN-F & $t=0.15$
& 0.4033\% & 10.9733\% & 0.5667\%
& 1.4481\% & 8.0830\% \\
& & $t=0.25$
& 0.6182\% & 17.9551\% & 0.9021\%
& 1.3594\% & 10.7680\% \\
& & Average
& 0.4361\% & 10.0176\% & 0.5768\%
& 2.4516\% & 7.6840\% \\
\addlinespace

& BS-PINN-L & $t=0.15$
& \textbf{0.3870\%} & \textbf{10.6389\%}
& \textbf{0.5466\%}
& \textbf{1.3051\%} & \textbf{7.5980\%} \\
& & $t=0.25$
& \textbf{0.6099\%} & \textbf{17.6945\%}
& \textbf{0.8895\%}
& \textbf{1.2637\%} & \textbf{10.0430\%} \\
& & Average
& \textbf{0.4186\%} & \textbf{9.7487\%}
& \textbf{0.5564\%}
& \textbf{2.3009\%} & \textbf{7.2790\%} \\

\midrule

$10\times100$ & PINN & $t=0.15$
& 0.3892\% & 10.1958\% & 0.5369\%
& 0.6188\% & 5.2450\% \\
& & $t=0.25$
& 0.6940\% & 16.7301\% & 0.9253\%
& 0.8807\% & 8.6490\% \\
& & Average
& 0.3777\% & 9.3397\% & 0.5391\%
& 0.9632\% & 5.5280\% \\
\addlinespace

& BS-PINN-F & $t=0.15$
& 0.3686\% & 9.7591\% & 0.5111\%
& 0.5424\% & 4.7340\% \\
& & $t=0.25$
& 0.6485\% & 17.0855\% & 0.9007\%
& 0.6558\% & 8.2370\% \\
& & Average
& 0.3412\% & 8.5830\% & 0.4637\%
& 0.8561\% & 4.7830\% \\
\addlinespace

& BS-PINN-L & $t=0.15$
& \textbf{0.3164\%} & \textbf{7.8883\%}
& \textbf{0.4266\%}
& \textbf{0.4567\%} & \textbf{3.8750\%} \\
& & $t=0.25$
& \textbf{0.5505\%} & \textbf{13.4869\%}
& \textbf{0.7393\%}
& \textbf{0.5735\%} & \textbf{6.6160\%} \\
& & Average
& \textbf{0.2920\%} & \textbf{6.9409\%}
& \textbf{0.3866\%}
& \textbf{0.7509\%} & \textbf{3.9650\%} \\

\bottomrule
\end{tabular}

\begin{tablenotes}[flushleft]
\footnotesize
\item Errors are reported as relative $L^2$
errors in percent. $E_{\bm{u}}$ denotes
the combined velocity error, and
$E_{\omega}$ denotes the vorticity error.
$E_p^{\mathrm{rec}}$ denotes the error in
pressure reconstructed from the predicted
velocity using the periodic pressure
Poisson equation, with pressure means
aligned at each snapshot.
Average denotes the arithmetic mean
of snapshot-wise relative errors at
$t=0,0.05,0.10,0.15,0.20,0.25$.
Bold values indicate the lowest error
among the three models for each
architecture, time, and error metric.
\end{tablenotes}

\end{threeparttable}
\end{table}

\begin{table}[htbp]
\centering
\small
\begin{threeparttable}
\caption{Double shear layer with $\rho=40$
and $\nu=0.01$. Relative $L^2$ errors (\%)
at representative time instants and averaged
over six evaluation snapshots.}
\label{tab:dsl_rho40_t025}

\begin{tabular}{lllrrrrr}
\toprule
Architecture & Activation & Time
& $E_u$ & $E_v$ & $E_{\bm{u}}$
& $E_{\omega}$ & $E_p^{\mathrm{rec}}$ \\
\midrule

$4\times50$ & PINN & $t=0.15$
& 2.0073\% & 37.5437\% & 2.4222\%
& 12.5838\% & 40.0540\% \\
& & $t=0.25$
& 2.0175\% & 48.9701\% & 2.6942\%
& 10.9696\% & 68.4255\% \\
& & Average
& 2.8176\% & 37.3307\% & 3.1878\%
& 21.6803\% & 53.4898\% \\
\addlinespace

& BS-PINN-F & $t=0.15$
& 1.1018\% & 26.0579\% & 1.4492\%
& 5.9005\% & 25.0813\% \\
& & $t=0.25$
& 1.1862\% & \textbf{39.7999\%}
& 1.8748\%
& 5.1175\% & 29.9423\% \\
& & Average
& 1.5874\% & 24.3486\% & 1.8972\%
& 12.8727\% & 28.5734\% \\
\addlinespace

& BS-PINN-L & $t=0.15$
& \textbf{1.0070\%} & \textbf{25.2810\%}
& \textbf{1.3595\%}
& \textbf{5.1366\%} & \textbf{16.5954\%} \\
& & $t=0.25$
& \textbf{1.0914\%} & 40.2154\%
& \textbf{1.8285\%}
& \textbf{4.4298\%} & \textbf{21.0864\%} \\
& & Average
& \textbf{1.3911\%} & \textbf{23.6678\%}
& \textbf{1.7118\%}
& \textbf{11.0086\%} & \textbf{21.6772\%} \\

\midrule

$7\times50$ & PINN & $t=0.15$
& 0.7499\% & 28.7016\% & 1.2800\%
& 3.6948\% & 34.9710\% \\
& & $t=0.25$
& 1.0001\% & 37.1139\% & 1.6832\%
& 4.2560\% & 44.2200\% \\
& & Average
& 0.9470\% & 28.3685\% & 1.4226\%
& 7.5578\% & 36.0970\% \\
\addlinespace

& BS-PINN-F & $t=0.15$
& 0.5218\% & 13.6686\% & 0.7184\%
& 2.2042\% & 8.9750\% \\
& & $t=0.25$
& 0.7288\% & 21.4101\% & 1.0681\%
& 1.9063\% & 11.2510\% \\
& & Average
& 0.6590\% & 12.8079\% & 0.8279\%
& 5.0573\% & 14.5130\% \\
\addlinespace

& BS-PINN-L & $t=0.15$
& \textbf{0.4886\%} & \textbf{12.5294\%}
& \textbf{0.6661\%}
& \textbf{2.0572\%} & \textbf{8.0210\%} \\
& & $t=0.25$
& \textbf{0.7062\%} & \textbf{20.0304\%}
& \textbf{1.0161\%}
& \textbf{1.6841\%} & \textbf{6.9710\%} \\
& & Average
& \textbf{0.6165\%} & \textbf{11.5977\%}
& \textbf{0.7667\%}
& \textbf{4.6870\%} & \textbf{11.3860\%} \\

\midrule

$10\times100$ & PINN & $t=0.15$
& 0.3842\% & 11.6050\% & 0.5687\%
& 0.9062\% & 10.0770\% \\
& & $t=0.25$
& 0.6334\% & 18.5073\% & 0.9257\%
& 1.3067\% & 13.9050\% \\
& & Average
& 0.3668\% & 10.8094\% & 0.5364\%
& 1.4173\% & 12.2250\% \\
\addlinespace

& BS-PINN-F & $t=0.15$
& \textbf{0.3610\%} & 9.9031\%
& 0.5083\%
& 0.4936\% & \textbf{6.2677\%} \\
& & $t=0.25$
& \textbf{0.6228\%} & 16.8096\%
& \textbf{0.8739\%}
& \textbf{0.6978\%} & \textbf{9.6629\%} \\
& & Average
& \textbf{0.3355\%} & 8.7730\%
& 0.4630\%
& 0.9226\% & \textbf{6.7347\%} \\
\addlinespace

& BS-PINN-L & $t=0.15$
& 0.3690\% & \textbf{9.6350\%}
& \textbf{0.5073\%}
& \textbf{0.4702\%} & 6.4980\% \\
& & $t=0.25$
& 0.6395\% & \textbf{16.5859\%}
& 0.8802\%
& 0.7382\% & 11.0404\% \\
& & Average
& 0.3390\% & \textbf{8.6044\%}
& \textbf{0.4610\%}
& \textbf{0.8764\%} & 7.0997\% \\

\bottomrule
\end{tabular}

\begin{tablenotes}[flushleft]
\footnotesize
\item Errors are reported as relative $L^2$
errors in percent. $E_{\bm{u}}$ denotes
the combined velocity error, and
$E_{\omega}$ denotes the vorticity error.
$E_p^{\mathrm{rec}}$ denotes the error in
pressure reconstructed from the predicted
velocity using the periodic pressure
Poisson equation, with pressure means
aligned at each snapshot.
Average denotes the arithmetic mean
of snapshot-wise relative errors at
$t=0,0.05,0.10,0.15,0.20,0.25$.
Bold values indicate the lowest error
among the three models for each
architecture, time, and error metric.
\end{tablenotes}

\end{threeparttable}
\end{table}

\paragraph{Long-time reconstruction of shear-layer roll-up under sparse velocity supervision.}

To examine the representation of instability-driven flow structures beyond the short-time interval, we consider a separate experiment over $t\in[0,8]$. In addition to the physics-informed constraints, both models use velocity supervision on a $32\times32$ spatial grid. The data-loss weight is set to $\lambda_{\mathrm{data}}=20$, assigning equal weights to the $u$- and $v$-component errors.
 This protocol differs from the DNS-referenced short-time tests and is therefore assessed separately through qualitative comparisons of the predicted flow patterns.

\begin{figure}[H]
    \centering
    \includegraphics[width=0.6\textwidth]
    {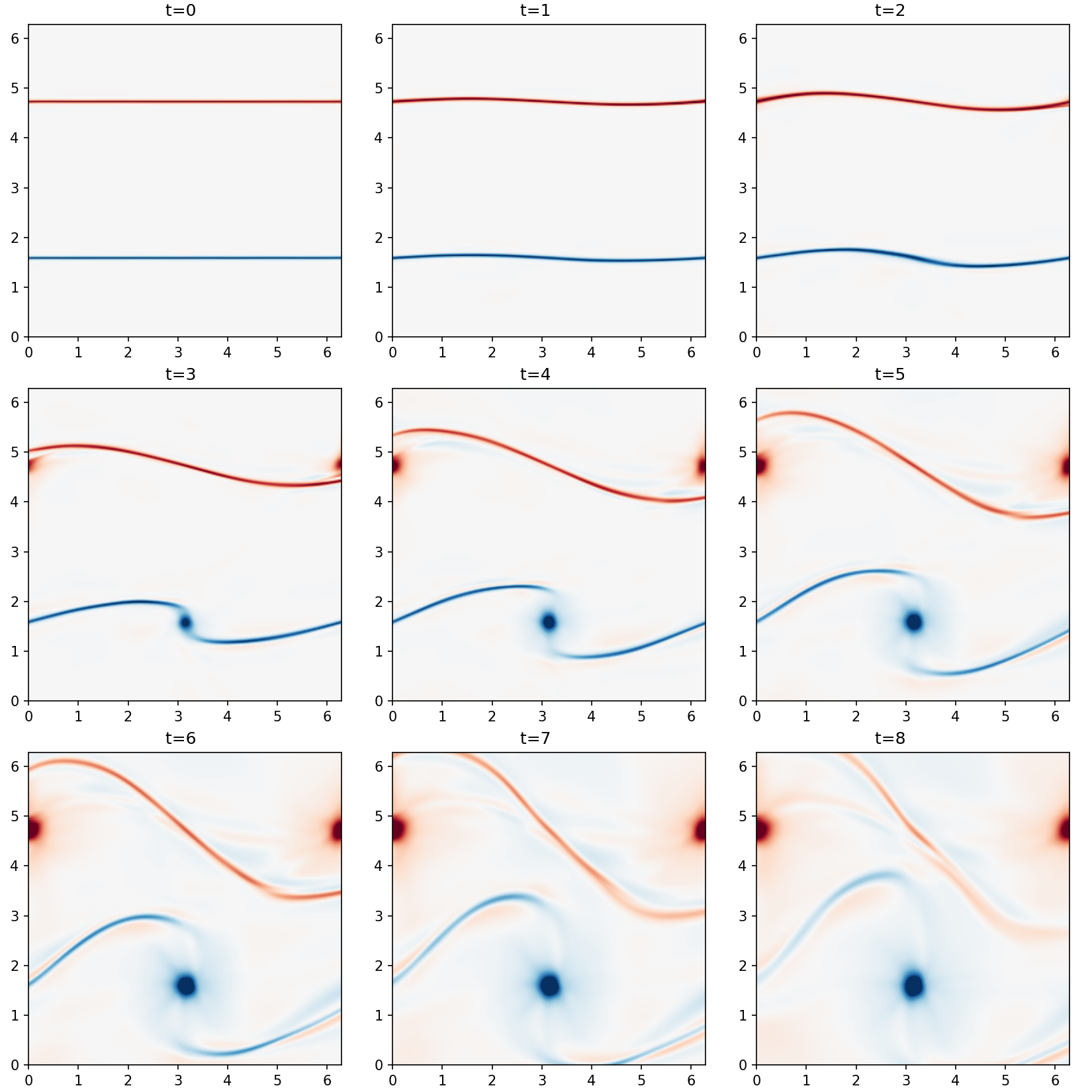}
    \caption{Predicted long-time evolution of the double shear layer at $t=0,1,\ldots,8$ using the conventional $\tanh$ PINN with velocity supervision on a $32\times32$ spatial grid. The late-time rolled-up patterns appear relatively diffuse.}
    \label{fig:dsl_long_tanh}
\end{figure}

\begin{figure}[H]
    \centering
    \includegraphics[width=0.6\textwidth]
    {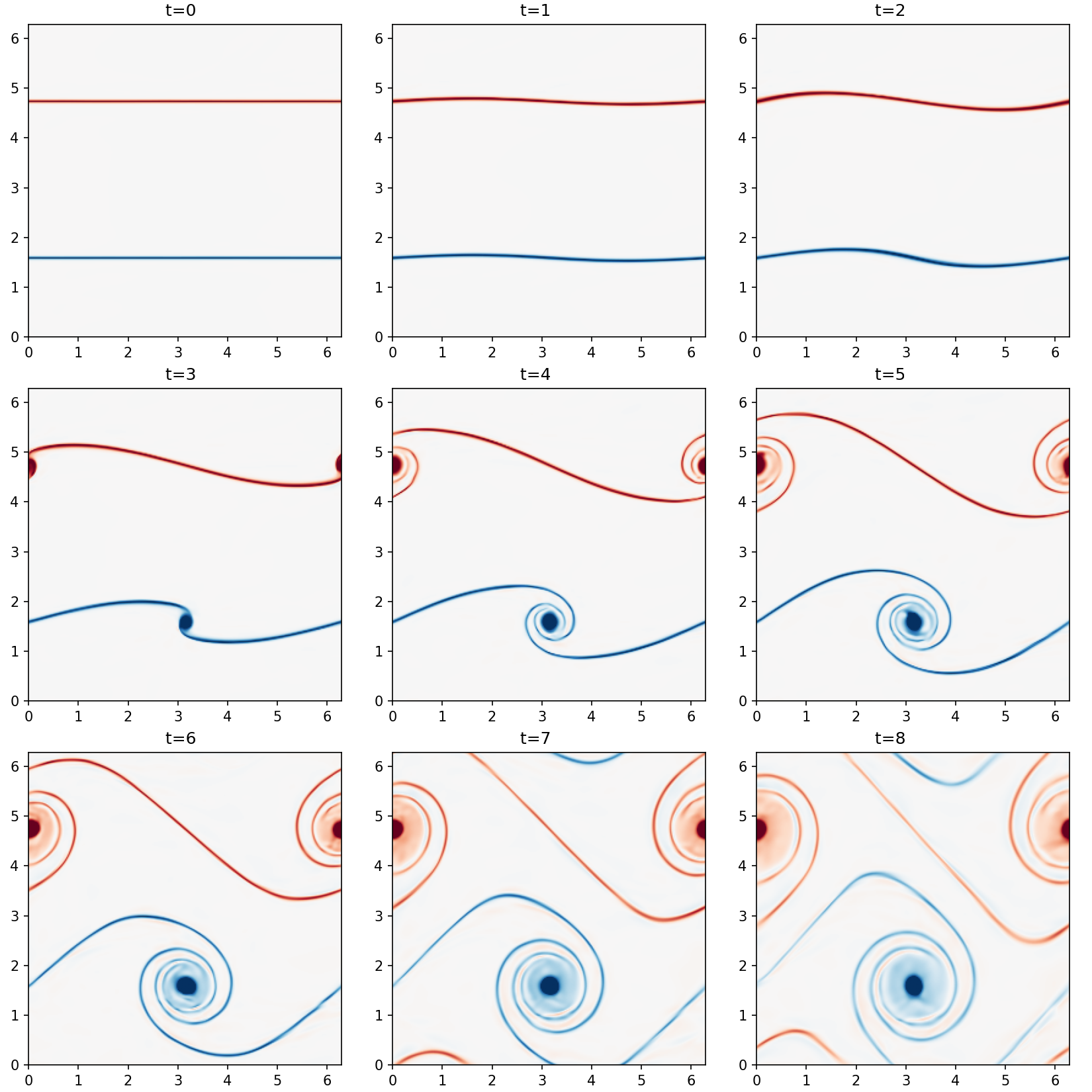}
    \caption{Predicted long-time evolution of the double shear layer at $t=0,1,\ldots,8$ using BS-PINN-F with velocity supervision on a $32\times32$ spatial grid. Distinct vortex cores and spiral roll-up patterns are visible in the later snapshots.}
    \label{fig:dsl_long_bs}
\end{figure}

Figures~\ref{fig:dsl_long_tanh} and~\ref{fig:dsl_long_bs} display the evolution predicted by the conventional $\tanh$ PINN and BS-PINN-F, respectively, at integer times from $t=0$ to $t=8$. Both models initially represent the two nearly straight shear layers and their progressive deformation. As the disturbance grows, the layers bend and begin to roll up, the development of coherent vortical structures becomes especially apparent in the later snapshots.

A clear distinction emerges after approximately $t=4$. BS-PINN-F produces identifiable vortex cores surrounded by increasingly pronounced spiral-shaped contours, and these rolled-up structures remain visible through $t=8$. By contrast, the corresponding structures produced by the conventional PINN are more diffuse, with less sharply defined cores and fewer distinguishable spiral windings. This visual difference indicates that the kernel-informed activation can represent the characteristic morphology of shear-layer roll-up more distinctly under the same sparse-supervision setting.

Importantly, no long-time DNS reference or quantitative error estimates are reported for this additional experiment. The comparison therefore supports a qualitative observation about the resolution of physically characteristic vortex-roll-up patterns, it does not establish quantitative superiority in the true long-time flow evolution.

\subsection{Discussion}
\label{subsec:results_discussion}

Taken together, the benchmark results indicate
that changing the hidden-layer nonlinearity
while retaining the governing Navier--Stokes
residuals can improve the accuracy of
incompressible-flow approximations.
Both kernel-inspired variants reduce the
reported combined velocity and pressure errors
in each of the 21 benchmark--architecture
comparisons. For the double-shear-layer
configurations, the pressure metric refers
to pressure reconstructed from the predicted
velocity. The gains are not confined to one
particular flow: they occur for a steady
analytical solution, a time-dependent wake,
a three-dimensional flow, and periodic flows
with smooth or localized spatial structures.

The largest gains are observed in several periodic-flow configurations. In the $4\times50$ Taylor--Green experiment, BS-PINN-L reduces the average combined velocity error by 69.1\% and the pressure error by 68.7\%, for the double shear layer with $\rho=10$ and a $10\times100$ network, the combined velocity-error reduction reaches 70.1\%. The double-shear-layer results are particularly informative because they connect these improvements to recognizable flow structures. In the DNS-referenced short-time tests, both BS-PINN variants reduce vorticity errors and BS-PINN-F closely reproduces the spatial localization of the two narrow shear bands. For instance, at $\rho=40$ with the $4\times50$ architecture, BS-PINN-L reduces the mean vorticity error from 21.6803\% to 11.0086\%. In the separate long-time, sparsely supervised experiment, BS-PINN-F also produces more distinct vortex cores and spiral roll-up patterns than the conventional $\tanh$ PINN, particularly over $t\in[4,8]$. These observations suggest that the kernel-informed activation can benefit not only global field errors but also the representation of characteristic shear-layer structures. The long-time evidence remains qualitative because a corresponding DNS-based error assessment is not available.

The learnable kernel is not consistently
superior to its fixed-scale counterpart.
BS-PINN-F gives the lowest mean combined
velocity error for each tested network size
in the Beltrami problem, and it also
outperforms BS-PINN-L on several Taylor--Green
and pressure metrics. The results therefore
point to the usefulness of the rational
kernel profile itself, while the additional
benefit of scale adaptation depends on the
specific problem and network configuration.

Finally, BS-PINN is a representation-informed
neural architecture, not a direct numerical
evaluation of a stochastic velocity representation
or a Biot--Savart integral. The whole-space
stochastic formulas motivate the activation shape,
but do not imply that a finite network exactly
implements the corresponding integral operator
on bounded or periodic domains.
The numerical evidence supports the effectiveness
of the proposed activation on the tested tasks,
further work is required to characterize its
approximation properties, parameter sensitivity,
training robustness, and long-time accuracy
quantitatively.

\section{Conclusion}

In this work, we investigated how the analytical structure of incompressible fluid flows can inform the design of physics-informed neural networks. Motivated by stochastic integral representations of the Navier--Stokes equations and the associated Biot--Savart law, we introduced a kernel-informed activation function derived from a regularized component of the Biot--Savart kernel. Based on this activation, we developed two variants, BS-PINN-F and BS-PINN-L, with fixed and learnable kernel scales, respectively. Both variants retain the conventional velocity--pressure formulation and physics-informed training framework. The proposed approach therefore incorporates information about the flow structure into the neural approximation space without modifying the governing equations or introducing additional physical constraints.

The proposed method was evaluated on five benchmark problems covering steady and unsteady flows, two- and three-dimensional configurations, and varying vortical structures. Across 21 benchmark--architecture comparisons, both BS-PINN variants consistently reduced the combined velocity error relative to the conventional $\tanh$-based PINN, with mean relative error reductions of 31.7\% and 33.2\% for BS-PINN-F and BS-PINN-L, respectively. Improvements were also observed in the reported pressure and vorticity errors. Beyond these aggregate metrics, the double-shear-layer experiments illustrate the ability of the kernel-informed representation to reproduce characteristic spatial flow structures. DNS-referenced short-time predictions resolve the narrow, oppositely signed shear layers with reduced vorticity error, while a separate long-time experiment with sparse velocity supervision shows more clearly defined vortex cores and spiral roll-up patterns for BS-PINN-F than for the conventional PINN through $t=8$. This long-time evidence is qualitative, since no corresponding DNS-based error estimates are reported. The results also demonstrate the applicability of the activation to the tested three-dimensional Beltrami flow.

An important observation is that learning the kernel scale does not consistently outperform fixing it. While BS-PINN-L provides additional flexibility and achieves the best accuracy in several configurations, BS-PINN-F remains competitive and is preferable for certain benchmarks and error metrics. This suggests that the kernel-informed nonlinear profile itself plays an important role in the observed improvements, whereas the benefit of scale adaptation depends on the flow configuration and network architecture.

The present construction should be understood as a representation-informed neural approximation rather than a direct implementation of the stochastic integral formula. The activation is motivated by a scalar profile of the regularized two-dimensional Biot--Savart kernel, and its numerical performance does not constitute a general approximation or convergence guarantee. Future work will focus on establishing theoretical approximation properties of kernel-informed networks, investigating the influence of kernel scales and training robustness, and extending the framework to more complex flow configurations and boundary geometries. More broadly, this study suggests that exploiting analytical solution structures offers a complementary direction for improving neural PDE solvers beyond conventional loss-function and sampling strategies.

\section*{Acknowledgments}
Zihao Guo was supported by the China Postdoctoral Science Foundation (Grant No.~2026M793346).

\newpage

\appendix
\renewcommand{\thesection}{Appendix \Alph{section}}

\section{Additional Numerical Results and Visualizations}
\label{representation}

\begin{figure}[H]
    \centering
    \includegraphics[width=\textwidth]{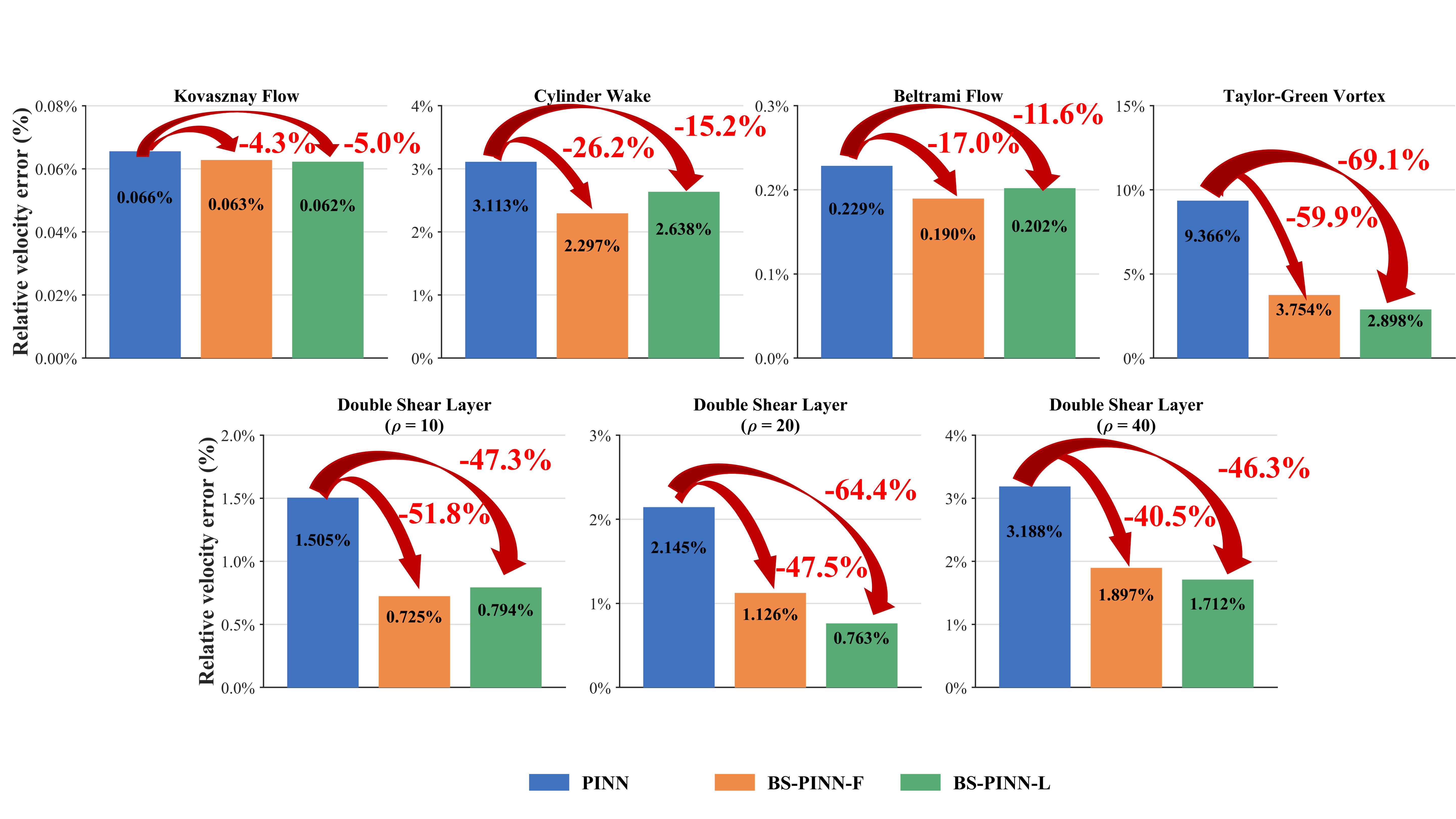}
    \caption{Comparison of combined velocity errors for the $4\times50$ architecture. Percentage annotations show the signed relative error changes with respect to the conventional PINN; negative values indicate error reductions.}
    \label{fig:velocity_450}
\end{figure}

\begin{figure}[H]
    \centering
    \includegraphics[width=\textwidth]{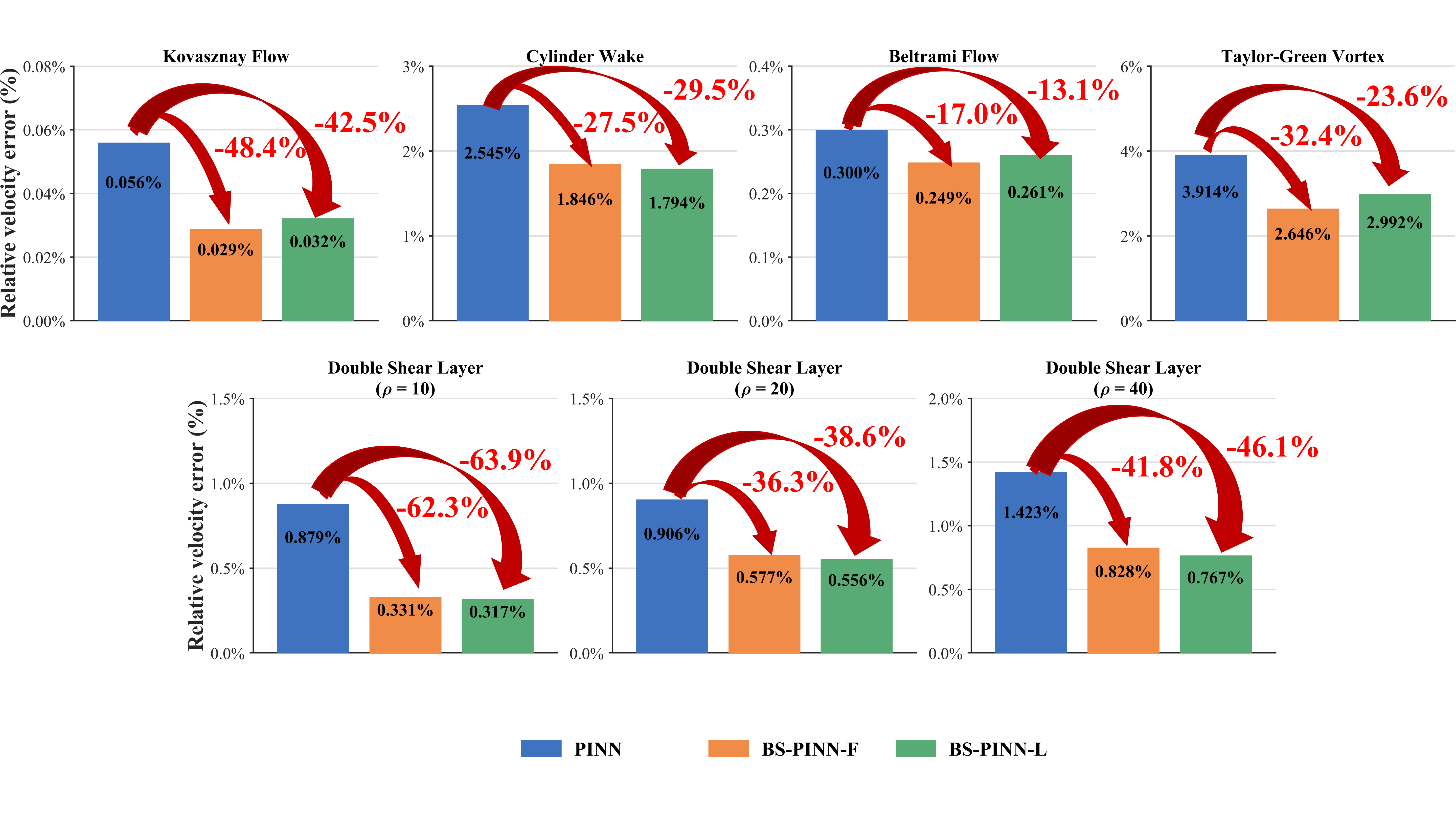}
    \caption{Comparison of combined velocity errors for the $7\times50$ architecture. Percentage annotations show the signed relative error changes with respect to the conventional PINN; negative values indicate error reductions.}
    \label{fig:velocity_750}
\end{figure}

\begin{figure}[H]
    \centering
    \includegraphics[width=\textwidth]{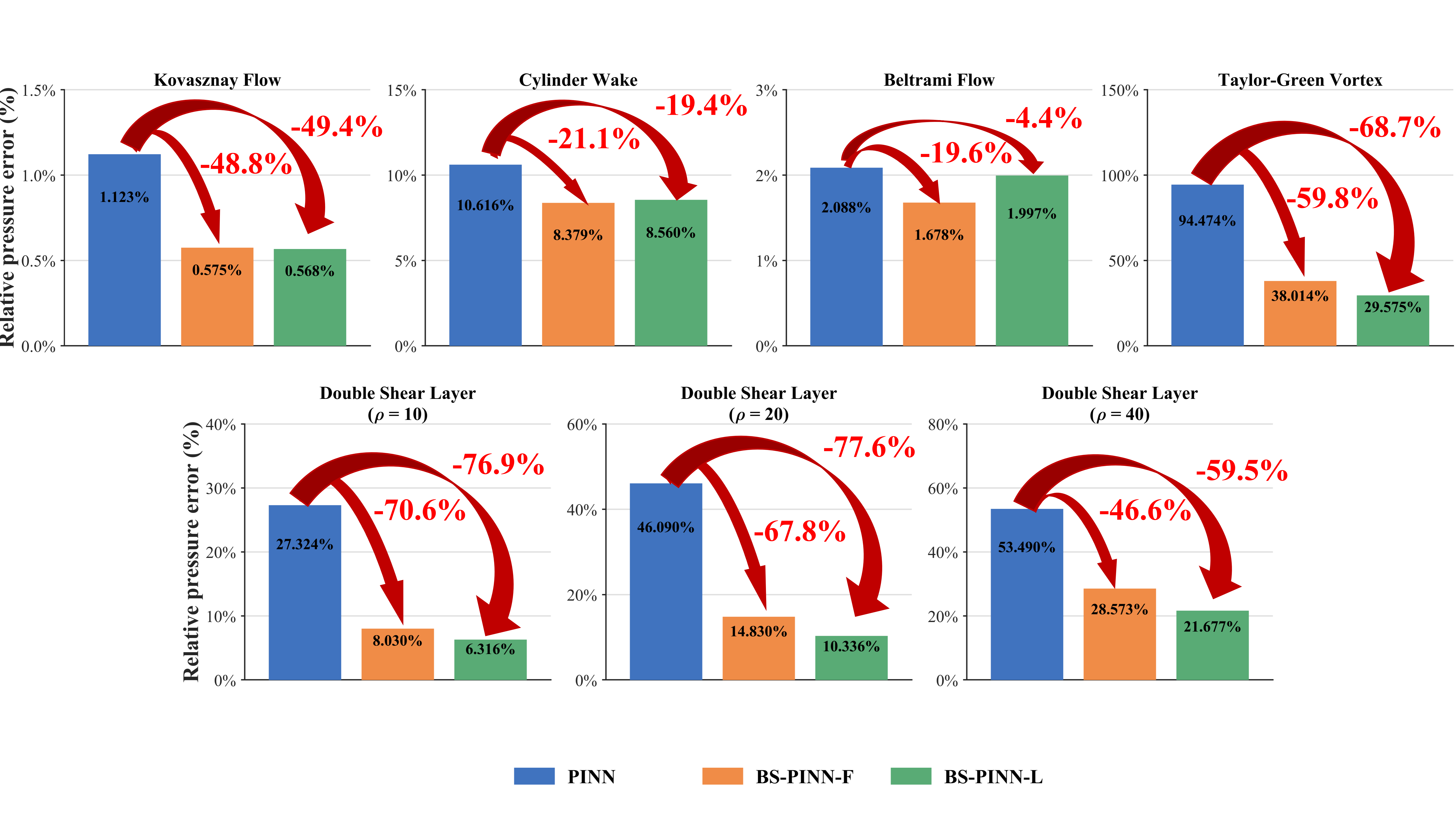}
    \caption{Comparison of reported pressure errors for the $4\times50$ architecture. Percentage annotations show the signed relative error changes with respect to the conventional PINN; negative values indicate error reductions.}
    \label{fig:pressure_450}
\end{figure}

\begin{figure}[H]
    \centering
    \includegraphics[width=\textwidth]{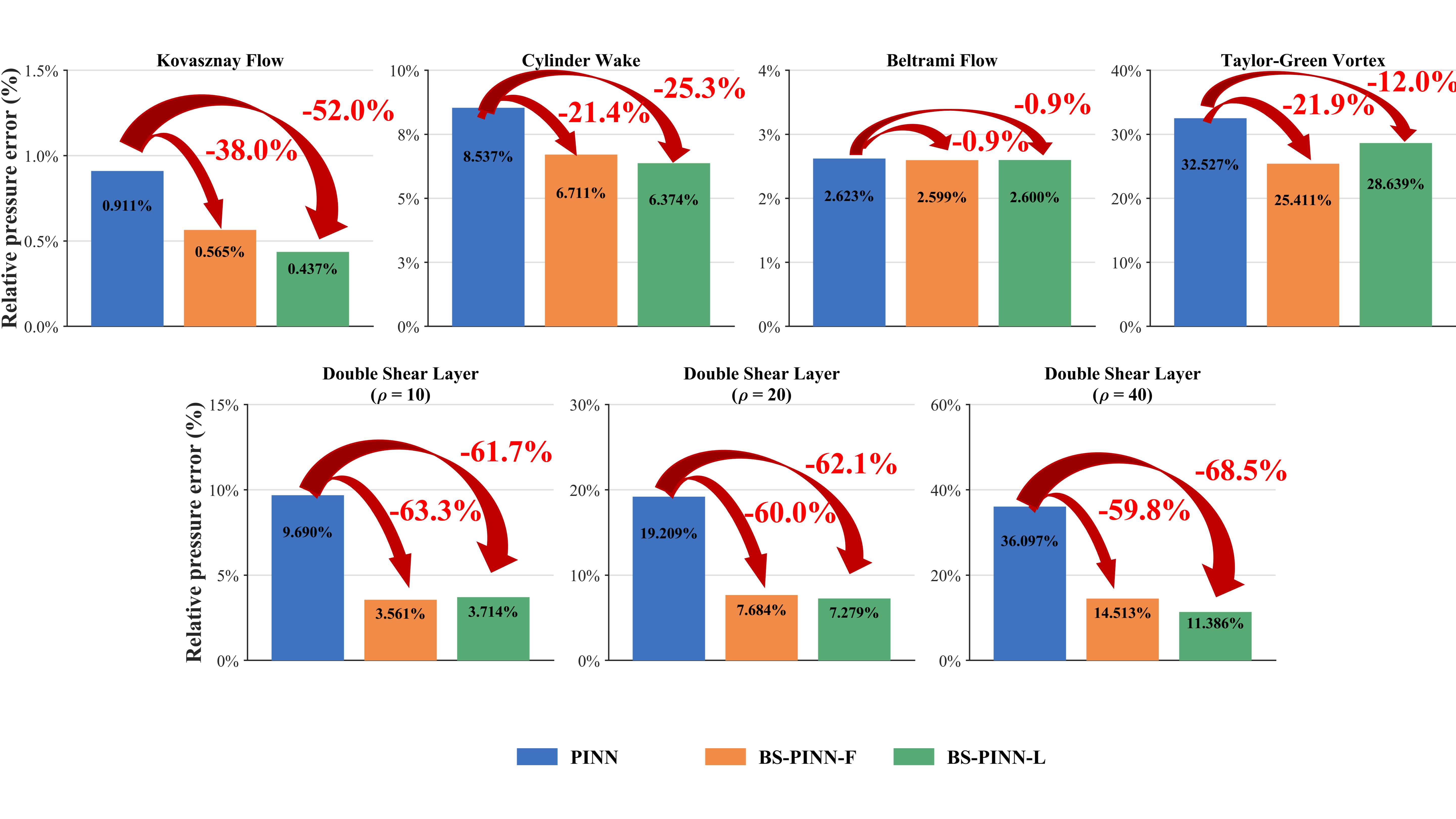}
    \caption{Comparison of reported pressure errors for the $7\times50$ architecture. Percentage annotations show the signed relative error changes with respect to the conventional PINN; negative values indicate error reductions.}
    \label{fig:pressure_750}
\end{figure}

\begin{figure}[H]
    \centering
    \includegraphics[width=0.8\textwidth]{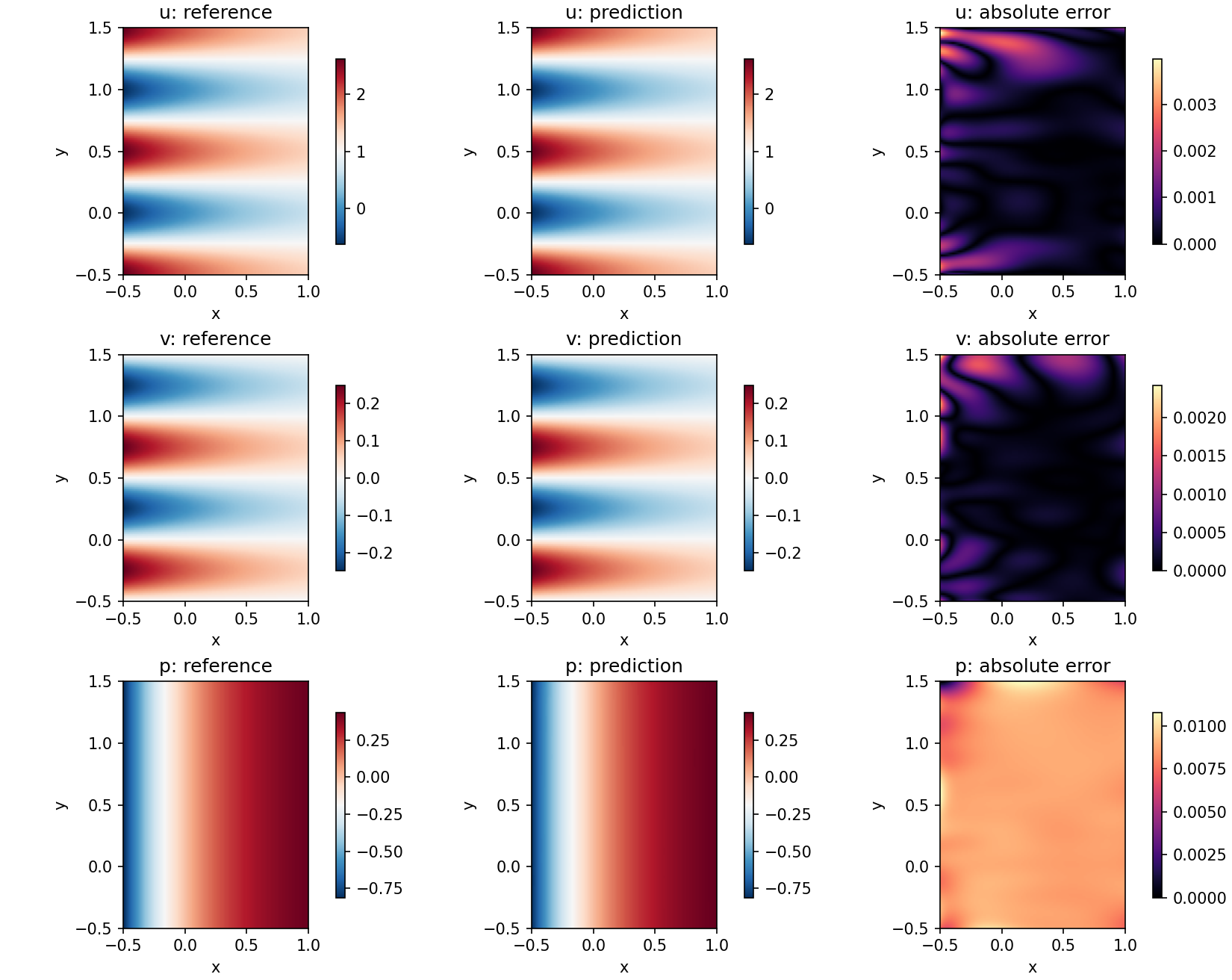}
    \caption{Kovasznay flow: reference solution,
    BS-PINN-F prediction, and pointwise absolute
    errors using the $4\times50$ architecture.
    The three rows show $u$, $v$, and $p$,
    respectively.}
    \label{fig:kovasznay450}
\end{figure}

\begin{figure}[H]
    \centering
    \includegraphics[width=0.8\textwidth]{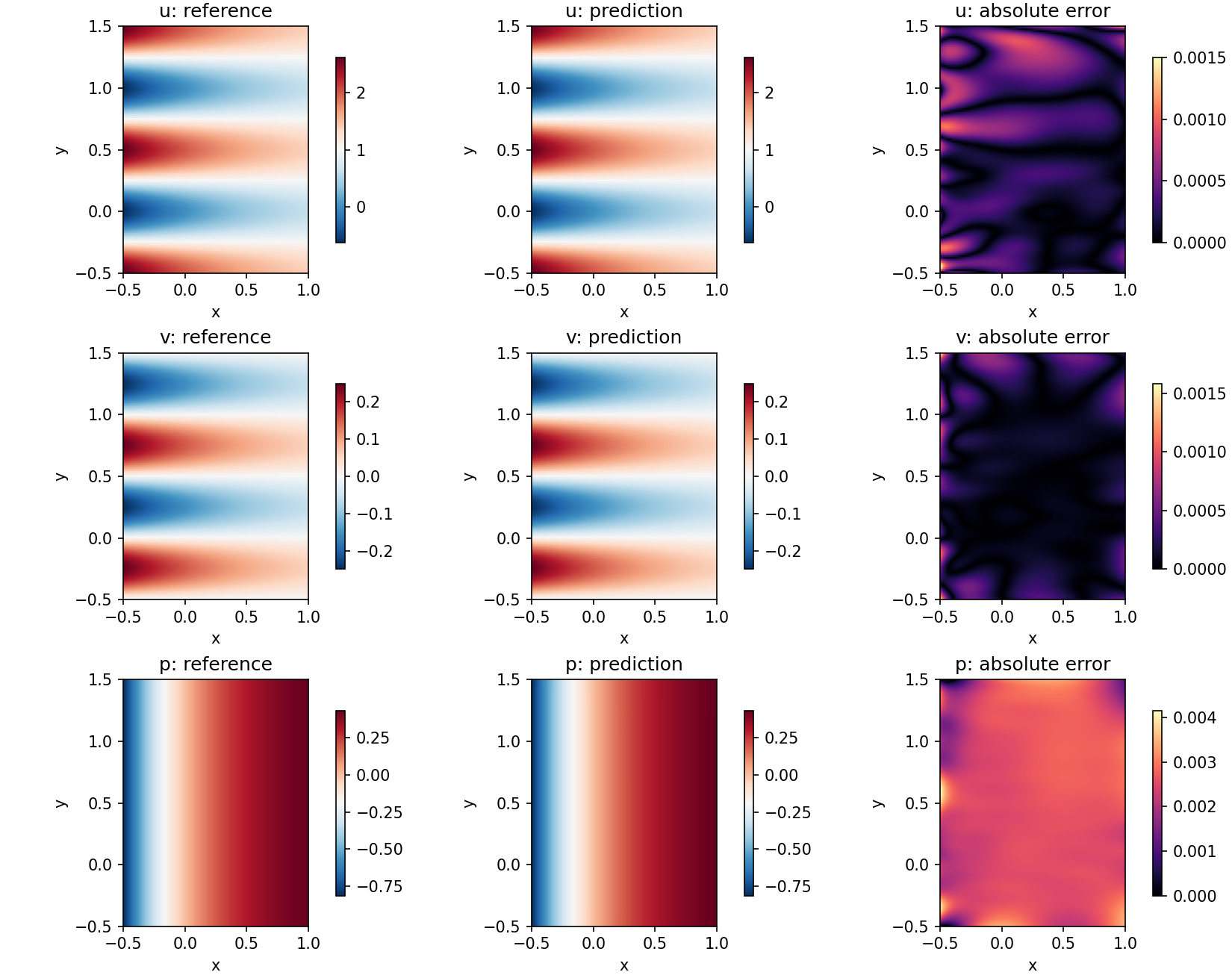}
    \caption{Kovasznay flow: reference solution,
    BS-PINN-F prediction, and pointwise absolute
    errors using the $7\times50$ architecture.
    The three rows show $u$, $v$, and $p$,
    respectively.}
    \label{fig:kovasznay750}
\end{figure}

\begin{figure}[H]
    \centering
    \includegraphics[width=0.8\textwidth]{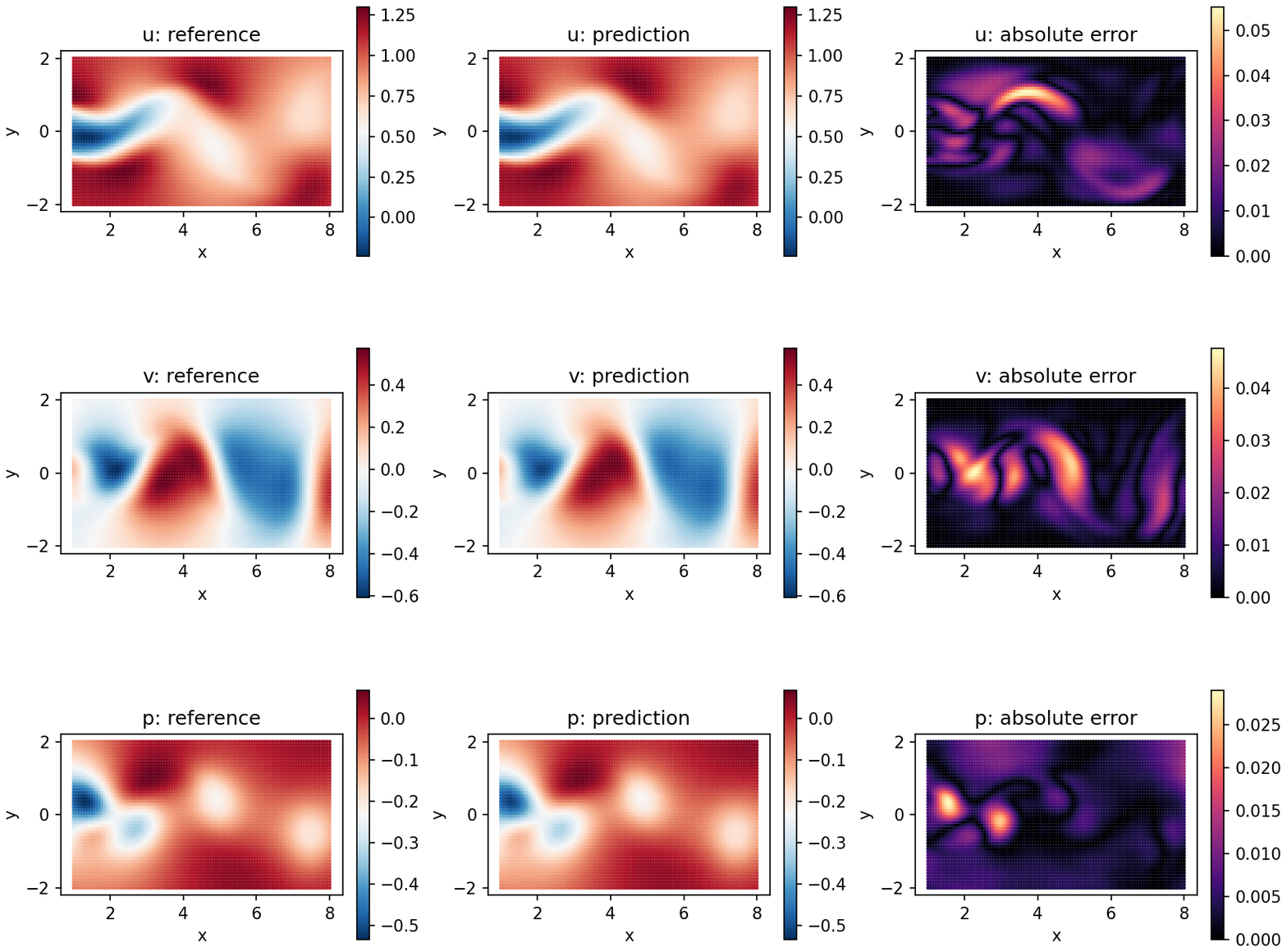}
    \caption{Cylinder wake: reference fields,
    BS-PINN-F predictions, and absolute errors
    at $t=3.5$ using the $4\times50$ architecture.}
\end{figure}

\begin{figure}[H]
    \centering
    \includegraphics[width=0.8\textwidth]{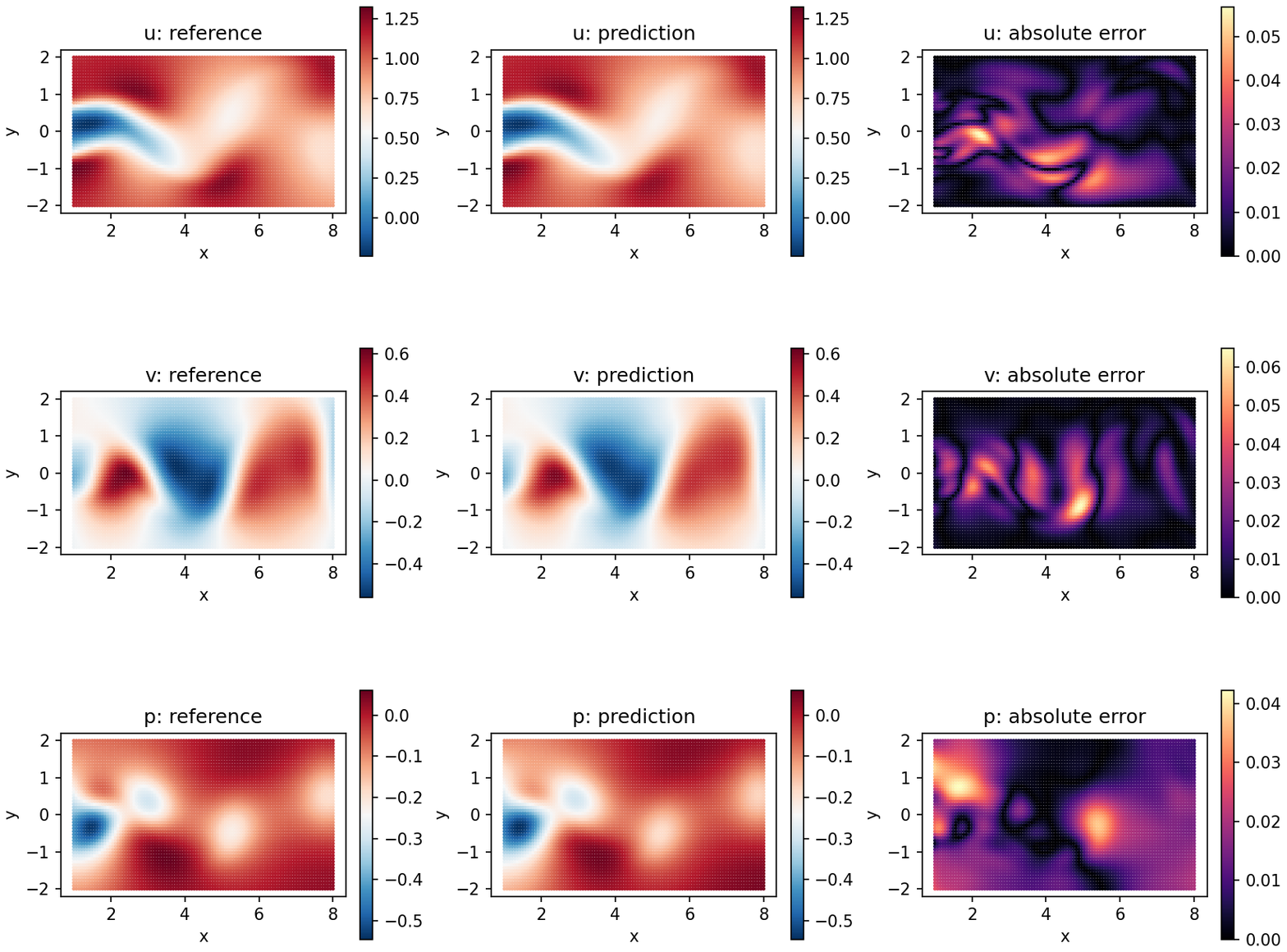}
    \caption{Cylinder wake: reference fields,
    BS-PINN-F predictions, and absolute errors
    at $t=6.9$ using the $4\times50$ architecture.}
\end{figure}

\begin{figure}[H]
    \centering
    \includegraphics[width=0.8\textwidth]{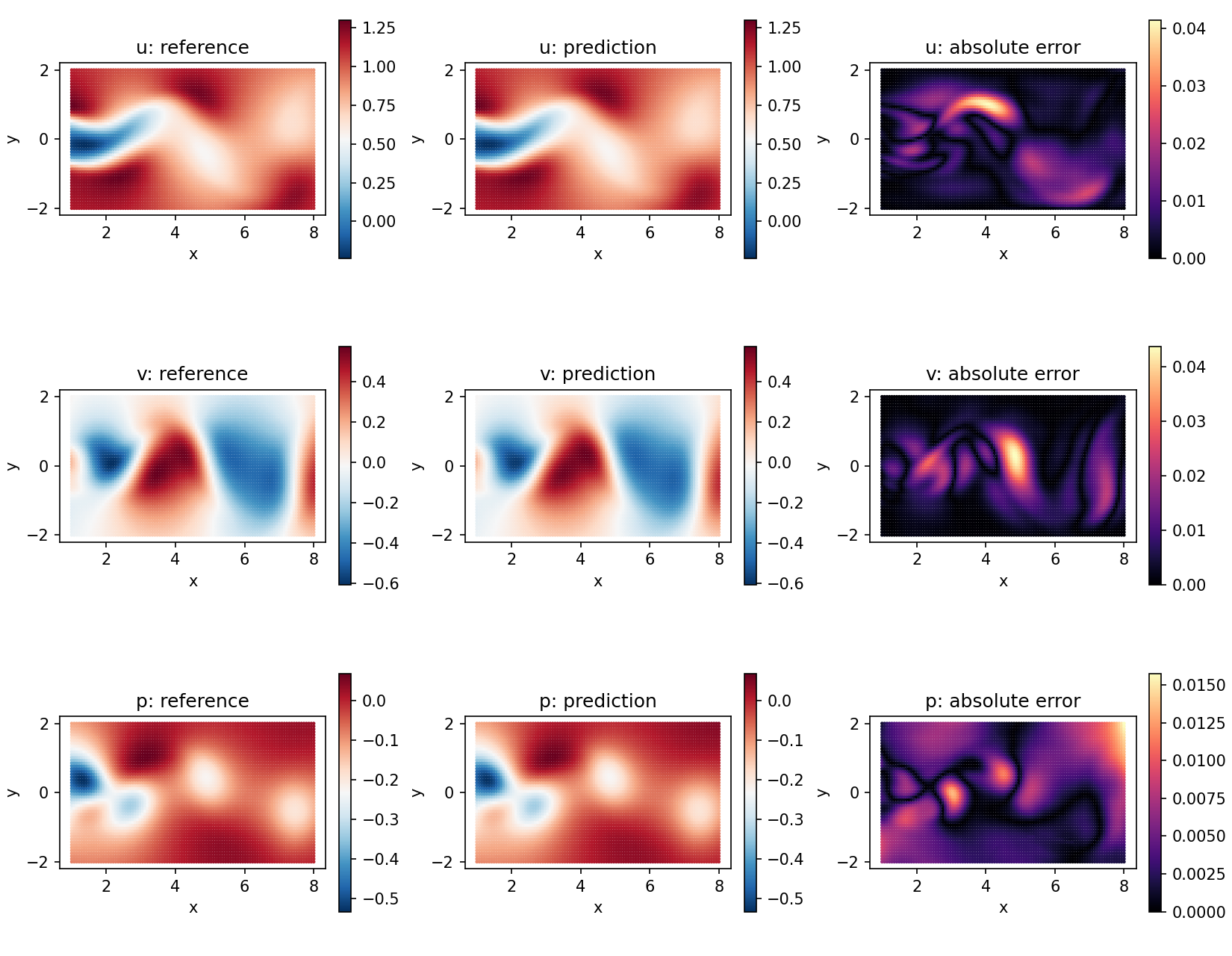}
    \caption{Cylinder wake: reference fields,
    BS-PINN-F predictions, and absolute errors
    at $t=3.5$ using the $7\times50$ architecture.}
\end{figure}

\begin{figure}[H]
    \centering
    \includegraphics[width=0.8\textwidth]{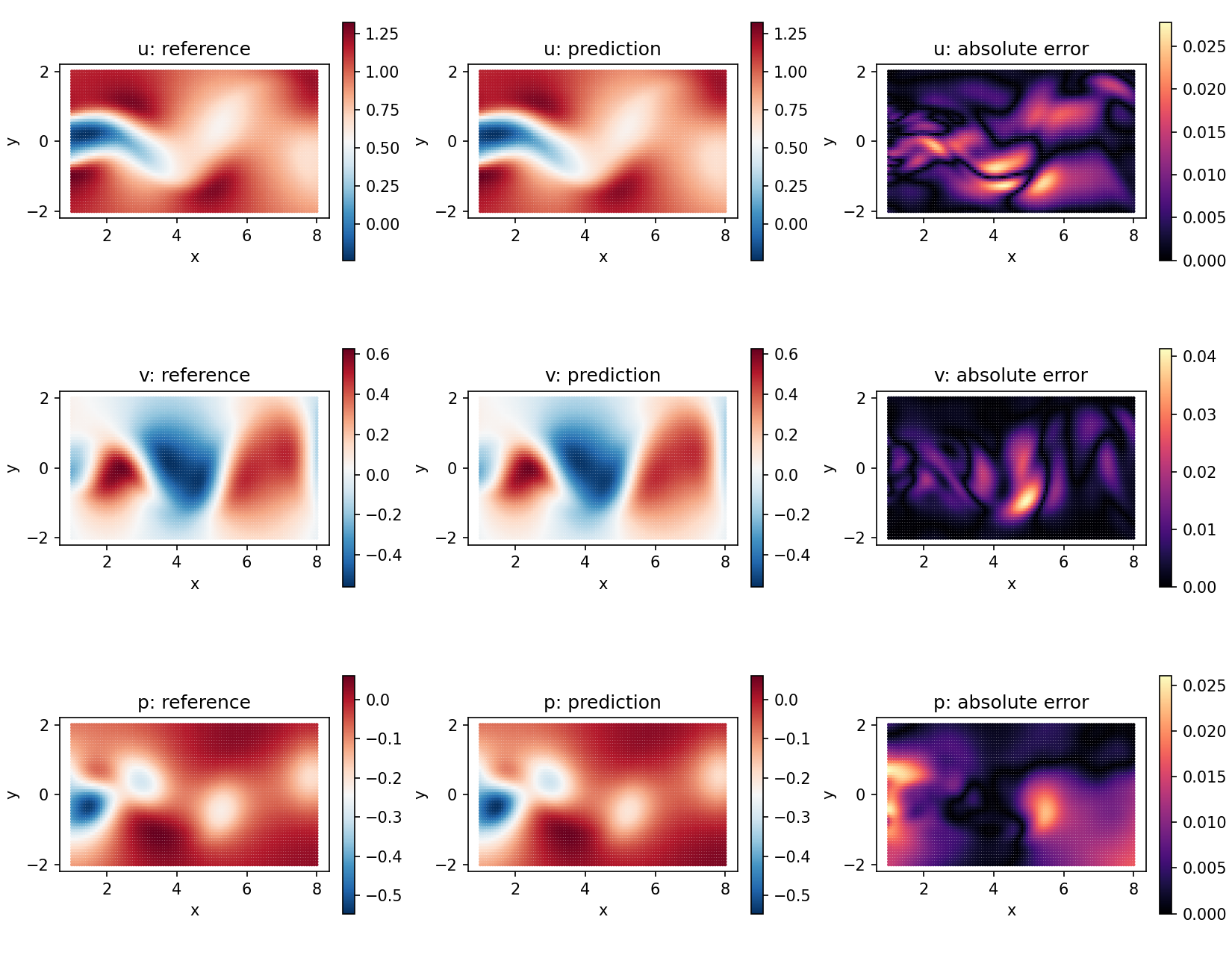}
    \caption{Cylinder wake: reference fields,
    BS-PINN-F predictions, and absolute errors
    at $t=6.9$ using the $7\times50$ architecture.}
\end{figure}

\begin{figure}[H]
    \centering
    \includegraphics[width=0.8\textwidth]{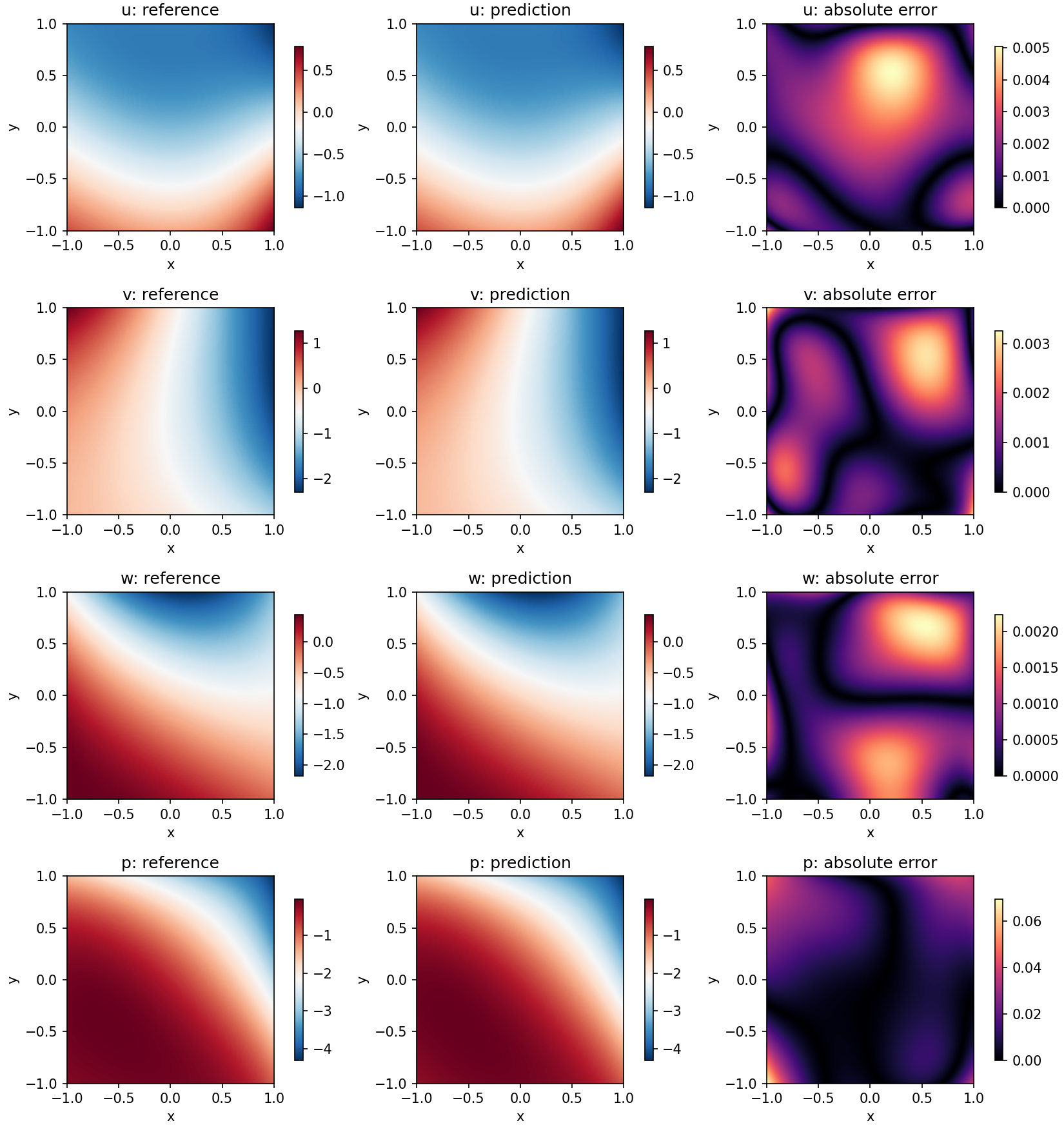}
    \caption{Three-dimensional Beltrami flow:
    reference fields, BS-PINN-F predictions,
    and absolute errors on the $z=0$ plane
    at $t=0.5$ using the $4\times50$ architecture.}
\end{figure}

\begin{figure}[H]
    \centering
    \includegraphics[width=0.8\textwidth]{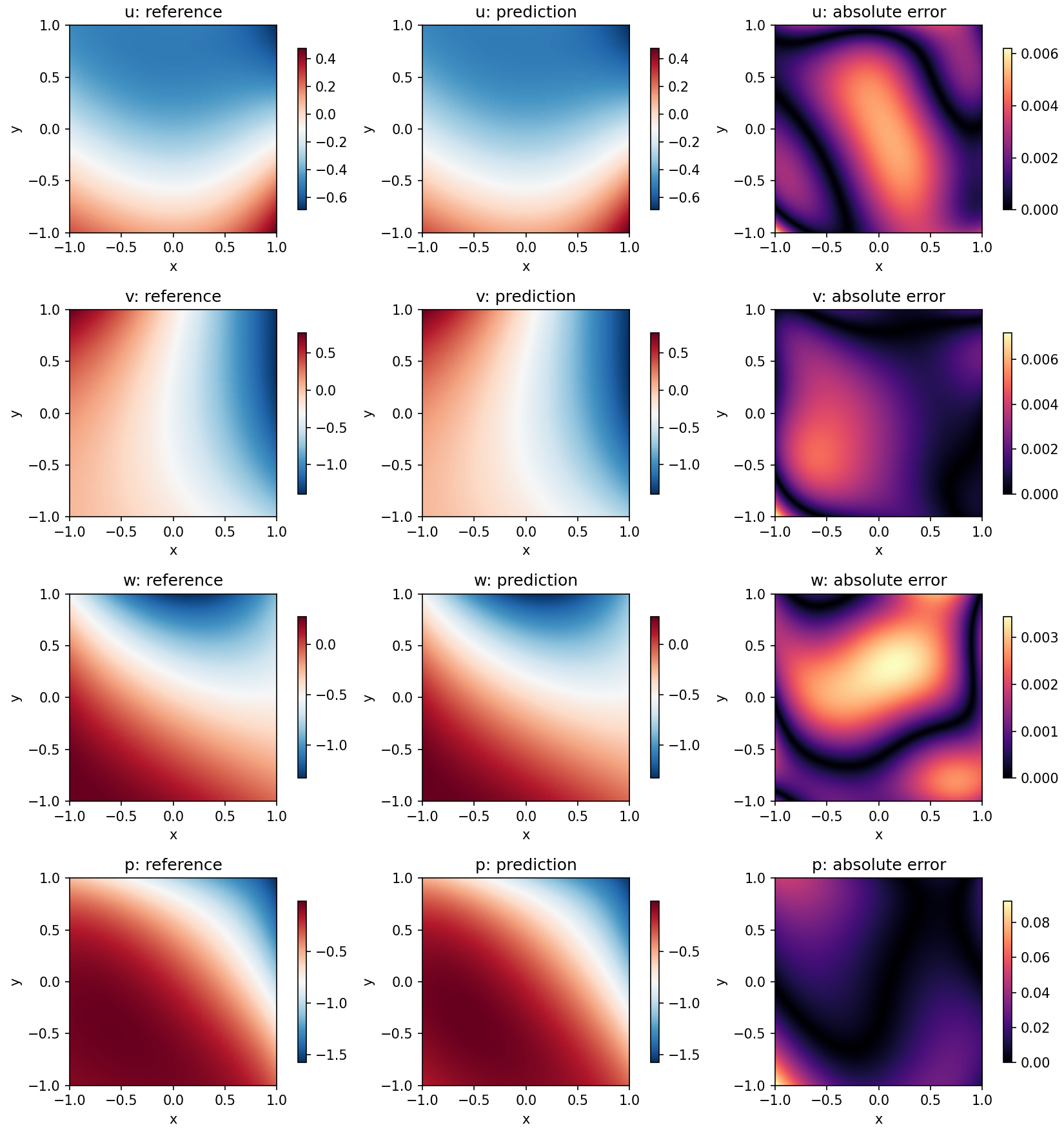}
    \caption{Three-dimensional Beltrami flow:
    reference fields, BS-PINN-F predictions,
    and absolute errors on the $z=0$ plane
    at $t=1.0$ using the $4\times50$ architecture.}
\end{figure}

\begin{figure}[H]
    \centering
    \includegraphics[width=0.8\textwidth]{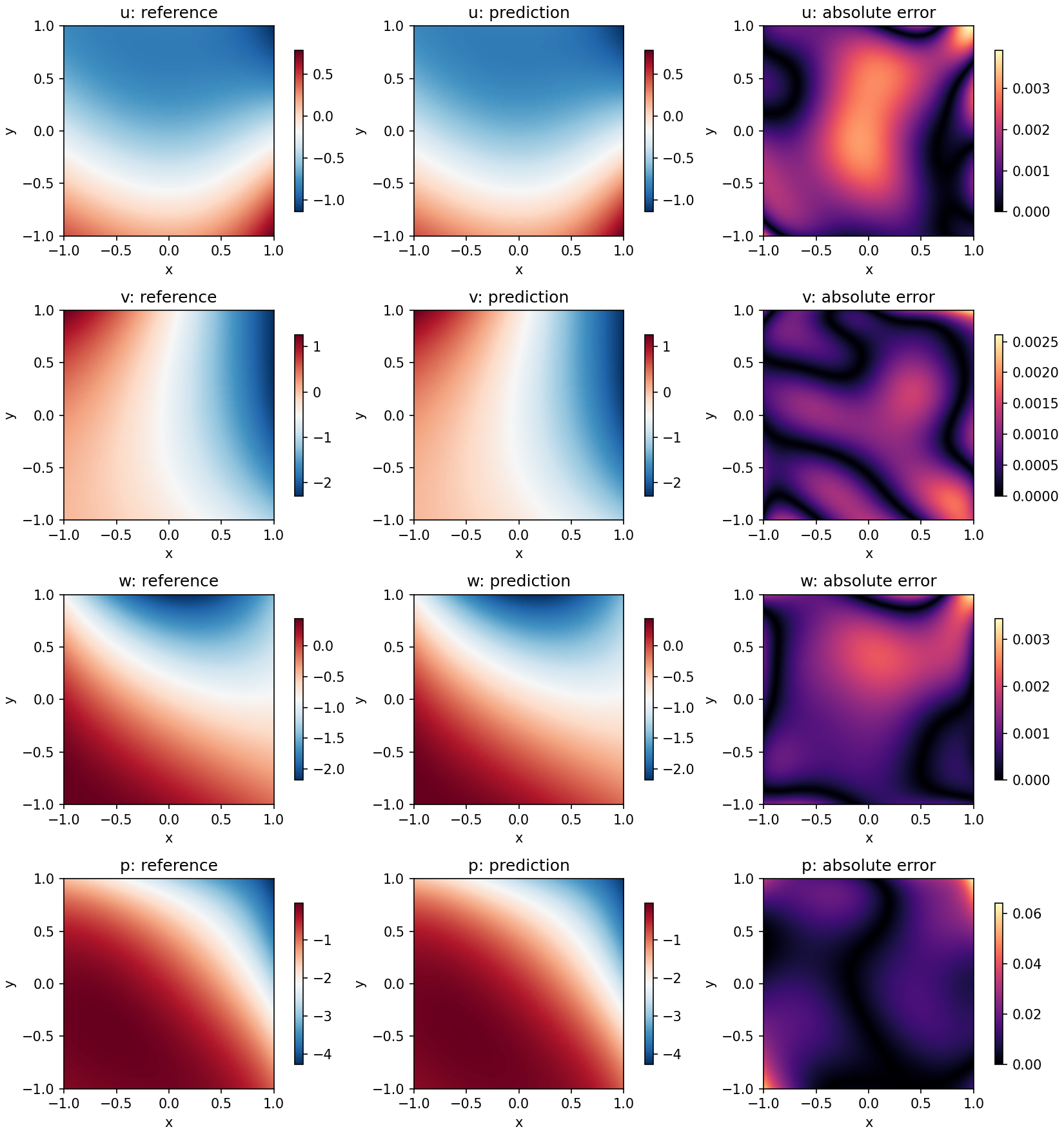}
    \caption{Three-dimensional Beltrami flow:
    reference fields, BS-PINN-F predictions,
    and absolute errors on the $z=0$ plane
    at $t=0.5$ using the $7\times50$ architecture.}
\end{figure}

\begin{figure}[H]
    \centering
    \includegraphics[width=0.8\textwidth]{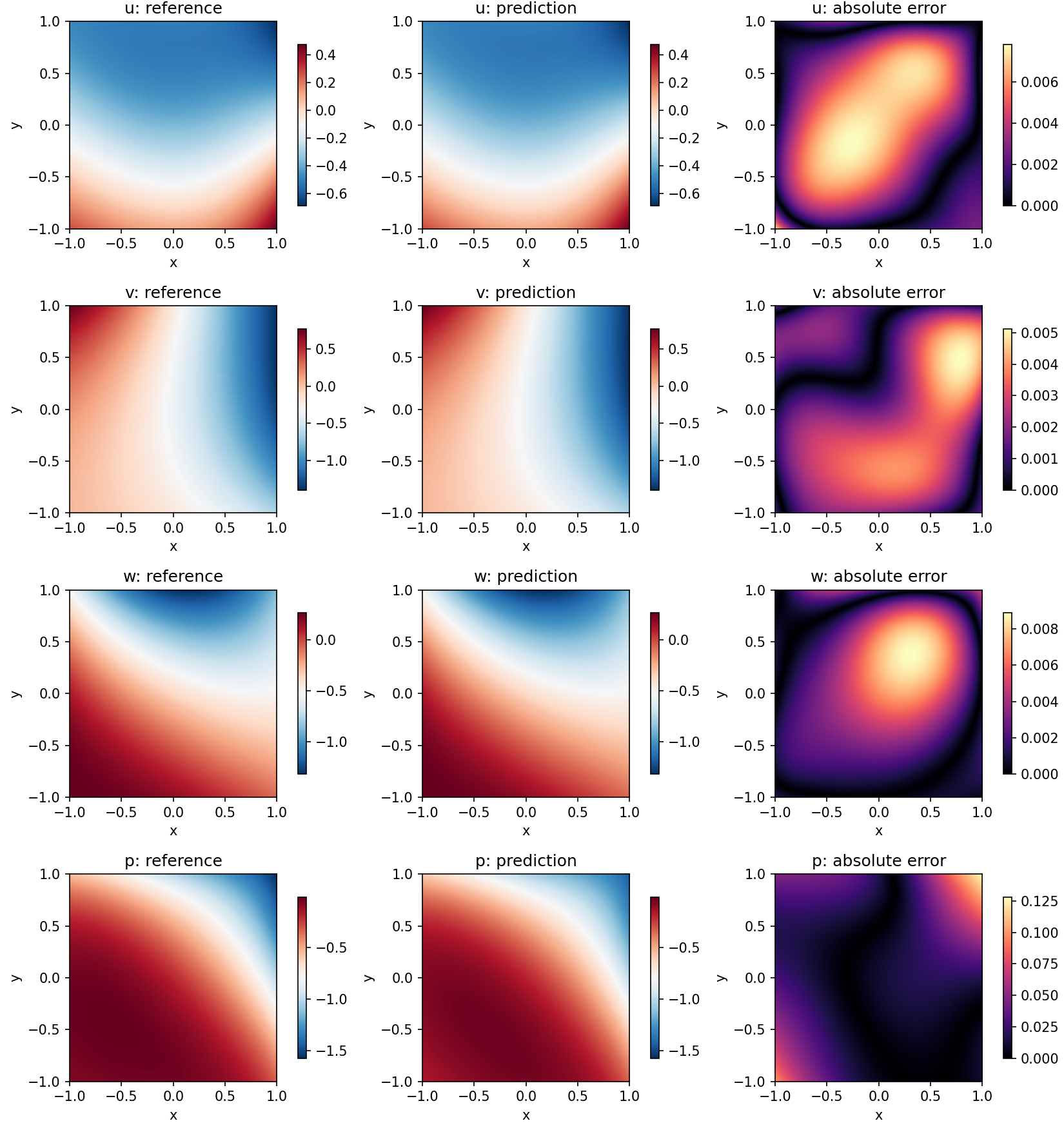}
    \caption{Three-dimensional Beltrami flow:
    reference fields, BS-PINN-F predictions,
    and absolute errors on the $z=0$ plane
    at $t=1.0$ using the $7\times50$ architecture.}
\end{figure}

\begin{figure}[H]
    \centering
    \includegraphics[width=0.8\textwidth]{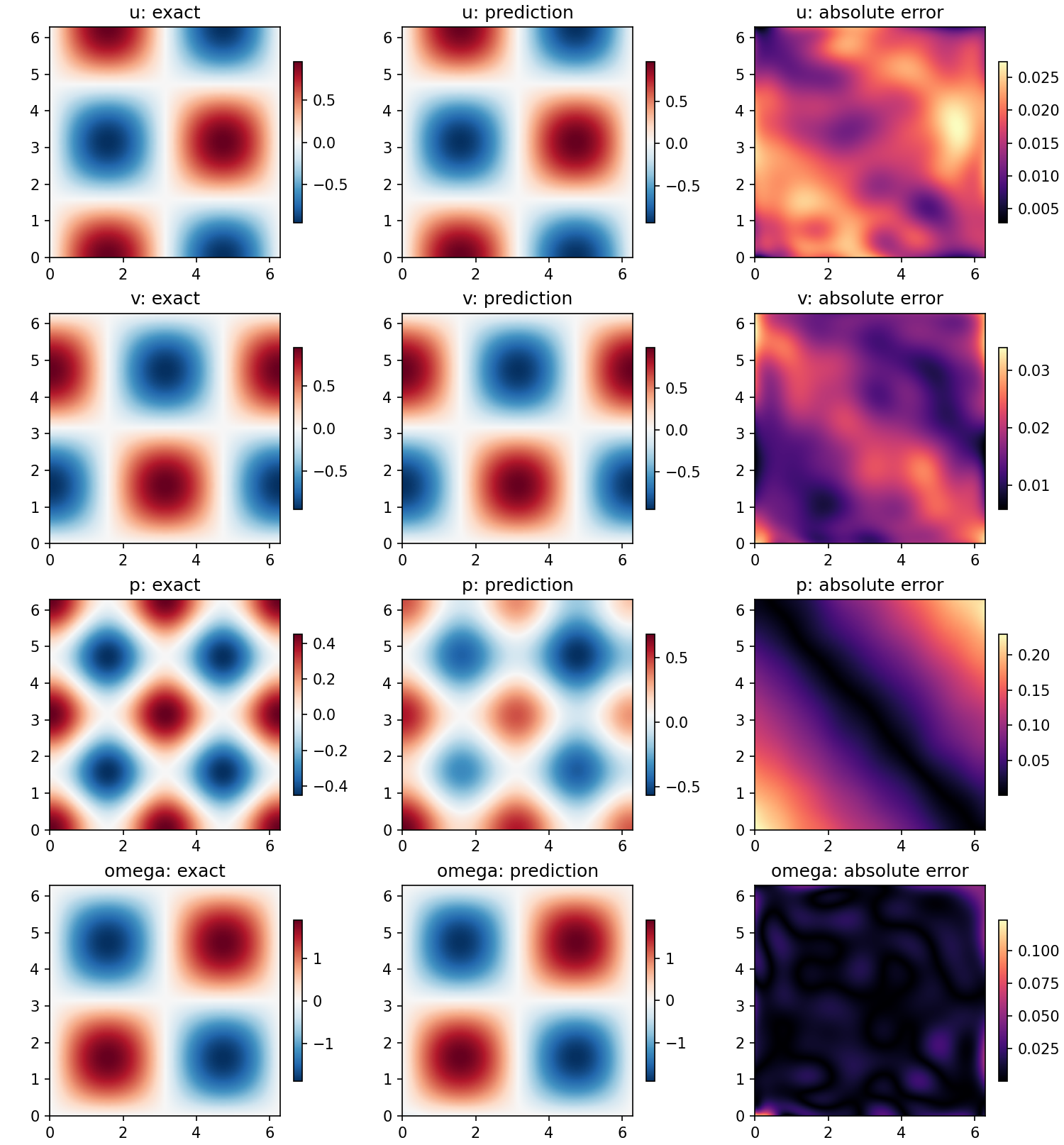}
    \caption{Taylor--Green vortex: reference fields,
    BS-PINN-F predictions, and absolute errors
    at $t=0.5$ using the $4\times50$ architecture.}
\end{figure}

\begin{figure}[H]
    \centering
    \includegraphics[width=0.8\textwidth]{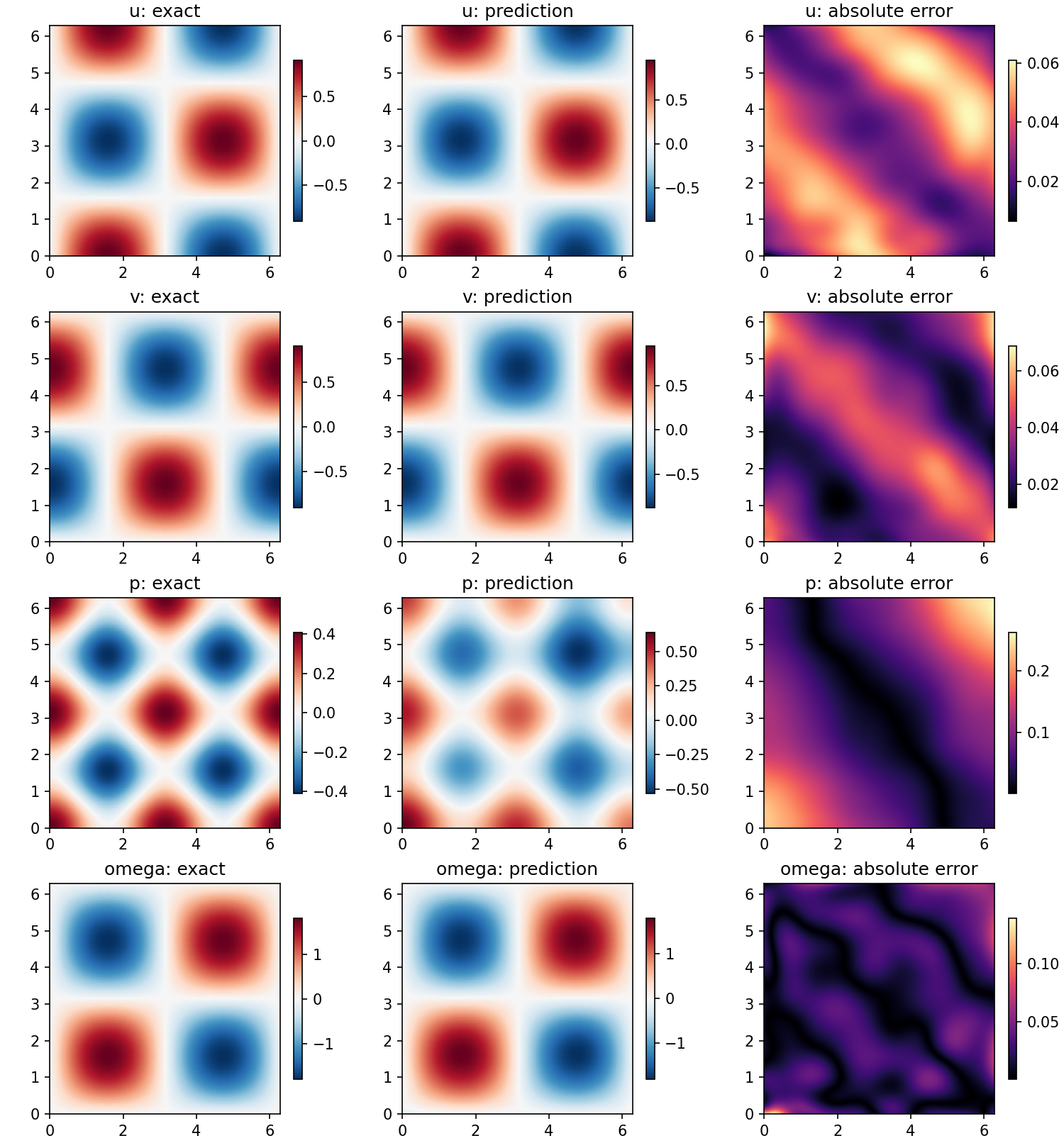}
    \caption{Taylor--Green vortex: reference fields,
    BS-PINN-F predictions, and absolute errors
    at $t=1.0$ using the $4\times50$ architecture.}
\end{figure}

\begin{figure}[H]
    \centering
    \includegraphics[width=0.8\textwidth]{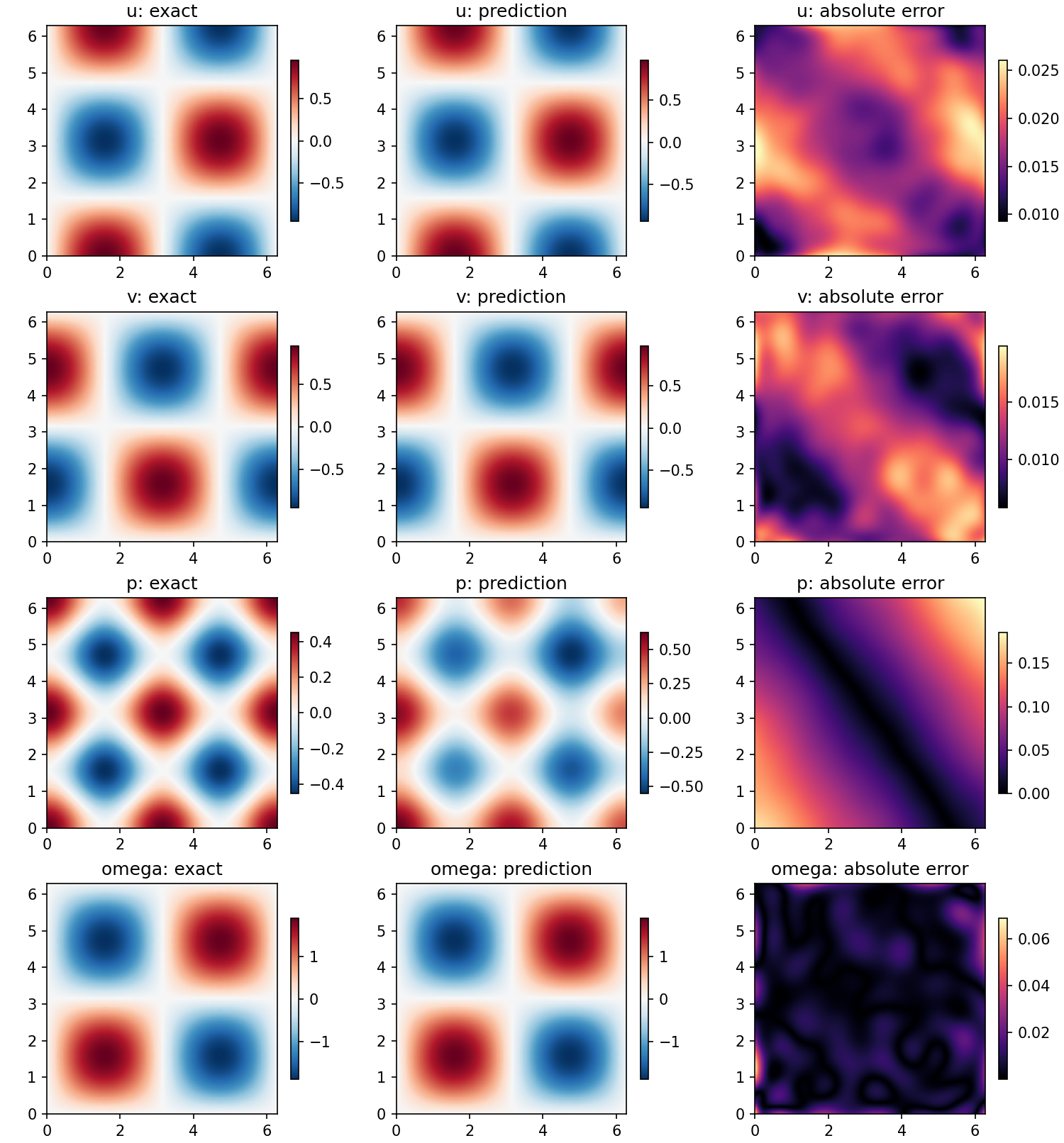}
    \caption{Taylor--Green vortex: reference fields,
    BS-PINN-F predictions, and absolute errors
    at $t=0.5$ using the $7\times50$ architecture.}
\end{figure}

\begin{figure}[H]
    \centering
    \includegraphics[width=0.8\textwidth]{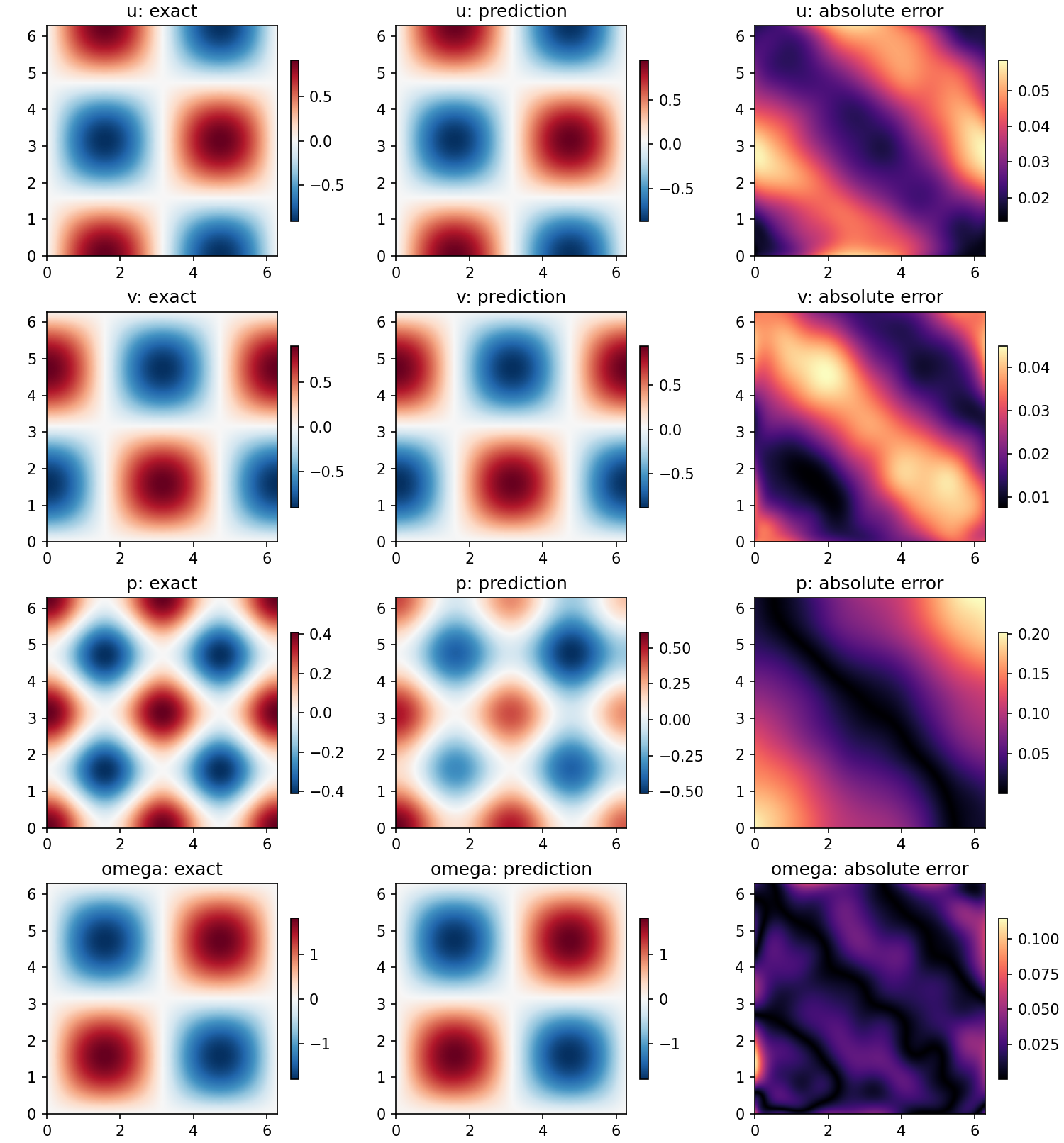}
    \caption{Taylor--Green vortex: reference fields,
    BS-PINN-F predictions, and absolute errors
    at $t=1.0$ using the $7\times50$ architecture.}
\end{figure}

\begin{figure}[H]
    \centering
    \includegraphics[width=0.8\textwidth]{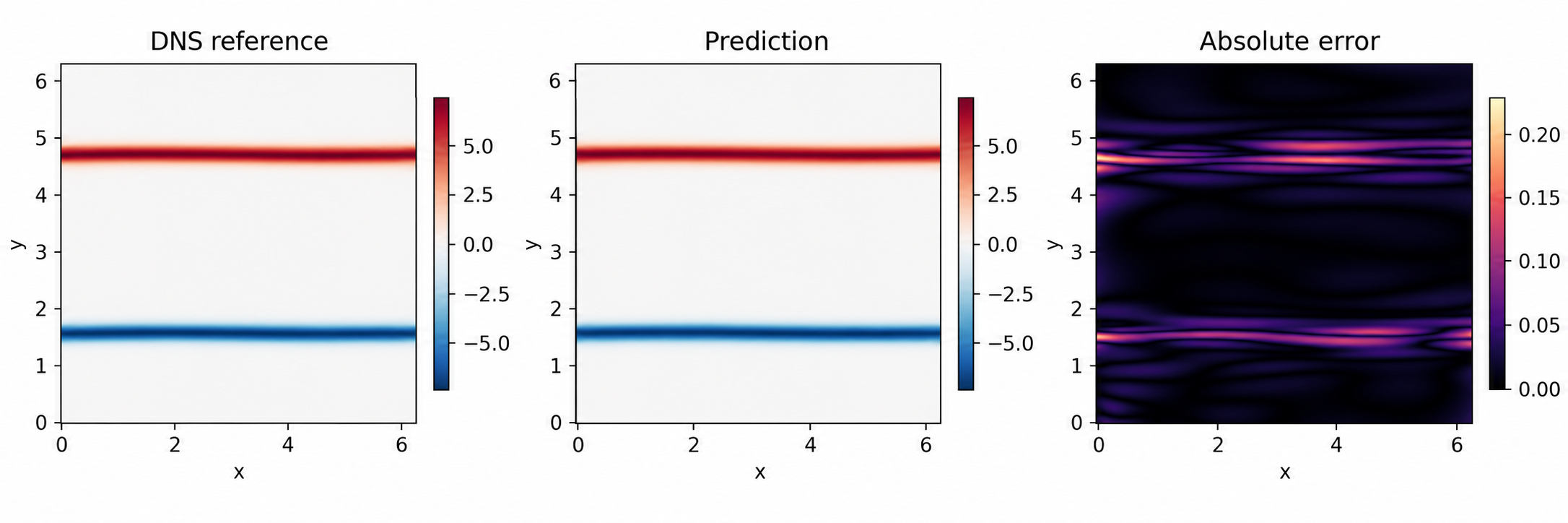}
    \caption{Double shear layer ($\rho=10$,
    $\nu=0.01$): DNS reference fields,
    BS-PINN-F predictions, and pointwise
    absolute errors at $t=0.25$ using
    the $4\times50$ architecture.}
\end{figure}

\begin{figure}[H]
    \centering
    \includegraphics[width=0.8\textwidth]{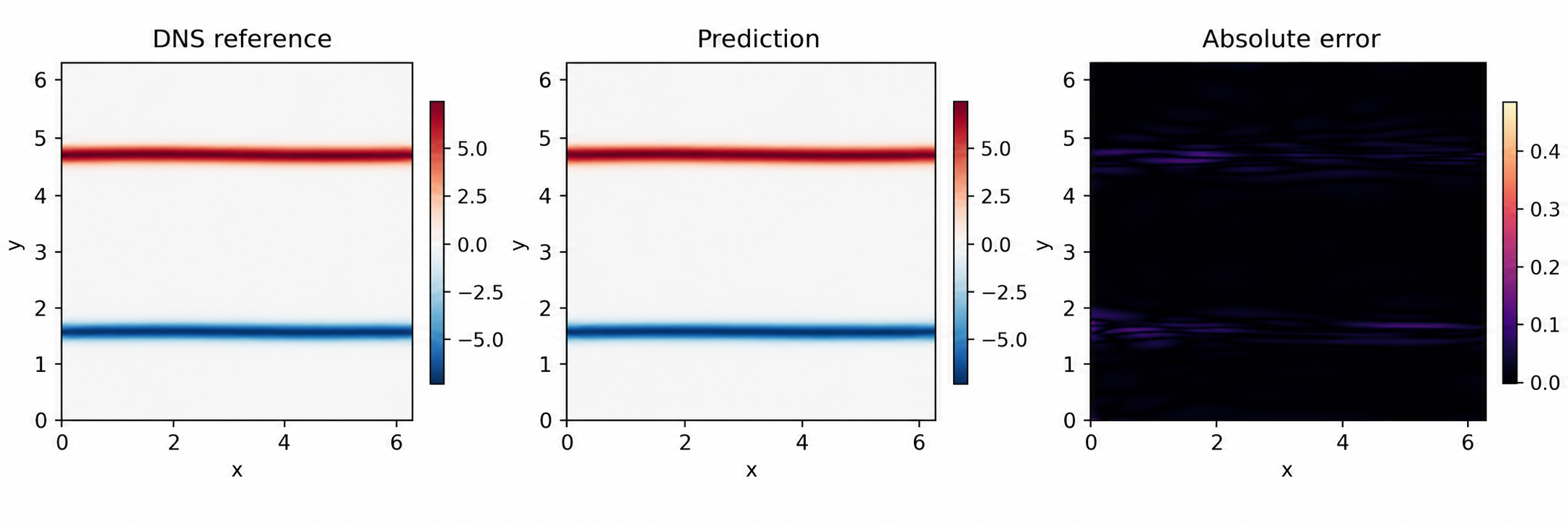}
    \caption{Double shear layer ($\rho=10$,
    $\nu=0.01$): DNS reference fields,
    BS-PINN-F predictions, and pointwise
    absolute errors at $t=0.25$ using
    the $7\times50$ architecture.}
\end{figure}

\begin{figure}[H]
    \centering
    \includegraphics[width=0.8\textwidth]{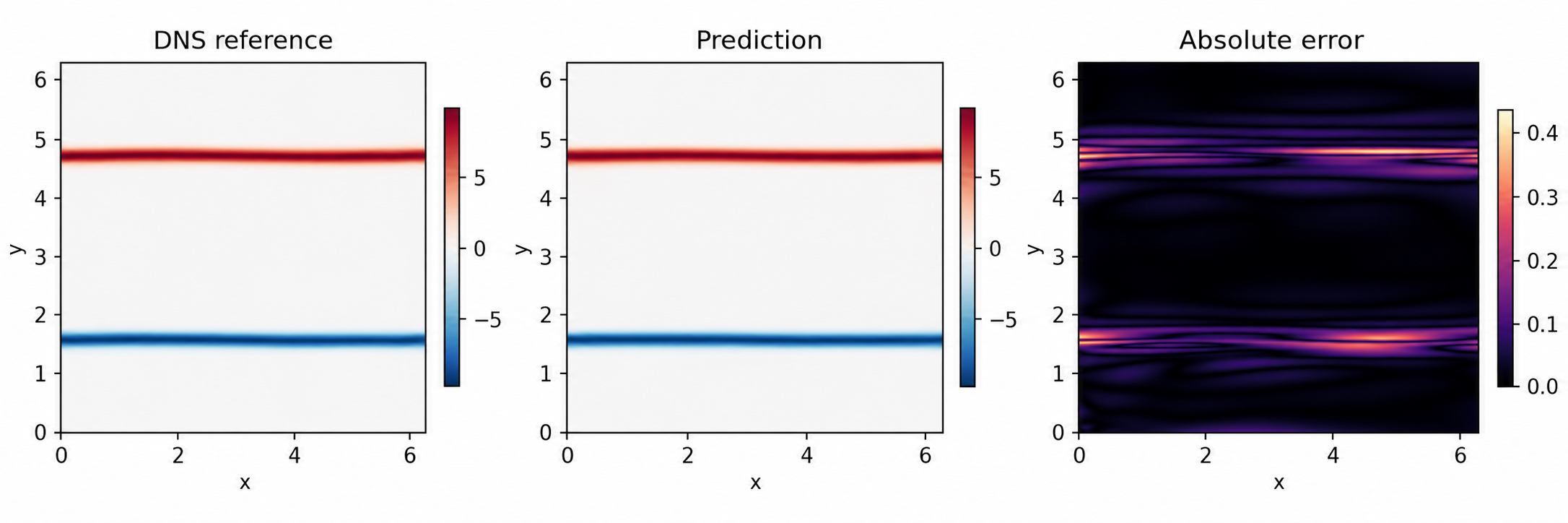}
    \caption{Double shear layer ($\rho=20$,
    $\nu=0.01$): DNS reference fields,
    BS-PINN-F predictions, and pointwise
    absolute errors at $t=0.25$ using
    the $4\times50$ architecture.}
\end{figure}

\begin{figure}[H]
    \centering
    \includegraphics[width=0.8\textwidth]{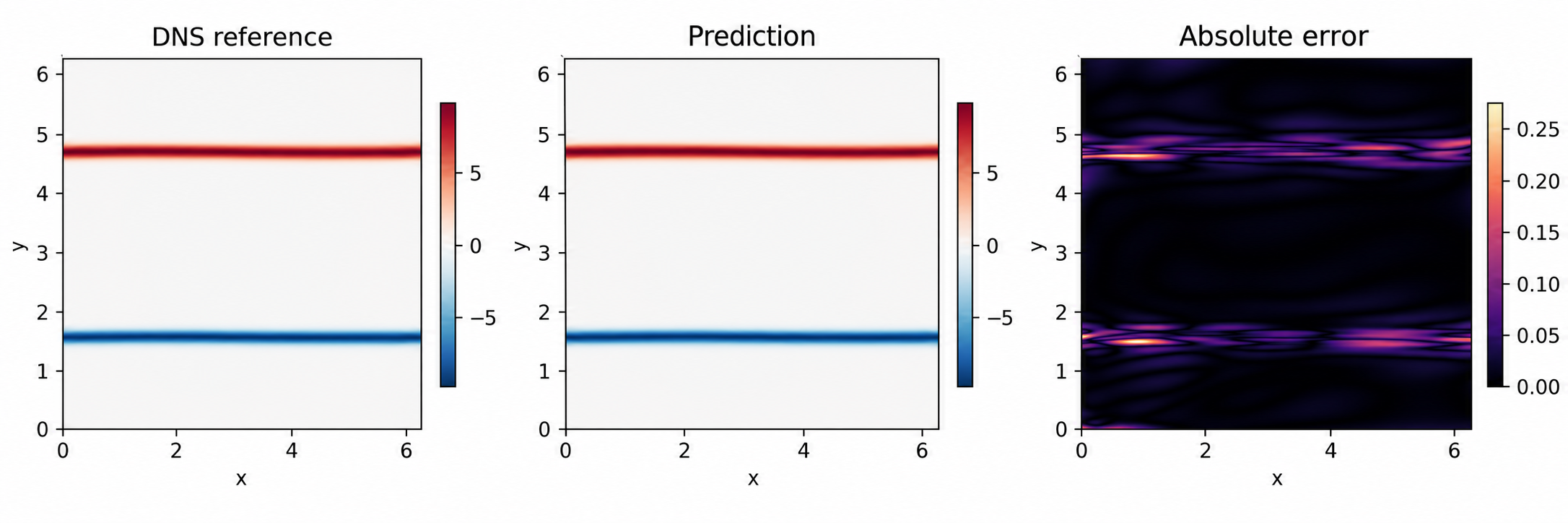}
    \caption{Double shear layer ($\rho=20$,
    $\nu=0.01$): DNS reference fields,
    BS-PINN-F predictions, and pointwise
    absolute errors at $t=0.25$ using
    the $7\times50$ architecture.}
\end{figure}

\begin{figure}[H]
    \centering
    \includegraphics[width=0.8\textwidth]{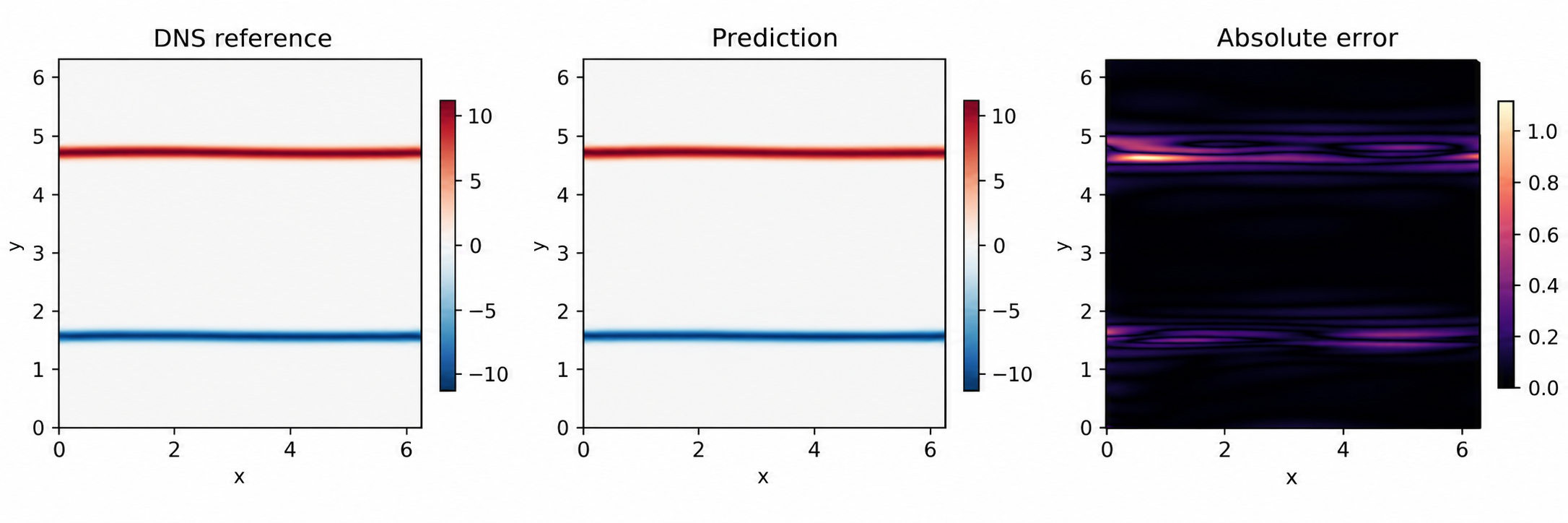}
    \caption{Double shear layer ($\rho=40$,
    $\nu=0.01$): DNS reference fields,
    BS-PINN-F predictions, and pointwise
    absolute errors at $t=0.25$ using
    the $4\times50$ architecture.}
\end{figure}

\begin{figure}[H]
    \centering
    \includegraphics[width=0.8\textwidth]{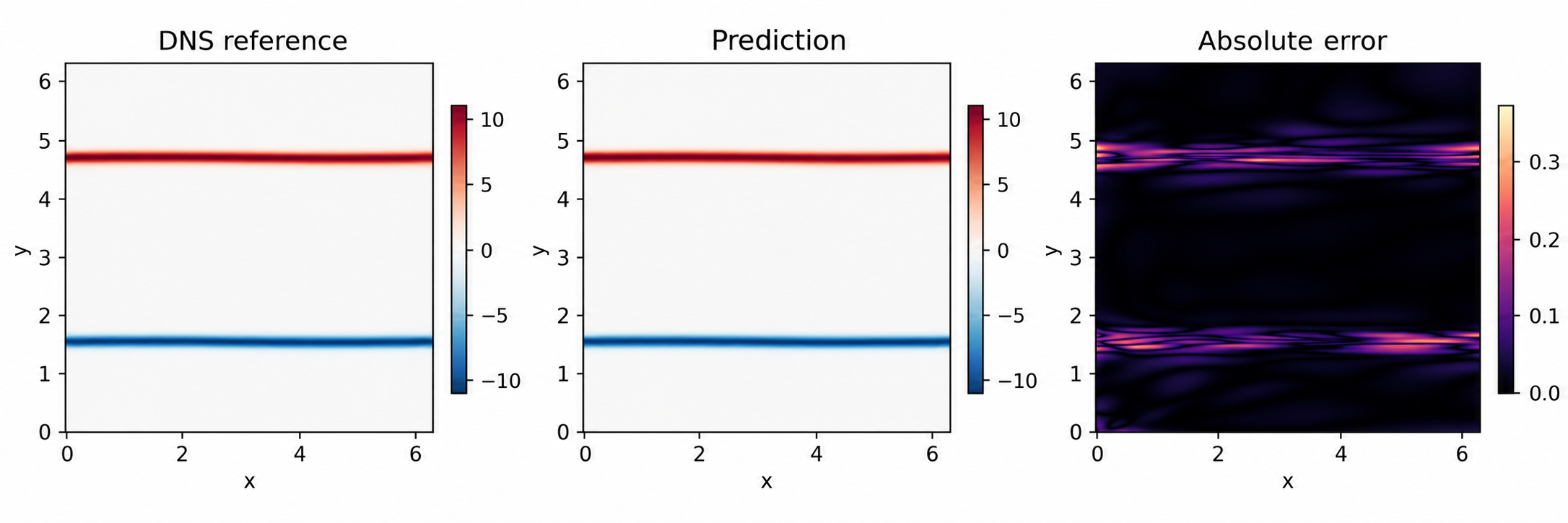}
    \caption{Double shear layer ($\rho=40$,
    $\nu=0.01$): DNS reference fields,
    BS-PINN-F predictions, and pointwise
    absolute errors at $t=0.25$ using
    the $7\times50$ architecture.}
\end{figure}

\section*{Declaration of generative AI and AI-assisted technologies in the manuscript preparation process}

During the preparation of this manuscript, the authors used ChatGPT (OpenAI) to assist with language editing and to generate the schematic illustrations of the five benchmark flow configurations in Table~\ref{tab:benchmark_overview}. These illustrations are intended solely for conceptual visualization and do not represent numerical simulation results. The authors reviewed and edited the AI-assisted content as necessary and take full responsibility for the final manuscript.

\end{document}